\documentclass[fleqn,usenatbib]{mnras}

\usepackage{newtxtext,newtxmath}

\usepackage[T1]{fontenc}

\DeclareRobustCommand{\VAN}[3]{#2}
\let\VANthebibliography\thebibliography
\def\thebibliography{\DeclareRobustCommand{\VAN}[3]{##3}\VANthebibliography}

\usepackage{graphicx}	
\usepackage{amsmath}	
\usepackage[dvipsnames]{xcolor} 
\usepackage[normalem]{ulem} 

\title[Double black holes in star clusters]{Dynamical double black holes and their host cluster properties}

\author[D. Chattopadhyay et al.]{
Debatri Chattopadhyay,$^{1,2,3}$\thanks{E-mail: dchattopadhyay@swin.edu.au}
Jarrod Hurley,$^{1,2}$
Simon Stevenson,$^{1,2}$
and Arihant Raidani$^{1,2}$
\\
$^{1}$Centre for Astrophysics and Supercomputing, Swinburne University of Technology, John St., Hawthorn, Victoria- 3122, Australia \\
$^{2}$ The ARC Centre of Excellence for Gravitational Wave Discovery,  OzGrav\\
$^{3}${Gravity Exploration Institute, School of Physics and Astronomy, Cardiff University, Cardiff, CF24 3AA, UK}
}

\date{Accepted XXX. Received YYY; in original form ZZZ}

\pubyear{2021}

\begin{document}
\label{firstpage}
\pagerange{\pageref{firstpage}--\pageref{lastpage}}
\maketitle

\begin{abstract}
We investigate the relationship between the global properties of star clusters
and their double black hole (DBH) populations.
We use the code {\tt NBODY6} to evolve a suite of star cluster models with an initial mass
of $\mathcal{O}(10^4)$M$_\odot$ and varying initial parameters. 
We conclude that cluster metallicity plays the most significant role in determining the lifespan of a cluster, while the initial half-mass radius is dominant in setting the rate of BH exchange interactions in the central cluster regions. 
We find that the mass of interacting BHs, rather than how frequently their interactions with other BHs occur,
is more crucial in the thermal expansion and eventual evaporation of the cluster. We formulate a novel approach to easily quantify the degree of BH-BH dynamical activity in each model.
We report 12 in-cluster and three out-of-cluster (after ejection from the cluster) DBH mergers, of different types (inspiral, eccentric, hierarchical) across the ten $N$-body models presented.
Our DBH merger efficiency is 3--4$\times10^{-5}$ mergers per M$_\odot$. 
We note the cluster initial density plays the most crucial role in determining the number of DBH mergers, with the potential presence of a transitional density point (between 1.2-3.8$\times10^3$M$_\odot$/pc$^3$) below which the number of in-cluster mergers increases with cluster density and above which the increased stellar density acts to prevent in-cluster BH mergers.
The importance of the history of dynamical interactions within the cluster in setting up the pathways to ejected DBH mergers is also discussed. 
\end{abstract}

\begin{keywords}
star cluster -- binary black hole -- gravitational waves
\end{keywords}



\section{Introduction}
\label{sec:introduction}

The three science runs by the advanced Laser Interferometer Gravitational-wave Observatory
\citep[LIGO;][]{TheLIGOScientificDetector:2014jea} and Virgo \citep{TheVirgoDetector:2014hva} since 2015 have catalogued in total over 50 binary black hole merger events \citep{LIGOScientific:2018mvr, Abbott_GWTC2:2020niy,TheLIGOScientificCollaboration:2021arXivGWTC3}. 
Analysis of the observed population of double black holes leads to a better understanding of their mass spectrum, spin magnitude and alignment and merger rate through cosmological history \citep{Abbott_Pop_GWTC2:2020gyp,TheLIGOScientificCollaboration:2021arXivGWTC3Pop}. 
The mass and spin spectra of these double black hole mergers can provide an indication of different evolutionary pathways for these mergers \citep{ Abbott_Pop_GWTC2:2020gyp, Bouffanais:2021wcr}, whether that be through isolated binary evolution \citep{Belczynski:2007,StevensonNature:2017tfq}, stellar triples \citep{Silsbee:2016djf,Antonini:2017ash} or quadruples \citep{Liu:2018vzk, Fragione:2020aki}, in dense stellar environments such as star clusters \citep{Banerjee:2010, RodriguezChatterjee:2016, Banerjee:2021} or in galactic nuclei \citep{ Antonini:2012ad, Stone:2016wzz}. 
A combination of observational evidence, theory and results of simulations exploring the various scenarios is required to develop the full picture of these merger histories. 

Very massive stars (approximately $100$\,M$_\odot$ or more on the zero age main sequence) are believed to become unstable to electron-positron pair production, which can lead to the star either being completely destroyed, leaving behind no remnant, or can, after a series of pulses, lead to the formation of a black hole (BH) less massive than roughly 40--50\,M$_\odot$ (with uncertainties) through core collapse \citep[see][for more discussion]{Fowler:1964, Bond:1984, Fryer:2000my, Heger:2001cd, WoosleyHegerWeaver:2002}. 
This phenomenon creates a dearth of stellar mass BHs between about 50--130\,M$_\odot$, termed as the (pulsational) pair instability or (P)PISN mass gap  \citep{WoosleyNature:2007, Belczynski:2017gds, Stevenson:2019rcw, Farmer:2020xne}. 
However, recent LIGO/Virgo observations of pre-merger BHs over 50\,M$_\odot$ has resulted in renewed analysis of the (P)PISN mass gap \citep{Abbott_GWTC2:2020niy,Wang:2021mdt, Baxter:2021swn, Edelman:2021fik}. 
In the dynamical formation channel (in dense systems such as star clusters and the disks of active galactic nuclei), more massive BHs can form through multiple generations of BH mergers, where a smaller BH merges with a companion, forming a more massive merger product.
This pathway can explain the occurrence of massive BHs in the (P)PISN mass gap \citep{Fishbach:2017,  Gerosa:2017kvu, Rodriguez:2017pec, Rodriguez:2019huv,  Kremer_multiple_gen:2020}. 
In particular, the observation of GW190521 \citep{Abbott_IMBH:2020tfl,Abbott:2020mjq}, a binary BH merger leading to the formation of a $\sim 140$\,M$_\odot$ intermediate mass BH (IMBH) remnant, has sparked further discussion on the possible formation scenarios.
Depending on the post-merger recoil kick, the remnant BH may get ejected out of the host cluster \citep{Merritt:2004xa, Herrmann:2007, HolleyBockelmann:2007eh, Abbott:2020mjq} 
or may possibly be retained in the cluster. 
The likelihood of the latter outcome is boosted by the belief that dynamical encounters lead to a BH spin distribution that is isotropic and biased towards low magnitude spins. 
For a merger involving BHs with low spin, 
the recoil velocity of the merger remnant can be lower than the escape velocity of a massive host cluster ensuring the BH will be retained inside \citep[see for e.g.][]{Miller:2002vg, RodriguezSpin:2016, Antonini:2019, Belczynski_Banerjee:2020bnq}{}. 
These in-cluster retained merger remnant BHs can not only occupy the void of the (P)PISN mass gap \citep{Gerosa:2017kvu, Fishbach:2017} but also undergo further mergers to become more massive \citep{Rodriguez:2019huv}. 

To further add to the mix, it has long been discussed that the Kozai-Lidov mechanism in three-body systems or binary-binary/single interactions can produce compact binary mergers at a higher eccentricity \citep{Wen:2002km, Antonini:2013tea, Samsing:2017rat, Rodriguez:2017pec, Martinez:2020} than through isolated binary evolution. 
Hence eccentric mergers are expected to be signatures of a dynamical environment. 
Evidence for GW190521 being a merger with significant eccentricity \citep[$e > 0.1$ at 10 Hz;][]{Gayathri:2020coq, Romero-Shaw:2020thy} has further increased the significance of studying BH mergers in dense stellar systems. 

The spin distribution of the binary BHs obtained from gravitational waves observations have long been expected to help infer the formation channels of such events, with increased detections finally assisting to resolve the relative abundance of mergers through the proposed dynamical vs. isolated channels \citep{Vitale:2015tea, Stevenson:2017dlk, Farr:2017uvj}. 
While dynamically formed BH pairs are predicted to have an isotropic spin distribution \citep{Sigurdsson:1993, Rodriguez:2015oxa, RodriguezSpin:2016}, isolated binary evolution is predominately expected to have the individual BHs spins aligned  \citep{Marchant:2016wow, StevensonNature:2017tfq, Gerosa:2018wbw} or in specific cases mis-aligned \citep[e.g.,][]{OShaughnessy:2017eks,Stegmann:2020kgb} with the binary orbital angular momentum. 
The analysis of the three LIGO/Virgo observing runs \citep{LIGOScientific:2018mvr, Abbott_GWTC2:2020niy,TheLIGOScientificCollaboration:2021arXivGWTC3Pop} rules out both extremes of perfectly aligned double BH spins and a completely isotropic spin distribution. 
Though there appears a slight preference for aligned spins, evidence of spin mis-alignment is also present \citep{Abbott_GWTC2:2020niy}, with about 12--44\% of binary BHs having their spins tilted by $>90^{\circ}$.
The third LIGO/Virgo observing run has shown indications of the presence of components of spin in the binary orbital plane, and hence binary BH spin precession about the orbital angular momentum \citep{Abbott_GWTC2:2020niy,Abbott_Pop_GWTC2:2020gyp,TheLIGOScientificCollaboration:2021arXivGWTC3Pop}. 
These two factors further emphasize the importance of the dynamical formation channel of binary BH mergers. 
The pre-merger progenitor LIGO/Virgo BHs have also shown evidence of small spin magnitudes \citep{Bavera:2020, Belczynski:2017gds}.

These observations have allowed the relative contributions of dynamical and isolated binary channels to be constrained. 
\citet{Bouffanais:2021wcr}, using a combination of young star cluster and isolated binary models, found 57--82\% of the LIGO/Virgo observations originating from dynamical sources, 
while \citet{Rodriguez:2021} conclude that the merger rate of double BHs from GWTC-2 LIGO/Virgo observations is entirely explainable through formation in dense star clusters.
Based on the observation of two eccentric binary BH mergers, \citet{Romero-Shaw:2021ApJL} argue that $\gtrsim 27\%$ of binary BHs may be formed dynamically.
\citet{Safarzadeh:2020ApJ} argued that dynamical formation contributes more than 50\% of LIGO/Virgo observations based on modelling of the distributions of BH spins.
Furthermore, \citet{Fragione:2021hhl} find that all but GW190521 can form in young open clusters. 
On the other hand, \citet{Tagawa:2021ofj} suggested the dominance of active galactic nuclei disks and nuclear star clusters in producing these observations rather than globular clusters. 
However, it is more likely that the entire population of the observed binary BH mergers contain contributions from multiple independent channels \citep[e.g.,][]{Bouffanais:2021wcr,Zevin_channel:2021}.
Future third generation gravitational wave detectors like the Cosmic Explorer \citep{CosmicE:2016mbw, Reitze:2019iox} and Einstein Telescope \citep{Punturo:2010zz, Broeck:2013rka, Maggiore:2019uih} are expected to shed more light on different formation channels as well as different generations of BH mergers \citep{Ng:2020}.

Exploring the dynamical formation channel---especially within star clusters---hence remains important in understanding the LIGO/Virgo observations of double BHs.
The dependence of in-cluster binary BH mergers on the cluster global properties has been studied using both Monte-Carlo and {\tt NBODY} simulations, which are generally in agreement with each other \citep{Rodriguez_MonteCarlo_NBODY:2016}. 
Even with multiple associated uncertainties pertaining to supernovae natal kicks and dynamical recoil kicks, more recent studies expect a larger fraction of BHs retained in star clusters \citep{2012Natur.490...71S, Morscher:2015, Rodriguez:2015oxa}, contrary to previous expectations of most BHs being ejected out \citep[e.g.][]{Sigurdsson:1993}. The dynamical age of the star cluster is especially important in regards to the
BHs. While young clusters may have a significant fraction of BHs, as the cluster matures to a post-core-collapse state the continuous evaporation of BHs leads to fewer BHs being retained inside the cluster. 
\citet{Banerjee:2010} showed with a host of {\tt NBODY6} cluster models with initial masses $\leq10^5$\,M$_\odot$ that intermediate-age star clusters form the ideal hub for double-BH mergers in contrast with old ($>4$\,Gyr)  globular clusters where most BH-BH mergers have already occurred or clusters which are too young ($<50$\,Myr) which are yet to achieve peak dynamical activity. 
The escaped binary BHs with significant eccentricity (a signature of the dynamical environment) at the moment of leaving the cluster can also merge within a Hubble time. 
Such out-of cluster mergers can still be considered of dynamical origin \citep{Anagnostou:2020eff}. 
It has also been noted that clusters with lower metallicity not only form more massive BHs \citep{Mapelli_IMBH:2016vca, Banerjee:2017}, but enhanced dynamical interactions further increase the rate of BH mergers hence creating even more massive BHs than through isolated binary evolution \citep{Rodriguez:2015oxa, DiCarlo:2019fcq}. 
Furthermore, as well as the cluster properties impacting the nature of the BHs produced, it is also found that the 
dynamical interactions of these stellar-mass BHs can affect the host cluster evolution \citep{Chatterjee_CoreCollapse:2013, Kremer_CoreCollapse:2019, Antonini:2019ulv}. 
Of the many uncertainties associated with binary stellar evolution, \citet{Chatterjee:2017A} concluded that although variations of the initial properties of the cluster
affected the number of BHs retained in the cluster, which in turn affected the cluster evolution lifetime, 
it mostly left the double BH dynamics unaffected (with the exception of assumptions on the initial mass fraction and stellar winds). 
It can hence be said that to fully understand the dynamical channel of double BH mergers, it is necessary to understand the properties of the star clusters which are ideal to host dynamically active populations of BHs.

In this paper, we explore the effect of a cluster's initial properties---metallicity, half-mass radius and initial binary semi-major axis distribution---on the evolution of the cluster, its double BH population and dynamical interactions within the cluster through a set of {\tt NBODY6} models.
In section.~\ref{sec:simulations} we describe our primary suite of models and their initial parameters.
In section.~\ref{sec:cluster_global_properties} we analyse the typical evolution of our base model, and investigate how the evolution of the cluster is affected by the initial choice of parameters. 
In section.~\ref{sec:black_hole_populations}, we describe some of the details of the populations of compact objects (black holes and neutron stars) in our cluster models.
We discuss the details of the double BH systems, including both their dynamics within the cluster, as well as their binary properties (mass, orbital eccentricity etc.) in section.~\ref{sec:double_black_holes}. 
We also describe double BH binaries which merge outside of the cluster mergers. 
We describe the details of several different double BH mergers and their remnant properties (mass, spin and recoil kick) from our models in section.~\ref{sec:in_cluster_mergers}. 
We summarise our findings in section.~\ref{sec:discussion} and consider some avenues for future studies. 

\begin{table*}
    \centering
    \hspace{-2.8cm}
    \resizebox{0.9\textwidth}{!}{\begin{minipage}{\textwidth}
    \begin{tabular}{lrrrrrrrrrrr}
    \hline
    Model &  Initial Mass & Total Stars. & Total Binaries  & Half-mass radii & Metallicity & Semi-major axis & Tidal ratio & Time Evolved & \% remaining & \multicolumn{2}{c|}{DBH mergers} \\
    (ID) & (M$_\odot$)& - & - & (pc) & (Z) & (distribution) & - &  (Myr) & - & (in-cluster) & (ejected)\\
    \hline
M-01  &  35503.25 & 55000 & 5000 & 3.08 & 0.01 & Sana12  & 0.068 &   9920 & 0.54 & 1 &0\\
M-02  & 34872.23 & 55000 & 5000 & 3.08 & 0.001 & Sana12  & 0.067 & 6361  & 0.78 & 0&0\\
M-03  &  35054.96 & 55000 & 5000 & 3.08 & 0.0005 & Sana12  & 0.066 & 5440 & 1.05 & 1&0\\
M-04a  &  35063.22 & 55000 & 5000 & 3.08 & 0.01 & DM91  &  0.067 & 11210  & 0.31 & 2&0\\
M-04b  &  34731.21 &55000 & 5000 & 3.08 & 0.01 & DM91  & 0.066 & 13000 & 5.36 & 1&0\\
M-05  &  35323.82 &55000 & 5000 & 1.54 & 0.01 & DM91  &   0.034 & 13000 & 8.35 & 4&0\\
\hline
M-06  &  6940.95 & 11000 & 1000 & 1.15 & 0.01 & Sana12  &  0.043 & 2472  & 1.22 & 0&0\\
M-07  &  48589.52 & 75000 & 5000 & 1.15 & 0.005 & DM91  &  0.022 & 9592  & 24.67 & 0&3\\
M-08  &  35195.93 & 55000 & 5000 & 1.54 & 0.01 & DM91  &  0.034 & 2782 &  49.45 & 3&0\\
M-09  & 35324.61 & 55000 & 10000 & 1.54 & 0.01 & DM91  & 0.033 & 403  & 70.68 & 0&0\\
M-10  &  35629.42 & 55000 & 5000 & 1.54 & 0.001 & DM91  &  0.034 & 2484   & 46.63 & 0&0\\
    \hline
    \end{tabular}
    \end{minipage}}
    \label{table:NBODY_models}
    \caption{{\tt NBODY6} models used in this work. `DM91' signifies the semi-major axis distribution prescription given by \citet{Duquennoy:1991}, while `Sana12' signifies the same from \citet{Sana:2012}. Tidal ratio signifies the initial ratio of the half-mass radius to the tidal radius and \% remaining =
    M$_\mathrm{f}$/M$_\mathrm{i}\times 100$, where M$_\mathrm{i,f}$ are initial and final cluster masses, respectively.}
\end{table*}

\section{Star cluster simulations}
\label{sec:simulations}

In this section, we outline our primary suite of models and their detailed specifications. 

\subsection{Models}
\label{subsec:models}

We present a set of eleven models simulated using the direct $N$-body code {\tt NBODY6} \citep{Aarseth:2003, 2012MNRAS.424..545N}. 
For these simulations we utilise the variant of {\tt NBODY6} referred to as {\tt NBODY7} \citep{Aarseth:2012MNRAS} 
that employs the Algorithmic Regularization method of \citet{Mikkola:1999MNRAS} for a consistent treatment of dynamically-formed multiple systems, as implemented (including post-Newtonian corrective terms) in \citet{Mikkola:2007ip}. 
In concert with modelling the $N$-body gravitational interactions, {\tt NBODY6/7} 
follows stellar \citep{Hurley:2000pk} and binary evolution \citep{Hurley:2002BSE} as described in \citet{Hurley:2000sz}. 
Each simulation is performed on a single node of the OzSTAR supercomputer at Swinburne University, making use of either a NVIDIA K10 or P100 graphics processing unit (GPU). 

Our focus is a primary ensemble of six models that each started with $N_\mathrm{tot} = 55\,000$ as the initial number of stars and $N_\mathrm{bin}=5\,000$ as the number of primordial binaries (meaning that $10\,000$ stars are binary members and the initial binary frequency is 10 per cent). 
These models are all evolved until they either reach an age of $13\,$Gyr, which we take as the approximate age of the universe, or they have less than 500 stars remaining. 
Whichever occurs first becomes the end-point. 
All of the models presented in this work are summarised in Table~\ref{table:NBODY_models} where M-01 is considered our base model 
and each of the subsequent five models in our primary set have one (or two) parameter(s) varied from M-01. 

The zero-age main sequence (ZAMS) masses of the stars, drawn from the \citet{Kroupa:2000iv} 
initial mass function (IMF), are in the mass range of 0.1--100\,M$_\odot$\footnote{Recently, a number of authors have examined the impact of uncertainties in the initial mass functions of star clusters (in particular, the possibility of a top-heavy initial mass function) on the evolution of star clusters and the populations of gravitational-wave sources they produce \citep{Haghi:2020ApJ,Weatherford:2021zdf,Wang:2021yzi}.}. 
For primordial binaries the drawn masses are combined to give the total binary mass and a mass-ratio is then selected from an uniform distribution before setting the component masses accordingly \citep{HurleySippel:2016}. This sampling method ensures that systems of $\geq$1 M$_\odot$ (since a primary of 0.1 M$_\odot$ cannot have a secondary of lower mass) have an uniform mass ratio ($q\leq$1) distribution. 
As a result, the initial cluster masses M$_\mathrm{i}$ for these models are $\approx 3.5 \times 10^4$ M$_\odot$, with slight variations induced due to sampling with different random number seeds. 

The initial positions and velocities of the cluster stars are initiated by assuming a density profile following  \citet{1911MNRAS..71..460P} and that the system begins in virial equilibrium \citep[for more details see][]{AarsethHW:1974}. 
The clusters are assumed to be orbiting in a 3-dimensional Milky Way-like  potential consisting of a point-mass bulge, an extended disc of uniform density \citep{MiyamotoNagai:1975} 
and a dark-matter halo of logarithmic potential \citep{Aarseth:2003}. 
The bulge and disc masses are taken to be $1.5 \times 10^{10} M_\odot$ and $5 \times 10^{10} M_\odot$ respectively, with disc scale-lengths $a = 4\,$kpc, $b = 0.5\,$kpc, as given by \citet{Xue:2008}. 
The halo is constrained by requiring that the combined mass of the bulge, disk and halo results in a circular velocity of $220\, {\rm km} \, {\rm s}^{-1}$ at $8.5\,$kpc from the Galactic centre \citep{Aarseth:2003}. 
Each of our model clusters is placed on a circular orbit in the Galactic plane, at a distance of 8.5\,kpc from the Galactic centre. 

The range of cluster metallicity, $Z$, explored in our models is $Z =0.01$, 0.001 and 0.0005, corresponding to $[\mathrm{Fe/H}]\approx -0.3$, -1.3 and -1.6, assuming solar metallicity to be $Z=0.02$. 
For our base model M-01 we take $Z = 0.01$ as typical of the upper end of the metallicity distribution of star clusters, represented by metal-rich Milky Way clusters including Liller 1 \citep{Stephens:2003gf}, NGC 6440 \citep{Ortolani:1994} and Palomer 8 \citep{Heitsch:1999},  
for example, as listed in \citep{Harris:2010}. 
Next, for M-02 we reduce the metallicity by a factor of 10 to $Z = 0.001$ to represent clusters of intermediate range metallicity, such as BH 261, NGC 6558 and Terzan 7 with $[\mathrm{Fe/H}]\approx-1.30$ \citep{Harris:2010}. 
Then for M-03 we use $Z = 0.0005$ guided by metal-poor 
clusters\footnote{Recently \citet{Larsen:2020Science} reported the most metal-poor globular cluster observed to date, with [Fe/H] = -2.9 so M-03 should more correctly be referred to as representative of a relatively metal-poor cluster.
}, 
with NGC 1904, Palomer 14 and NGC 5986 as some examples 
\citep{Harris:2010}. 
Given that our primary focus is investigating the effect of cluster global properties on the core BH dynamics 
and that lower $Z$ clusters are expected to produce more massive BHs \citep{Hurley:2000pk,Belczynski:2010} this motivates the metallicities chosen for models M-02 and M-03. 
However, it is interesting that the massive ($\approx 10^6$ M$_\odot$) and metal-rich cluster Liller 1 shows evidence of high dynamical activity through the observation of excessive gamma emission and is postulated to host a population of millisecond pulsars as a result \citep{Saracino:2015}. 
Thus we shouldn't rule out the possibility of interesting results from our models with higher $Z$.

The majority of our models have the initial half-mass radius, $R_\mathrm{h}$, set to be approximately $3\,$pc. 
However, to study the effect of the stellar density on the resultant BH binary population, we have evolved model M-05 with a smaller initial half-mass radius of $R_\mathrm{h} \sim 1.5\,$pc. In Table~\ref{table:NBODY_models} we also show the ratio of the initial half-mass radius to tidal (or Jacobi) radius, calculated as in \citet{Madrid:2012}, and it can be seen that all of our models start as tidally under-filled.
The issue of sampling noise has been investigated by two models M-04a and M-04b with different seeds for the random number generator used in {\tt NBODY6} while all other parameters remain the same 
(see Table~\ref{table:NBODY_models}). 
We utilise two orbital period distributions for the primordial binaries of our models: 
the \citet{Duquennoy:1991} distribution (referred to as ``DM91" hereafter and used in models M-04a, M-04b and M05) 
and the \citet{Sana:2012} distribution (referred to as  ``Sana12"  hereafter and used in models M-01, M-02 and M-03). 
For all models we set the eccentricities of the binaries following the process described in \citet{GellerH:2013} which is guided by observations of the young open cluster M35 \citep{Meibom:2005}. 

As shown by \citet{Sana:2012}, about 56\% of O-stars in their sample of 71 Milky Way systems used to constrain their distribution are in a binary, and using the Sana12 distribution will increase the massive star binary fraction (relative to DM91) by decreasing their intrinsic orbital separation (since wider binaries are easily disrupted). 
As O-stars are typical progenitors of BHs, a higher proportion of close massive primordial binaries can be expected to affect the resultant dynamics of the cluster. 
We can investigate this by comparing models M-01 and M-04a (or M-04b) which have the same setup aside from the orbital period distribution. 

We have evolved some additional models to help investigate the cluster evolutionary trends further, focusing on the early evolution when the cluster BH population is forming. 
One parameter we investigate here is in the initial $N$ of the models: 
M-06 starts with $N_\mathrm{tot} = 11\,000$ (a factor of five less than for M-01) while M-07 starts with $N_\mathrm{tot} = 75\,000$ 
(and also a smaller half-mass radius). 
With model M-09 we look at the effect of an increased primordial binary fraction, starting with $N_\mathrm{bin}=10\,000$. 
These models are also summarised in Table~\ref{table:NBODY_models} and we will discuss their particulars in more detail when they are utilised within the text. 
In general we were not concerned with evolving these models to completion, although M-06 with the smaller $N$ did evolve quickly to completion.  
M-08 is identical to M-05 with a different seed and only evolved to about 50\% of its mass remaining. 
Model M-10 is the same as for models M-05 and M-08 aside from a lower metallicity of $Z = 0.001$. 

\subsection{Wind mass loss and remnant prescriptions}
\label{subsec:winds_and_sne}

Compared to the stellar evolution algorithm described in \citet{Hurley:2000pk} our models take advantage of two important updates. 
Firstly, compact object masses are assigned according to the prescription from \citet{Belczynski:2008mr}. Secondly, for mass-loss from stellar winds we use the updated model by \citet{Vink:2001} described in \citet{Belczynski:2010} which is referred to as the {\it Vink et al. model}. 
The updated mass loss for massive stars rich in hydrogen has the form of metallicity dependant power law as $\Dot{M} \propto Z^{\alpha}$, with $\alpha$ being a function of the effective temperature of the star. 
For lower mass stars we follow the mass loss prescription described in \citet{Hurley:2000pk}.
The \citet{Vink:2001} model reduces the wind mass loss for stars with lower metallicity.  
While the remnant masses for stars with ZAMS mass less than 30\,M$_\odot$ do not change through the \citet{Vink:2001} prescription, stars with higher masses produce more massive pre-supernova objects, experiencing a larger fallback and hence becoming heavier BHs \citep{Belczynski:2010}. 
The stellar and binary evolution updates that have been made to {\tt NBODY6} and {\tt NBODY7} are also described in \citet{Banerjee:2019jjs}

\begin{figure}
    \centering
    \includegraphics[width=0.35\textwidth]{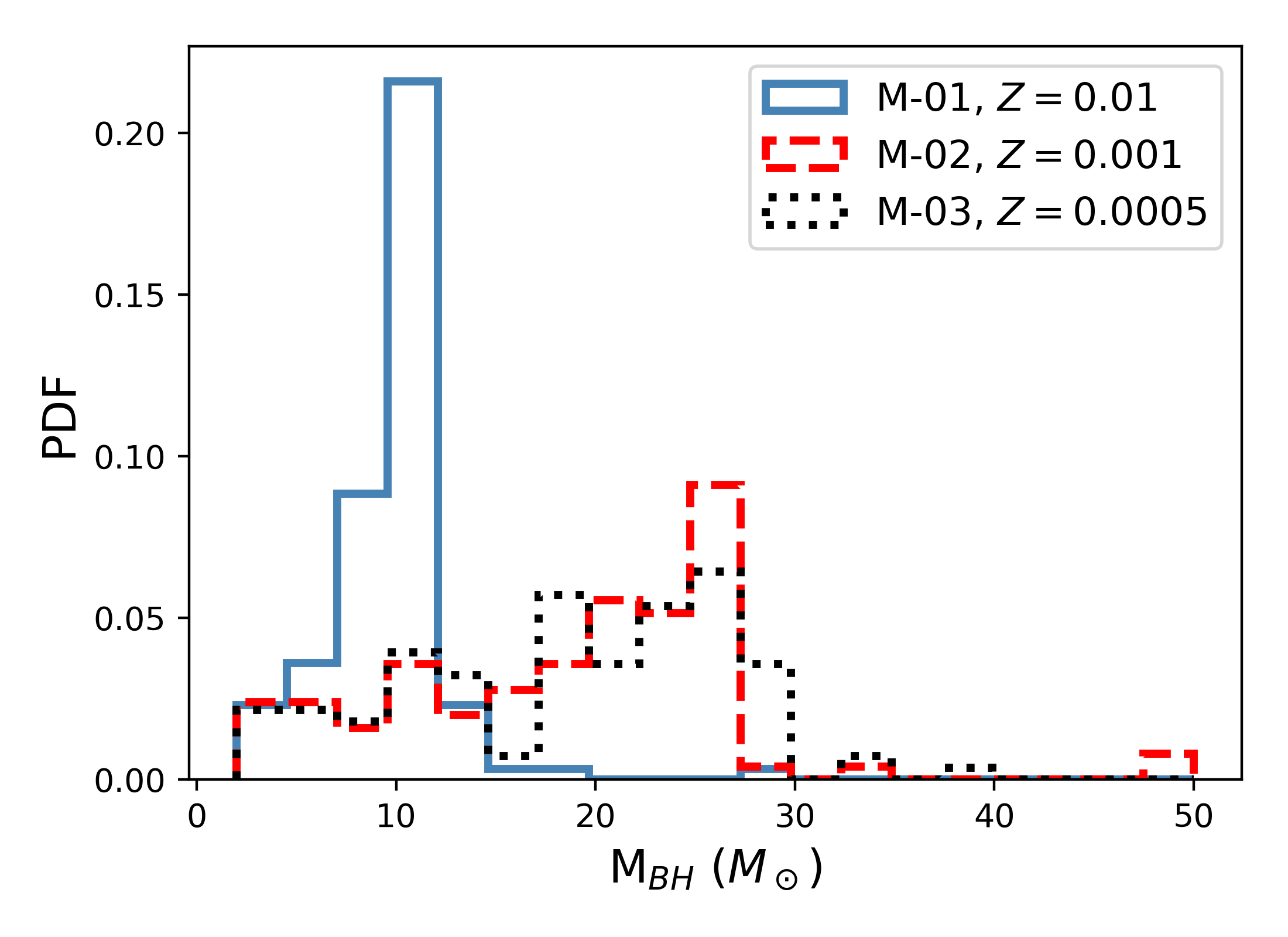}
    \caption{The initial mass distribution of BHs formed from single and binary evolution in models M-01, M-02 and M-03 with metallicities $Z=0.01,0.001,0.0005$, respectively. }
    \label{fig:bh_ini_mass}
\end{figure}

The initial mass distribution of the BHs right after formation through binary and stellar evolution for different metallicities for our three primary models is shown in Fig.~\ref{fig:bh_ini_mass}. The peak of the distribution gets biased towards more massive BHs for lower metallicities.

Compact object progenitors are usually expected to go through a phase of supernova explosion prior to the formation of the remnant. 
The asymmetry of the explosion mechanism results in the remnant receiving some amount of recoil kick, especially for neutron stars \citep{GunnOstriker:1970, Helfand:1977, Lyne:1994}. 
Though the expected neutron star retention fraction for massive globular clusters with $\approx10^6$ initial stars is about 10--20\%,  \citet{Pfahl:2001df} used Monte Carlo simulations in which the neutron star kicks were drawn from either a Maxwellian or Lorentzian \citep{Paczynski:1990} distribution to show that typically $\lesssim 10$\% of the neutron stars are retained in  clusters under these assumptions.
Hence, there is a need to revise the neutron star kick prescription in clusters. 
The neutron star natal kick can be as high as 1000\,km/sec \citep{Arzoumanian:2002,Chatterjee:2005ApJ}, though \citet{Verbunt:2017zqi} show that a significant fraction of neutron stars may have velocity kicks $< 60$\,km/sec. 
It is generally accepted that the core collapse supernova mechanism has on an average a higher natal kick  distribution \citep[Maxwellian $\sigma \approx 265$ km/sec from][]{Hobbs:2005}{} as discussed in \citet{Fryer:2012}
compared to electron capture \citep{Nomoto:1984, Podsiadlowski:2004, Gessner:2018ekd} and ultra stripped supernovae \citep{Tauris:2013, Tauris:2015, Muller:2018utr}  with typically velocities $< 30$\,km/sec \citep{Pfahl:2002, Podsiadlowski:2004, Suwa:2015saa}. 
On the other hand, observations of Galactic low mass X-ray binaries shed some light on the possible BH birth kicks which are expected to be lower than for neutron stars \citep{Repetto:2012, Repetto:2015kra, Mandel:2015eta}.
Uncertainties still remain in correlating their mass and orbital properties to the ejection mechanism (and hence kick) as discussed by \citet{Janka:2013hfa} and  \citet{Sukhbold:2015wba}. 
Though \citet{Belczynski:2001uc} showed a mass dependant BH kick distribution, \citet{Repetto:2012} found BH kicks to be similar to neutron stars. 

It is usually assumed that the BH kick is scaled down by the relative amount of material ejected during the supernova that falls back onto the proto-remnant \citep{Fryer:2012}, thus increasing the final mass of the remnant and decreasing its natal kick.

For simplicity, we do not differentiate between types of supernova kicks in our models. 
Combining the fact that we evolve small clusters with low escape velocities, together with the uncertainties of compact object kicks and their retention fraction in star clusters, we use
a flat distribution of the natal kick between 0--100\,km/sec for neutron stars, so that about 10\% of the neutron stars are retained in the cluster. 
The BHs in our models use the same kick distribution which is further lowered by fallback mass scaling. 

This means that for BHs where all of the ejected material is accreted back onto the BH (so that the resultant BH mass is the same as the mass of the star at the time of the supernova) the outcome will be a natal kick that is zero. This occurs for stars that had a carbon-oxygen core-mass of $7.6 \,$ M$_\odot$ or greater at the time of the supernova. For lower-mass stars the kick chosen from the flat distribution between 0--100\,km/sec is scaled linearly by the fractional amount of ejected mass that falls back onto the remnant. 
This scaling process is described further in \citet{Fryer:2012} and \citet{Banerjee:2019jjs}

\section{Cluster Global Properties}
\label{sec:cluster_global_properties}
In this section we examine the evolution of the global properties (e.g., mass and radius) of our clusters with time. We begin by focusing on our fiducial model M-01 (section~\ref{sec:model1}). 
We then in turn examine the role of the initial cluster metallicity (section~\ref{sec:metallicity}), 
the initial distribution of binary orbital periods (section~\ref{sec:porb}) and the initial half-mass radius (section~\ref{sec:Ini_Rh}) on the evolution of star clusters. We briefly describe some additional models in section~\ref{sec:add_mods}.

\subsection{The evolution of model M-01}
\label{sec:model1}

Model M-01 acts as our base model. 
It started with $45\,000$ single stars and $5\,000$ binaries, with a metallicity of $Z = 0.01$, a half-mass radius of $3.08\,$pc and used the Sana12 distribution to set the initial orbital periods of the binaries. 
The model was evolved to an age of just over $10\,$Gyr when less than 1\% of the $3.5 \times 10^4 \,  M_\odot$ initial mass remained 
(see Table~\ref{table:NBODY_models} for a summary of the model).

\begin{figure}\centering
\includegraphics[width=0.35\textwidth]{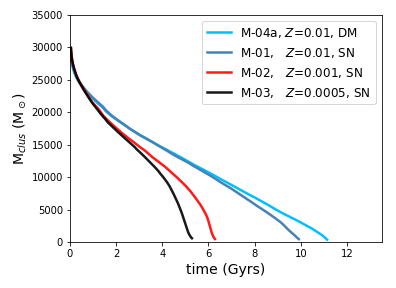}
\includegraphics[width=0.35\textwidth]{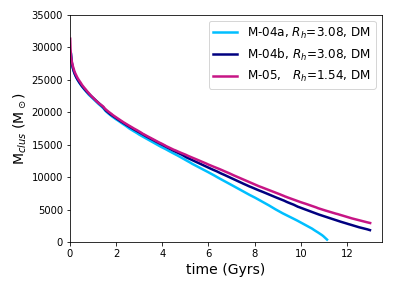}
\caption{
Total mass of the model clusters as a function of time. The impact of metallicity on the cluster lifetime
is depicted in the upper panel. The lower panel shows the mass evolution of models with varying initial cluster half-mass radii.}
\label{fig:Mclus_Grid}
\end{figure}
\begin{figure}\centering
\includegraphics[width=0.35\textwidth]{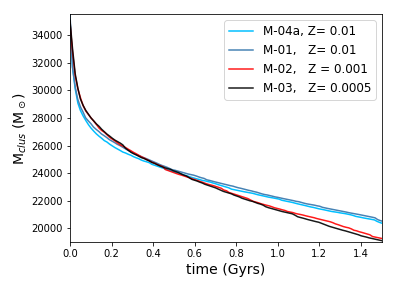}
\includegraphics[width=0.35\textwidth]{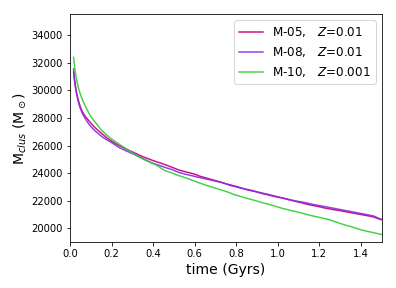}
\caption{
Total mass of the model clusters as a function of time, focused on the early evolution. The upper panel corresponds to the upper panel of Figure~\ref{fig:Mclus_Grid}. The lower panel shows some additional models in which the metallicity is varied (see Table~\ref{table:NBODY_models} for details).}
\label{fig:MclusIni_Zgrid1}
\end{figure}
\begin{figure}\centering
\includegraphics[width=0.45\textwidth]{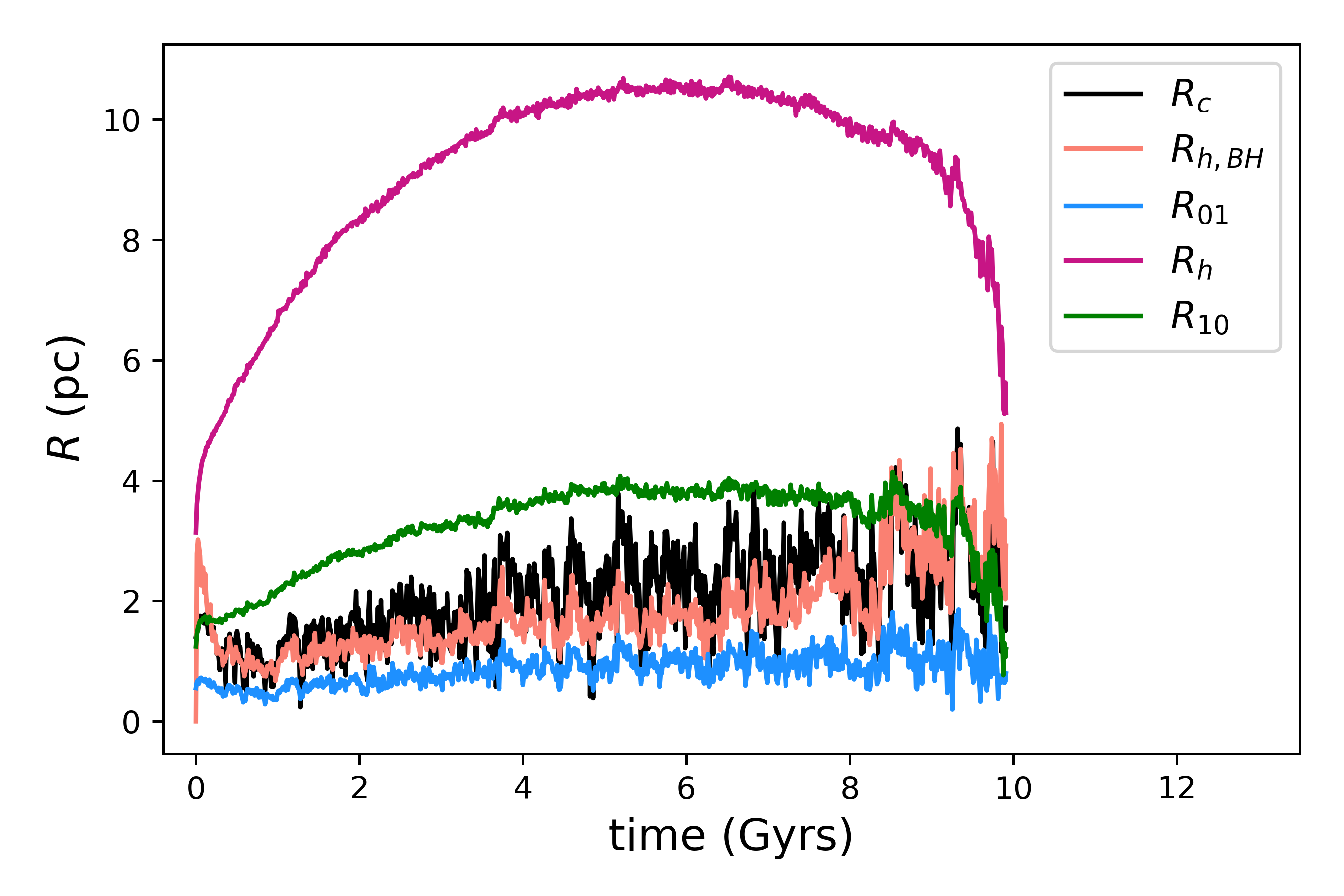}
\caption{Evolution of characteristic cluster radii including $R_c$ (core radius), $R_{h,\mathrm{BH}}$ (BH half mass radius), $R_{01}$ (1\% Lagrangian radius), $R_{10}$ (10\% Lagrangian radius) and $R_{h}$ (half-mass radius) with time for model M-01.}
\label{fig:R_M01}
\end{figure}

Figure~\ref{fig:Mclus_Grid} shows the evolution of the cluster mass ($M_\mathrm{clus}$) across the lifetime of the model and 
Figure~\ref{fig:MclusIni_Zgrid1} provides a close-up of the early evolution. 
The decrease of cluster mass throughout the evolution of the model is a combined effect of stellar evolution (stellar winds and supernovae), tidal stripping by the external Galactic potential and dynamical interactions imparting kinetic energy to the stars to escape the cluster. 
Depending on the stage of a cluster in its evolutionary pathway, one or more of these processes becomes the dominant factor for cluster mass loss \citep{Lamers:2010, Madrid:2012}. 
As expected, mass-loss owing to stellar evolution dominates the initial stage. 
For M-01 we find that 302 NSs and 120 BHs form within the first 100\,Myr with a combined $6\,040\, M_\odot$ of mass lost through stellar winds and supernovae. 
We also note that of these only 37 of the NSs are retained in the cluster at 100 Myr which means that a further $390\, M_\odot$ is lost from the cluster as a result of remnants that are ejected after receiving a velocity kick. 
All of the 120 BHs are retained 
(at least initially: the remnant properties of the clusters will be discussed in more detail in Sections~\ref{sec:black_hole_populations} and \ref{sec:double_black_holes}). 
Subsequent to this early phase of compact object formation, the dynamical evolution of the cluster becomes the primary driver for determining the loss of mass from the cluster, as we will discuss below. 

An ongoing process for a star cluster is that of energy equipartition driven by a quest for thermal equilibrium, a state which the cluster can strive for but never fully achieve in reality \citep{Giersz:1997}. 
The cumulative exchanges of energy between stars within this relaxation-driven process results in low-mass stars heating up and moving preferentially to the outer cluster regions while heavier stars sink towards the centre, which we observe as mass-segregation. 
 
In Fig.~\ref{fig:R_M01} we show the evolution of the 50\%, 10\% and 1\% Lagrangian radii (the radii of successive spheres containing a certain portion of the cluster mass), which are $R_{\rm h}$, $R_{10}$ and $R_{01}$ respectively, for M-01. 
We also show the density-weighted core radius, $R_{\rm c}$, commonly used in $N$-body simulations \citep{Casertano:1985} 
as well as the half-mass radius of the BH population, $R_{\rm h,BH}$ (which we will discuss later). 
We clearly see that the early phase of mass-loss dominated by stellar evolution produces an overall expansion of the inner regions of the cluster. 
This is most pronounced for $R_{\rm h}$ but also evident for $R_{10}$. 
The expansion of $R_{\rm h}$ is then sustained by the equipartition of energy process where the flow-on effect of mass-segregation is to create a density contrast and a tendency for the inner regions of the cluster to contract while the outer regions expand to conserve energy \citep{Meylan:1996yx}. 
This expansion also drives stars towards the tidal boundary of the cluster and results in the loss of mass as stars escape into the field of the host galaxy.  
As the cluster mass decreases with time so does the extent of the tidal boundary and eventually we see a turnover in $R_{\rm h}$ and a subsequent decrease. 
For model M-01 this turnover (or peak) in the $R_{\rm h}$ evolution occurs at an age of about $6\,$Gyr when the cluster has 30\% 
of the initial mass remaining. 
The continuous tidal and occasional shock stripping by the host galaxy becomes further accelerated as the cluster loses more mass 
and we see that the cluster moves quickly towards complete dissolution from about $9\,$Gyr onwards. 

In theory the removal of energy from the core leads to a phase of core collapse where $R_{\rm c}$ drops to small values and a deep gravitational potential well develops. 
This enhances dynamical interactions in the central regions and results in the formation of new binaries or the hardening of existing binaries 
\citep{Lemaitre:1955VA, Spitzer:1971, Aarseth:1974} which provides 
an energy source to halt the collapse and even cause a period of core expansion \citep{Heggie:1975, Hut:1983}. 
As such, the gravitational activity and the pattern of thermal energy that is generated can result in a series of $R_\mathrm{c}$ expansions and contractions. 
We see this oscillatory and even noisy behaviour of $R_\mathrm{c}$ 
in Fig.~\ref{fig:R_M01} and there are signs of local minima (near $1.5\,$Gyr for example). 
However, we do not see any sign of a clear deep core-collapse phase. 
In fact we would have to say that on average $R_\mathrm{c}$ increases over time. 
This is also true of $R_{01}$ which is less noisy in behaviour but shows no signs of a significant contraction phase. 
The lack of a clear core-collapse phase is not unexpected for models with a substantial primordial binary population that can provide a centralised heating source from the beginning of the evolution \citep{Vesperini:1994,Heggie_Hut:2003, Fregeau:2007, Chatterjee_CoreCollapse:2013, Kremer_CoreCollapse:2019}.  
This is accentuated in our models by a sizeable BH population that quickly sinks towards the centre of the cluster (if the progenitor massive stars were not there already) and pushes low-mass main-sequence stars and white dwarfs outwards. 
In Fig.~\ref{fig:R_M01} we see that the half-mass radius for the BHs sits well within $R_{10}$ and even inside $R_\mathrm{c}$. 
Thus the core becomes a hub of activity for these massive bodies such that the action of single and binary BHs inflates the central regions and restricts the possibility of deep collapse. 

\begin{figure}\centering
\includegraphics[width=0.35\textwidth]{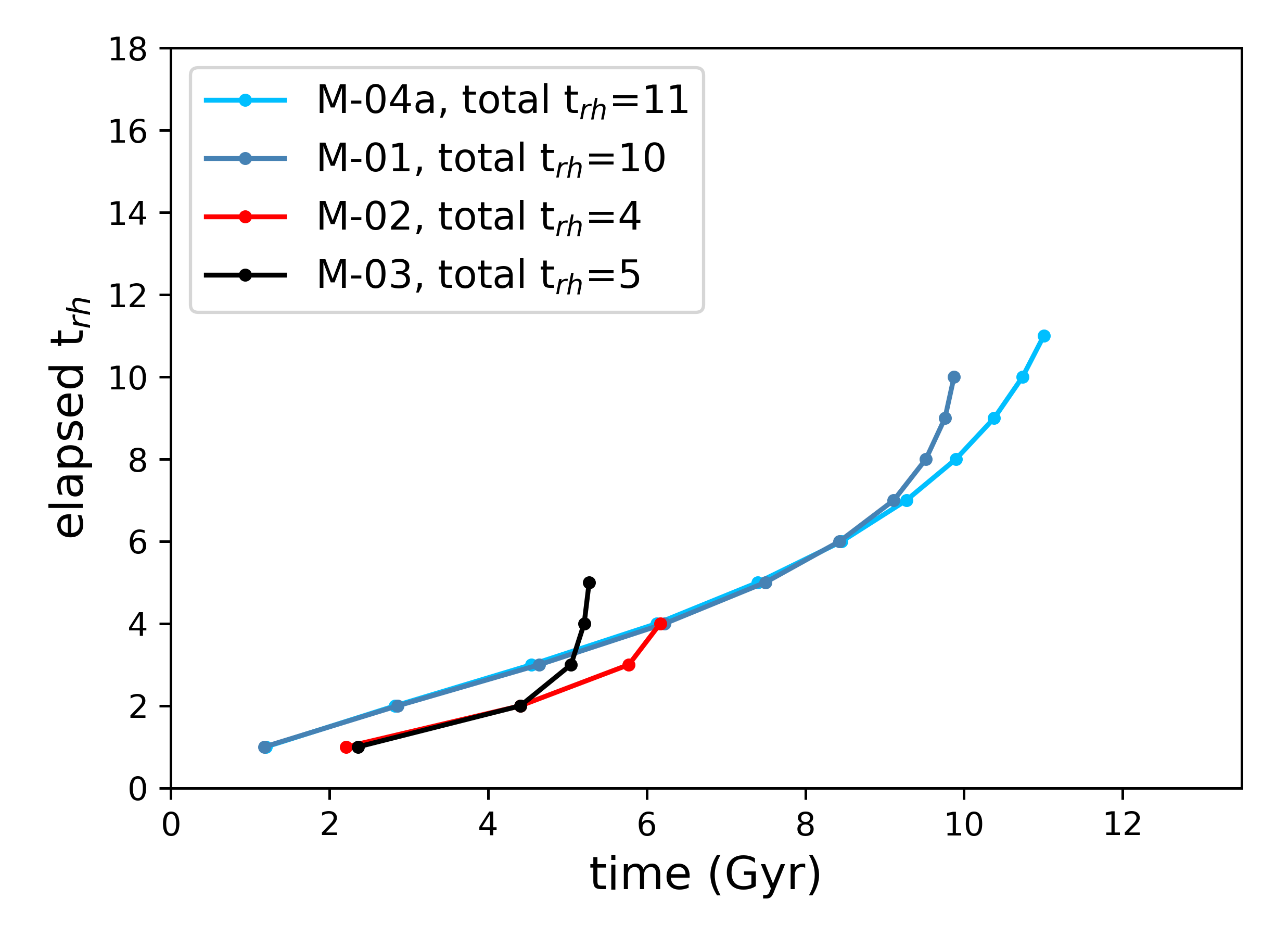}
\includegraphics[width=0.35\textwidth]{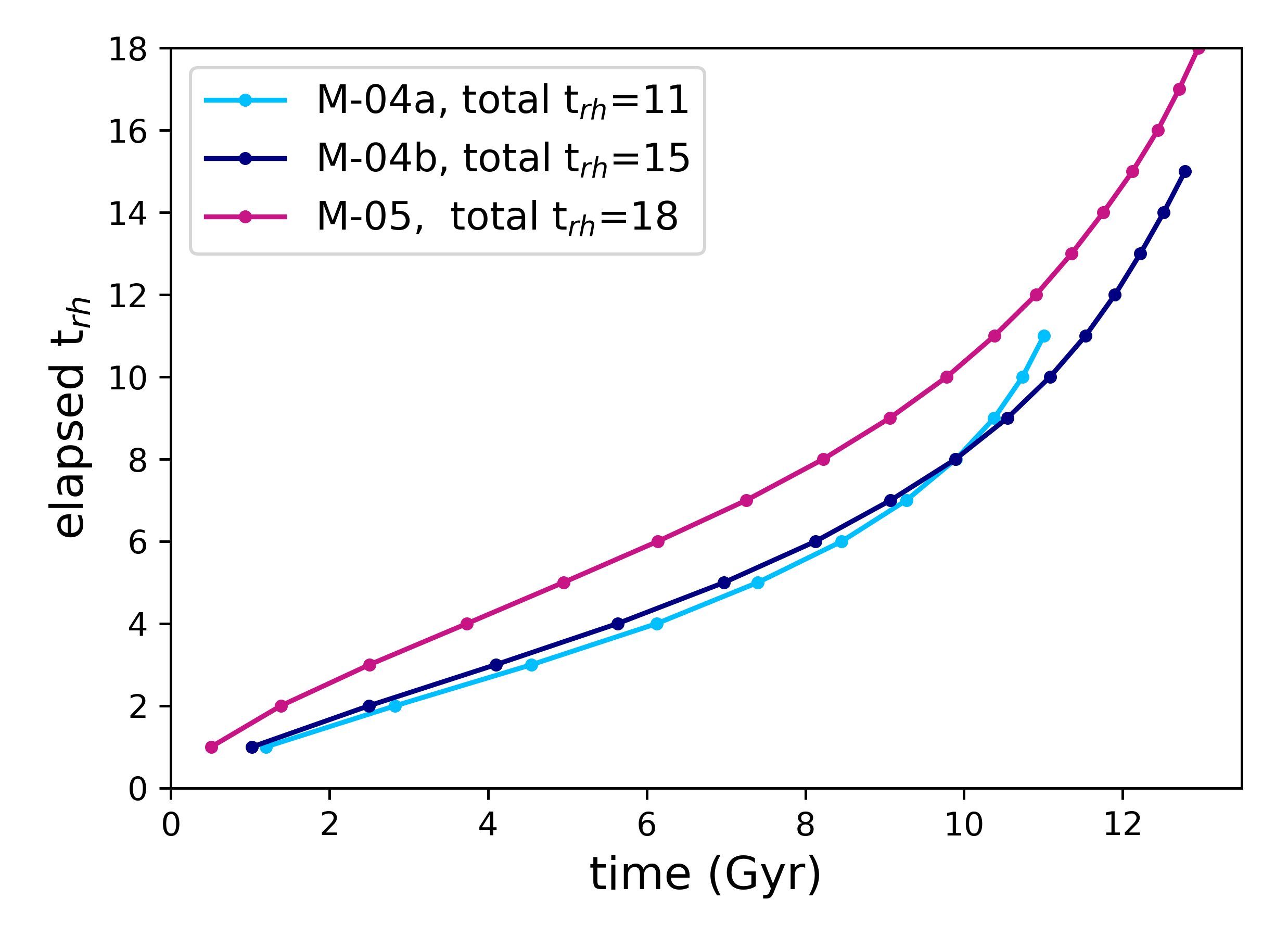}
\caption{The number of elapsed $t_\mathrm{rh}$ with respect to physical time are shown for a grid of cluster models with varying metallicity (top panel) and initial half-mass radius (bottom panel).}
\label{fig:ttrh}
\end{figure}

To measure the dynamical timescale of a cluster, and to use it as a reference timescale for different evolutionary phases,
the concept of relaxation time ($t_\mathrm{relax}$) is often used where this is defined as the time required by a particle to be deflected perpendicularly from the direction of its initial velocity 
(lose the memory of its initial conditions) 
while traversing through the cluster potential. The relaxation time is a local quantity and will be shorter in the dense core than in the sparser outer regions. 
It is common to take the half-mass relaxation time, $t_{\rm rh}$, 
as the typical or average relaxation time for the cluster.
The relaxation time at the half mass radius is related to the cluster mass ($M_\mathrm{clus}$ ), number of particles ($N$) and half mass radius $R_h$ as
\begin{equation}
    \label{eqn:trelax}
    t_\mathrm{rh} = K \times \frac{N}{\psi\mathrm{ln}_{10}\Lambda} \times 
    \sqrt{\frac{R_{h}^3}{GM_\mathrm{clus}}} \quad .
\end{equation}
The quantity $\mathrm{ln}_{10}\Lambda$ is called the Coulomb logarithm (which includes a dependence on $N$) and $\psi$ depends on the mass spectrum of the cluster particles, with $\psi=1$ for clusters with equi-mass bodies. 
The term $K$ is the constant of proportionality. 
While there are multiple forms of Eqn.~\ref{eqn:trelax}, derived from \citet{Spitzer:1971}, we use the form as described in equation\,19 of \citet{Hurley:2000sz} with $\Lambda = 0.4N$, $\psi = 1$ and the constant $K/\sqrt{G} = 0.894$. 
In this paper we use the \citet{Hurley:2000sz} definition of t$_\mathrm{rh}$ in Myrs, where $R_\mathrm{h}$ is measured in parsecs and $M_\mathrm{clus}$ is in solar masses. 
However, we note that other definitions of t$_\mathrm{rh}$ that use $\psi>1$ for a broader mass spectrum, will result in a shorter calculated relaxation time \citep{Antonini:2019ulv}.

As $R_\mathrm{h}$ increases over the first half of the cluster evolution, as we have seen for model M-01, $t_\mathrm{rh}$ also increases. 
This is despite the cluster losing mass because $R_h$ remains the dominant factor in Eqn.~\ref{eqn:trelax}. 
For model M-01 $t_\mathrm{rh}$ is about $300\,$Myr at the onset and has increased to about $1\,600\,$Myr by the time ($2.5\,$Gyr) that the cluster has lost half of its initial mass. 
It is at a similar value when $R_{\rm h}$ reaches its peak value of $10.7\,$pc (after $6\,$Gyr) before decreasing steadily as the cluster heads towards dissolution. 
We show the number of elapsed half-mass relaxation times as a function of time in Fig.~\ref{fig:ttrh} where we see that M-01 has reached the equivalent of 10 half-mass relaxation times at the end-point 
(noting that we have used $9.9\,$Gyr when $N = 500$ as the end-point rather than the later dissolution time shown in Table~\ref{table:NBODY_models}).

\subsection{Metallicity}
\label{sec:metallicity}

\begin{figure}\centering
\includegraphics[width=0.35\textwidth]{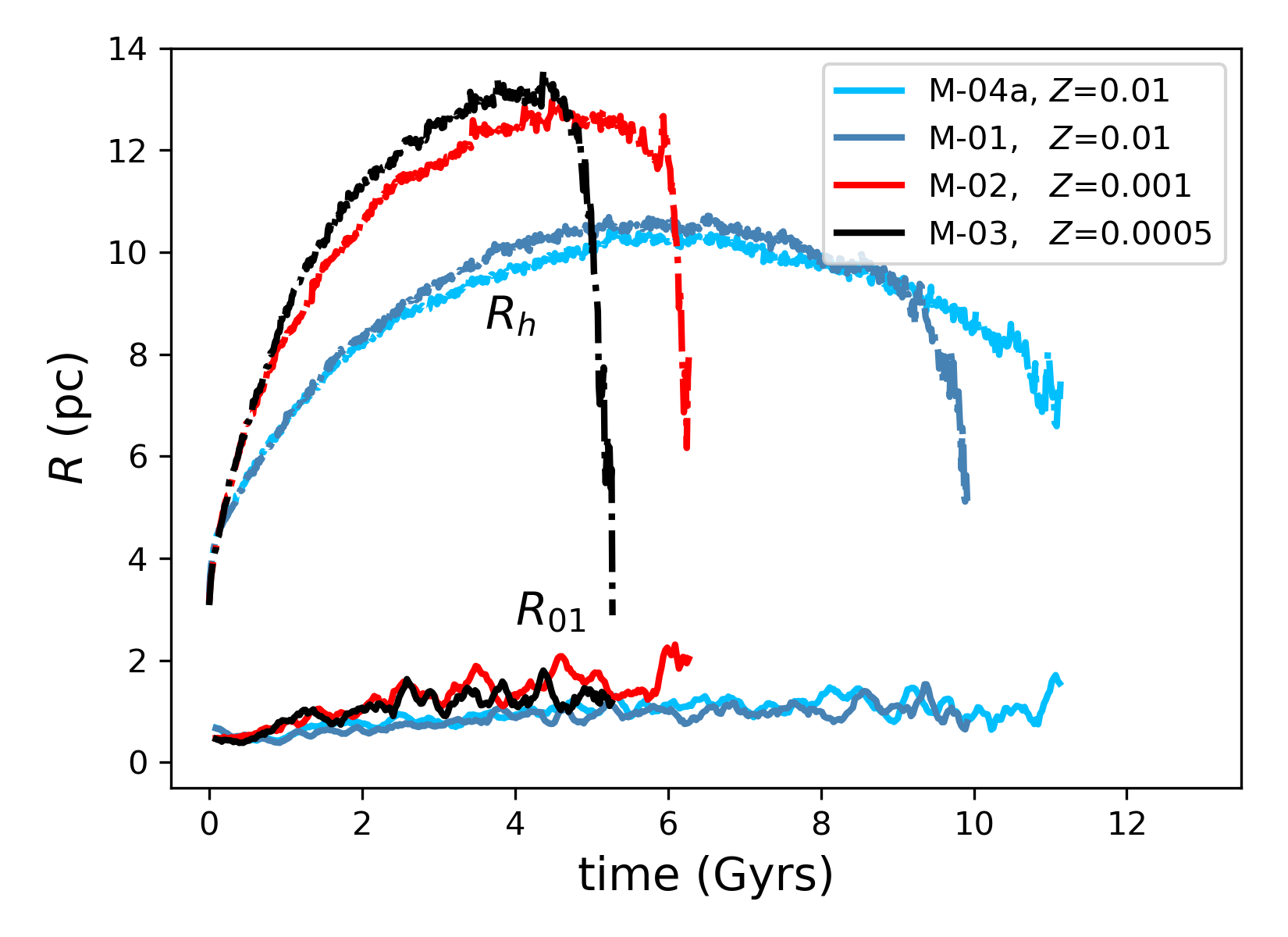}
\includegraphics[width=0.35\textwidth]{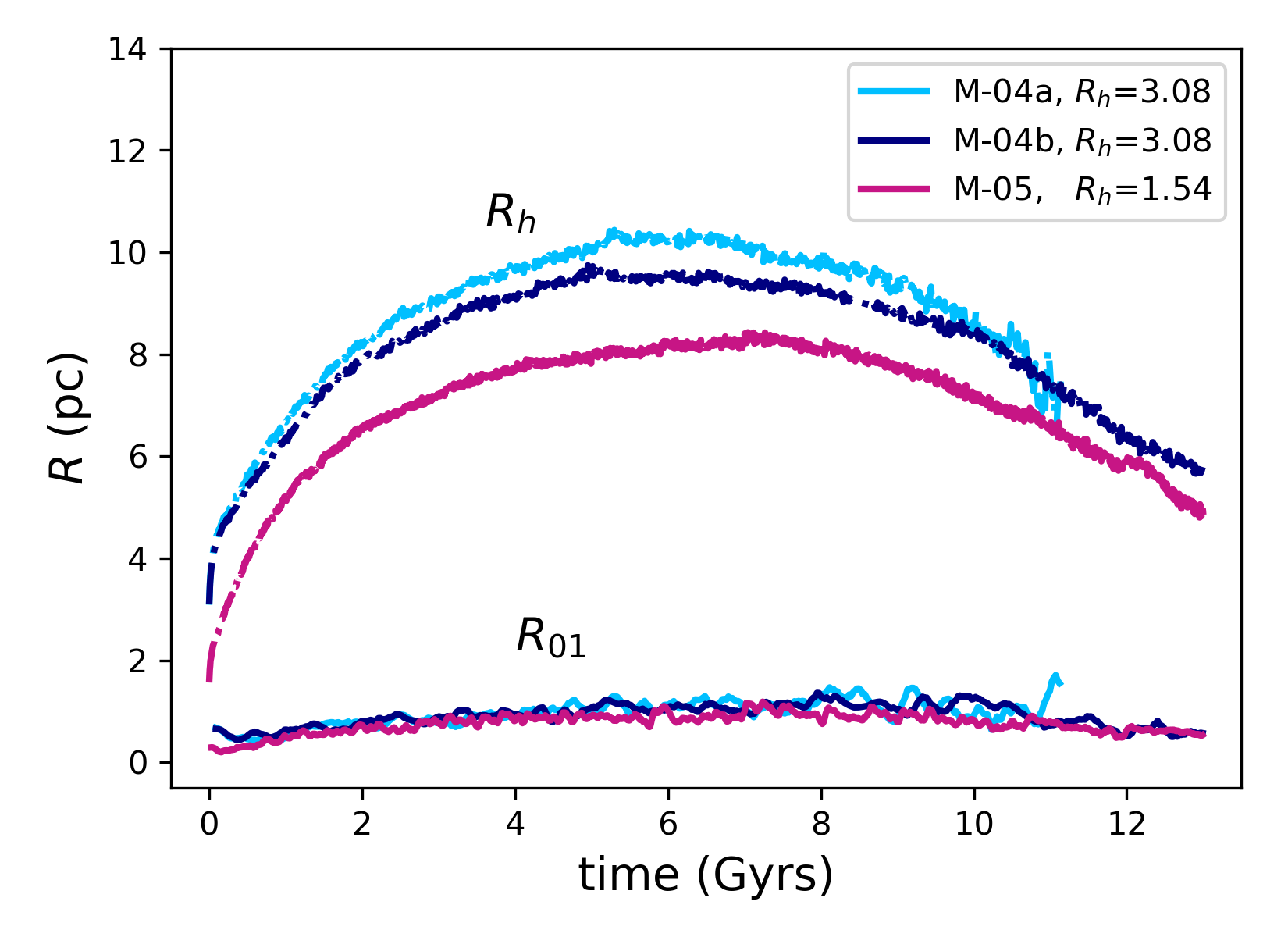}
\caption{Evolution of $R_\mathrm{01}$ and $R_\mathrm{h}$ as a function of time. The top panel shows the impact of metallicity, whilst the bottom panel shows the impact of the initial half-mass radius.}
\label{fig:R_Grid}
\end{figure}

We next look at the differences that arise as a result of the metallicity of the stellar populations by comparing models M-01 ($Z = 0.01$), M-02 ($Z = 0.001$) and M-03 ($Z = 0.0005$). 
The evolution of cluster mass is shown in the upper panels of Figs.~\ref{fig:Mclus_Grid} and 
\ref{fig:MclusIni_Zgrid1} for these three models, with the evolution of key radii compared in Fig.~\ref{fig:R_Grid}. 

The primary observation from Fig.~\ref{fig:Mclus_Grid} is that the lower $Z$ models M-02 and M-03 evolve on a faster timescale than the metal-rich M-01. 
The cluster mass loss rate in the first few hundred million years is slower for M-02 and M-03 (see Fig.~\ref{fig:MclusIni_Zgrid1}) due to reduced stellar winds \citep{Mapelli:2014yna}. 
However, shortly afterwards the metal-poor M-02 and M-03 overtake the metal-rich M-01 in mass loss, at ages of approximately $350$ and $500\,$Myr, respectively (see Fig.~\ref{fig:MclusIni_Zgrid1}) and end up with lifetimes about half that of model M-01.

The differences in evolution timescales can be traced back to differences in the masses of the BH populations and the effect that this has on the cluster structure. 
M-01, M-02 and M-03 have the same initial half mass radius (and thus density). 
However, we see that very quickly the low-$Z$ models exhibit enhanced growth of $R_\mathrm{h}$: the peak $R_\mathrm{h}$ reached for M-02 and M-03 is $12.8\,$pc and $13.4\,$pc, respectively, with both more than 20\% larger than the peak of $10.6\,$pc obtained by M-01. 
Moreover, M-02 and M-03 take only about $5$ and $4\,$Gyr, respectively, to reach their peak compared to $6\,$Gyr for M-01. 

\begin{figure}\centering
\includegraphics[width=0.35\textwidth]{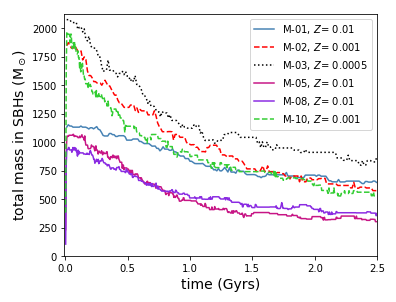}
\caption{The total mass in single BHs as a function of time for the first 2.5\,Gyr is shown for two grids of models with varying metallicity.  
}
\label{fig:SBHmass}
\end{figure}

\begin{figure}\centering
\includegraphics[width=0.35\textwidth]{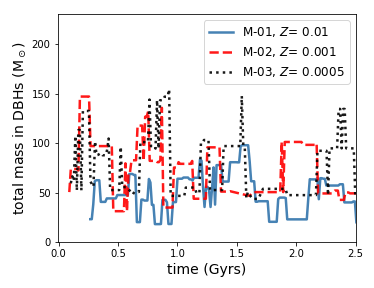}
\includegraphics[width=0.35\textwidth]{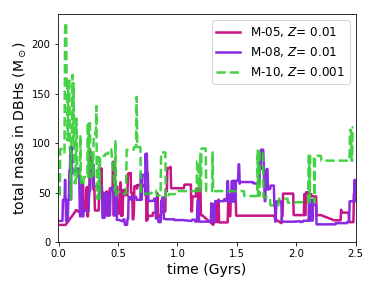}
\caption{The total mass in double BHs as a function of time for the first 2.5\,Gyr is shown for two grids of models with varying metallicity. The upper panel shows the model set with initial R$_\mathrm{h}=3.08$\,pc and the lower panel with initial R$_\mathrm{h}=1.54$\,pc. 
}
\label{fig:DBHmass}
\end{figure}

The divergence in $R_\mathrm{h}$ evolution sets in shortly after formation of the BHs 
where the low-$Z$ clusters form more massive BHs \citep{Maeder:1992, Hurley:2000pk, Mirabel:2016msh}. 

This has already been demonstrated by Fig.~\ref{fig:bh_ini_mass} but we further emphasise the fact by showing the mass in single BHs as a function of time for a range of models in Fig.~\ref{fig:SBHmass}. 
We see that particularly at early times (up to about $0.5\,$Gyr) the 
metal-rich clusters only carry about half the amount of mass in single BHs compared to their metal-poor counterparts, irrespective of their initial density. 

Even more importantly the low-$Z$ clusters form more massive DBHs and form them earlier than in M-01: 
the upper panel of Fig.~\ref{fig:DBHmass} shows that on average M-02 and M-03 have more than 1.5 $\times$ the mass in DBH systems than M-01. 
In fact we find that the average mass of the DBHs in M-02 and M-03 is roughly double that in M-01 over the cluster lifetime. 
This results in enhanced core dynamical activity for M-02 and M-03 which causes increased thermal energy injection into the cluster halo, expanding the cluster and accelerating the mass-loss across the tidal boundary. 
Although the increased frequency of massive-BH dynamical interactions in metal-poor clusters is initially 
more efficient at feeding thermal energy to the cluster core and halo, the increased mass-loss decreases this efficiency in the long run. 
This is because heightened mass-loss reduces both the gravitational potential well of the core and the rate of dynamical friction due to less BHs being retained in the cluster over an $\mathcal{O}$(Gyr) timescale. 

This increased expansion for the low-$Z$ clusters is clearly evident when we look at the evolution of $R_\mathrm{h}$ and  $R_\mathrm{01}$ in Fig.~\ref{fig:R_Grid}.  
Interestingly, the increased expansion rate for M-02 and M-03 compared to M-01 onsets at an earlier time for $R_\mathrm{h}$ ($\approx$ 0.20 Gyr) than that for $R_\mathrm{01}$ ($\approx$ 0.75 Gyr). 
This reflects the local evolutionary timescales and the battle in the central regions between the tendency for gravitational collapse and the thermal energy generated through interactions with the DBH population. 
Overall expansion prevails but the extent is reduced and the onset delayed for $R_\mathrm{01}$ as a result of this battle. 
This same interplay of gravity and the generation of thermal energy causes the cluster core and hence $R_\mathrm{01}$ to alternatively contract and dilate. 
These central oscillations appear to be more extreme for M-02 and M-03 (top panel of Fig.~\ref{fig:R_Grid}) 
although we should be careful not to read too much into the behaviour towards the end of the evolution when $N$ is greatly reduced and the results become noisy owing to low number statistics. 

\begin{figure}\centering
\includegraphics[width=0.34\textwidth]{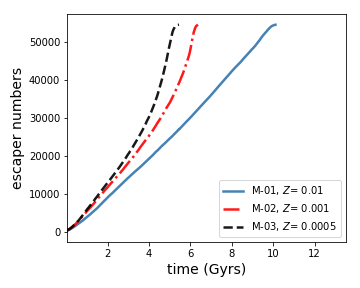}
\includegraphics[width=0.35\textwidth]{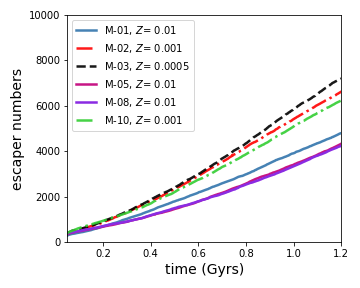}
\includegraphics[width=0.35\textwidth]{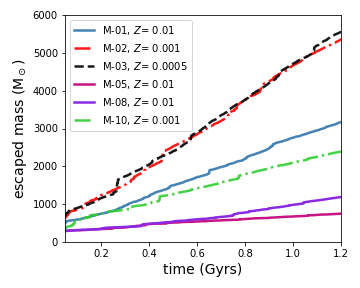}
\caption{Cumulative number of escapers from a cluster as a function of time. The top panel shows the variation as a function of metallicity for the life of the clusters, the middle panel focuses on the first $\sim 1$\,Gyr for selected models, whilst the bottom panel shows the mass loss for the first $2$\,Gyr. 
}
\label{fig:esc}
\end{figure}
\begin{figure}\centering
\includegraphics[width=0.35\textwidth]{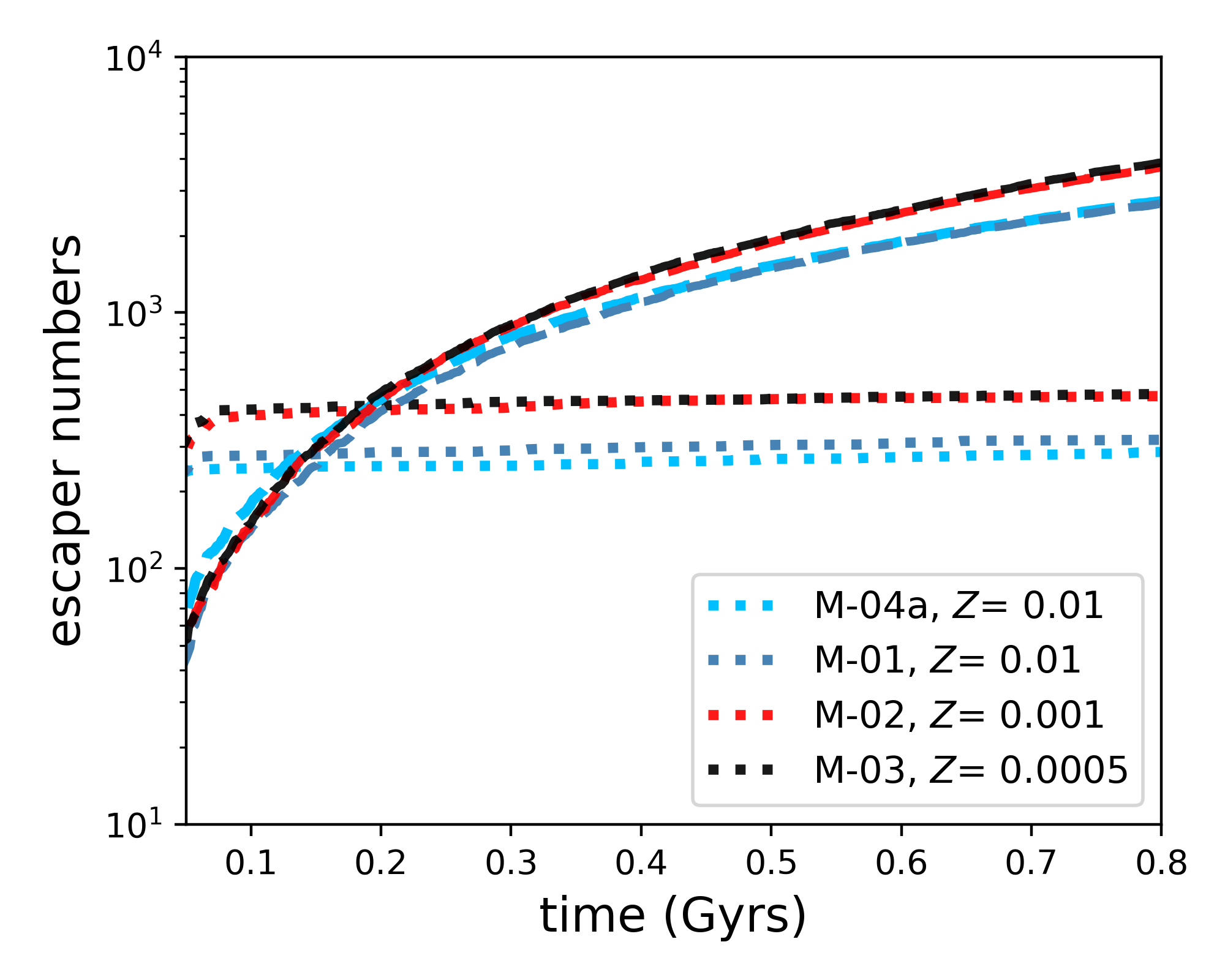}
\includegraphics[width=0.35\textwidth]{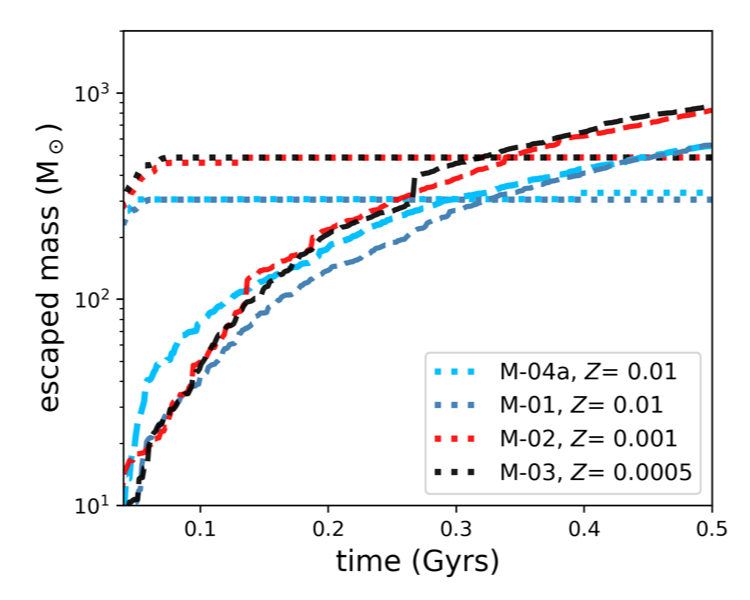}
\caption{The top plot shows the cumulative number of escapers split into subcategories of evaporation (dashed lines) and ejection (dotted lines) from a cluster as a function of time. The bottom plots shows the cumulative mass in these two types of escapers. and only the first $800\,$Myr and $500\,$Myr (respectively) are shown in order to provide sufficient resolution for the important early stages. 
}
\label{fig:esc_ejec_eva}
\end{figure}


Enhanced $R_\mathrm{h}$ expansion in metal-poor clusters not only leads to more mass-loss through tidal stripping 
\citep[referred to as evaporation:][]{Gieles:2011}{}
but can also lead to an increase in stellar ejections through a combination of the velocity kicks given to stars in interactions with the more sizeable DBHs and the lower escape velocity. 
We refer to stars which leave the cluster through either of these two mechanisms as escapers.
Given that the external tidal field remains the same for all of our models, the rate of escape of stars from the cluster can also reflect the internal heat-up and hence dynamical activity of the cluster. 
Fig.~\ref{fig:esc} shows the cumulative number and mass of escaped bodies from the star cluster models where we clearly see a greater number of;  as well as more mass in the escapers 
at any given age in metal-poor M-02 and M-03 than in M-01. 
The stars (and their mass) that escape from the cluster through natal kicks or recoil velocities and those that are stripped away by the host galaxy are differentiated in Fig.~\ref{fig:esc_ejec_eva}. 
While the evaporation from the cluster appears to increase with time (the more a cluster loses its mass, its tidal radius and escape velocity are reduced and it is easier for the host galaxy to strip more stars), the number of ejected systems reach 200--400 within the first 100\,Myr, and then onwards there is very little increase.  
For all models shown in Fig.~\ref{fig:esc_ejec_eva}, about 80\% of the cumulative ejected systems at 1.5\,Gyr have already escaped within the first 100\,Myr, while the cumulative number ejected by 4\,Gyr is only about 1.1-1.2$\times$ that at 1.5\,Gyr. 
We also observe that for both evaporation and ejection, metal-poor clusters have a more rapid rate of mass loss than for clusters with $Z=0.01$ (aside from the evaporation rate at very early times). For ejection, it needs to be remembered that metal-poor stars tend to evolve faster than their metal-rich counterparts of the same mass. Though this difference is negligible for masses $>10$\,M$_\odot$, the initial mass function being bottom heavy ensures that the metal-poor clusters start to lose mass through stellar evolution more rapidly from an early stage.
This is also manifested in the finding that M-02 and M-03 have shorter lifetimes than M-01. 
The rate of cluster evolution can be measured in the number of elapsed half-mass relaxation times $t_\mathrm{rh}$ at a particular age. 
As we see from Fig.~\ref{fig:ttrh}, M-01 completes three half-mass relaxation times by around 4 Gyr; whereas at the same physical time M-02 and M-03 have completed only two. 
To reach the mature state of $<500$ stars remaining, M-01 requires 10 $t_\mathrm{rh}$ while M-02 and M-03 take only 4 and 5 $t_\mathrm{rh}$, respectively. 
In terms of the physical time-scale, to reach 
the same point 
M-01 with $Z=0.01$ takes about $9.9\,$Gyr, while M-02 and M-03 require about $6.3\,$Gyr and  $5.3\,$Gyr, respectively. 
Thus, using the half-mass relaxation timescale as a measure, we see that the dynamical evolution of the low-$Z$ clusters has not been accelerated. 
Instead, the presence of the more massive BH population has dramatically truncated the overall lifetime and the dynamical lifetime of the low-$Z$ clusters. 
As we shall see in section~\ref{sec:double_black_holes} this is not at the expense of enhanced dynamical activity for the very centrally-condensed BH population.

A related quantity is the escape velocity $V_\mathrm{esc}$.
Though the initial cluster density of M-01, M-02 and M-03 is the same (all starting with $R_\mathrm{h} = 3.08\,$pc), very quickly
the $V_\mathrm{esc}$ distributions become lower for the lower cluster metallicities. 
This is a direct consequence of the enhanced expansion of the inner regions of the clusters of lower $Z$ that have produced the more massive BHs. 
The difference between the low-$Z$ and high-$Z$ escape velocities increases over time, fuelled by the snowball effect of it being easier to escape, thus more mass is lost and the $V_\mathrm{esc}$ reduces further.

We can utilise some of our additional models to see whether or not this behaviour with metallicity holds up in general and across a wider range of cluster initial conditions. 
Model M-05 from our main set has the same metallicity ($Z = 0.01$) as M-01 but an initial $R_{\rm h}$ reduced by half to $1.54\,$pc and thus an increased initial stellar density. 
These two models will be compared in more detail in Sec.~\ref{sec:Ini_Rh} below. 
For now we can pair M-05 with M-08, which has the same initial conditions aside from a different random number seed, and compare to M-10 which again is the same other than a lower metallicity of $Z = 0.001$, 
noting that M-08 and M-10 were not evolved until late times as we were only interested in their early evolution. 
The early evolution of $M_\mathrm{clus}$ for these models with initial $R_\mathrm{h}=1.54\,$pc is shown in the lower panel of Figure~\ref{fig:MclusIni_Zgrid1} where we see the same metallicity trend as we saw for the models with larger initial $R_{\rm h}$, namely that over time the low-$Z$ M-10 is losing more mass than the higher-$Z$ M-05 and M-08. 
This is also the case when we look at the escaper numbers 
in the lower panel of Fig.~\ref{fig:esc}, where we see that M-10 has a similar cumulative number of escaped stars as M-02, while M-05 and M-08 are similar to M-01. 
While the models of higher initial density have a slightly lower rate of ejected stars in general, the overriding factor in determining this appears to be $Z$. 
Furthermore, we see the same pattern of increased early DBH behaviour for low-$Z$ when we look at the lower panel of Fig.~\ref{fig:DBHmass} 
where the 
net DBH mass for M-10 remains 1.5-2$\times$ higher than that of M-05 and M-08. 
Thus it appears that the importance of $Z$ on the cluster evolution and dissolution time is even more paramount than initial density.

Our results here echo \citet{Banerjee:2017} who also used the updated wind prescription \citep{Vink:2001, Belczynski:2008mr} and found a similar impact of BHs on the dynamical evolution in models of different metallicities. 
\citet{Banerjee:2017} also noted that for the same initial density, clusters of lower $Z$ had a shorter lifespan due to both a heightened rate of dynamical interactions and the presence of more massive BHs. 
In contrast,  \citet{Sippel:2012}, who used the original \citet{Hurley:2000pk} wind prescription, concluded that cluster $Z$ did not significantly affect the evolution of $R_\mathrm{h}$. 
This too is the result of having BHs of lower mass than in our current models with the \citet{Vink:2001} stellar wind prescription for massive stars.
Overall, our grids of models of different $Z$ and different initial $R_\mathrm{h}$ show that the factor of more massive BHs (as present in the low $Z$ models) is dominant when it comes to heating the cluster, more so than two-body BH interactions as will be discussed further in the following sections. 
\begin{figure}\centering
\includegraphics[width=0.35\textwidth]{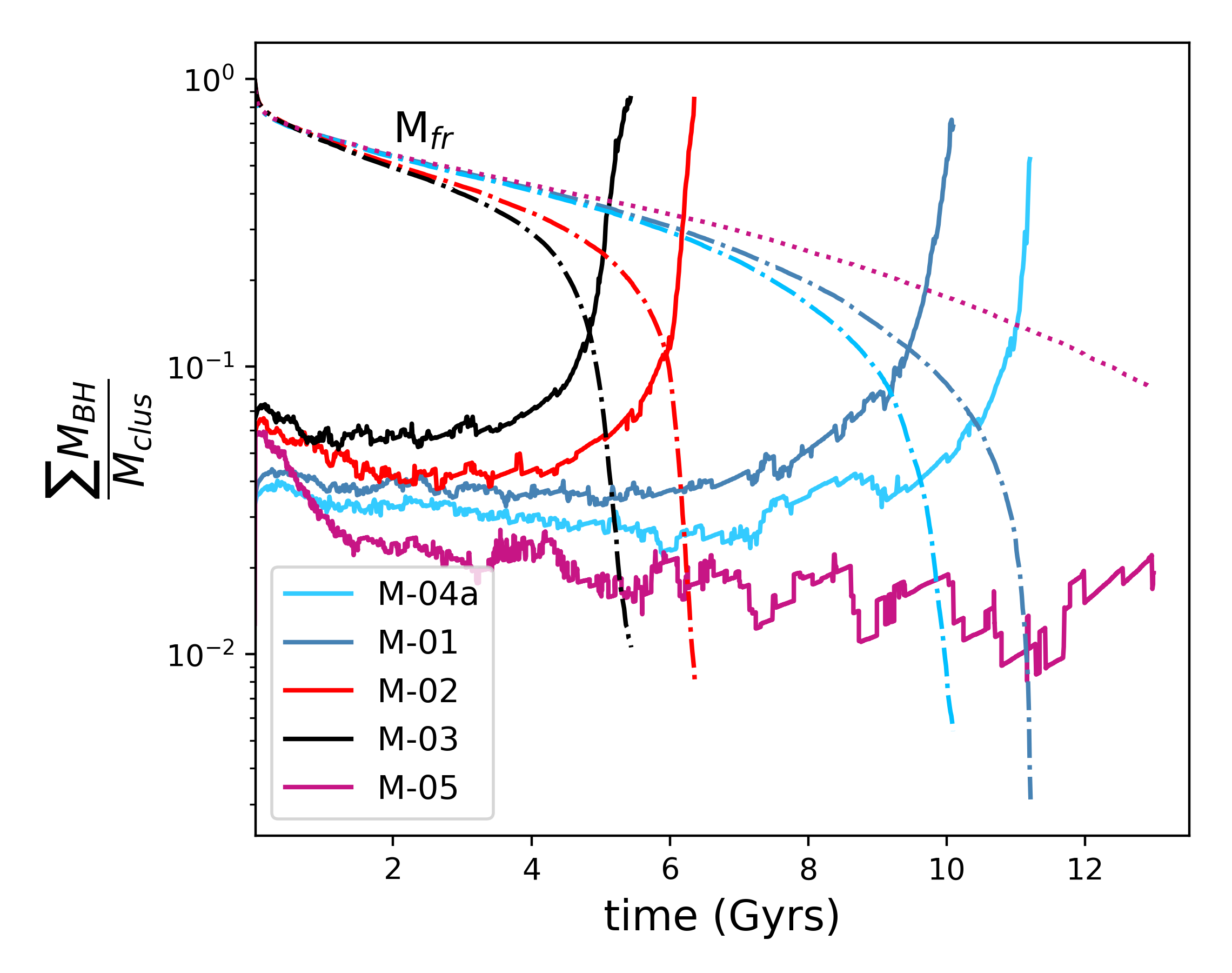}
\caption{The mass ratio of the in-cluster BHs to the cluster mass over time is shown by the solid lines. The dotted lines signify fractional mass left in the cluster $M_\mathrm{fr}=M_\mathrm{clus}/M_\mathrm{ini}$, where $M_\mathrm{ini}$ is the initial mass of the cluster.}
\label{fig:MassFraction}
\end{figure}

On a similar note, 
\citet{Mapelli:2014yna} showed the trend of early dissolution of metal-poor clusters and attributed this to the higher density of the cluster core within the first few hundred Myr of the cluster's lifetime. It was argued that since larger stellar winds in metal-rich clusters (with the expectation that this corresponds to greater net stellar evolution mass-loss) somewhat quench the central gravitational potential, their core density remains smaller. Higher core density in a low metallicity cluster ensures more three-body interactions, which injects more heat into the halo and aids in its early evaporation. 
On the contrary, we clearly show that the net mass removed from the clusters due to stellar evolution is always larger in metal-poor clusters (see bottom panel Fig.~\ref{fig:esc_ejec_eva}) when the full extent of the IMF is considered.
Furthermore, the amount of mass held in BHs (both singly and in binaries) for the first few Gyr is also always higher in the metal-poor clusters. We will also show in section~\ref{sec:DBH_dynamics} that the rate of DBH interactions themselves effects the cluster expansion to a smaller extent. Indeed, under an impulse approximation the hardening rate of a binary is independent of the mass of the binary but rather depends on its local mass density and velocity dispersion (see \citealp{Mapelli_MaxwellDemon:2018gcb} for discussion). 
For our clusters of different metallicity we observe that the core density for the first few hundred Myr of cluster evolution is only slightly high for metal-poor clusters and over the first Gyr there is not much difference at all on average. 
We thus argue in a similar vein as before that it is the mass of the BHs --- specifically the width of the BH mass function --- that play the key role in injecting more energy into the halo in the cases of metal-poor clusters rather than the higher rate of three-body interactions. The broader mass function shortens the mass segregation timescale in metal-poor clusters. It is the (mass dependent) kinetic energy that gets drawn out of the binaries per interaction, rather than the number of interactions, that becomes more important in determining the amount of thermal energy pumped onto the halo which affects the cluster's lifetime.

We also show the variation of the fractional mass in the BHs residing inside the cluster as a function of time in Fig.~\ref{fig:MassFraction}. All clusters shown in Fig.~\ref{fig:MassFraction} have an initial BH mass fraction of less than the critical 10\% suggested by \citet{Breen_Heggie:2013}, however the cluster mass loss is balanced in such a way that the BH mass fraction still remains nearly constant through most of the cluster evolution. 
It is only at late times, when the clusters are close to the point of dissolution and only have of order 100 bound bodies remaining, that the mass fraction in BHs trends towards unity. 
The metal-poor clusters, with initially more massive BHs, do show slightly more tendency to move towards BH domination at an earlier stage of cluster evolution (in terms of the fractional cluster mass remaining). 
The dense cluster M-05 appears to eject BHs more rapidly, resulting in a lower BH mass fraction throughout its evolution \citep[similar to][]{Gieles:2021NatAsG}.
M-05 is terminated at 13\,Gyr, at which point its BH population does not even account for 10\% of the cluster's net mass.

\subsection{Initial Orbital Period}
\label{sec:porb}

\begin{figure}\centering
\includegraphics[width=0.35\textwidth]{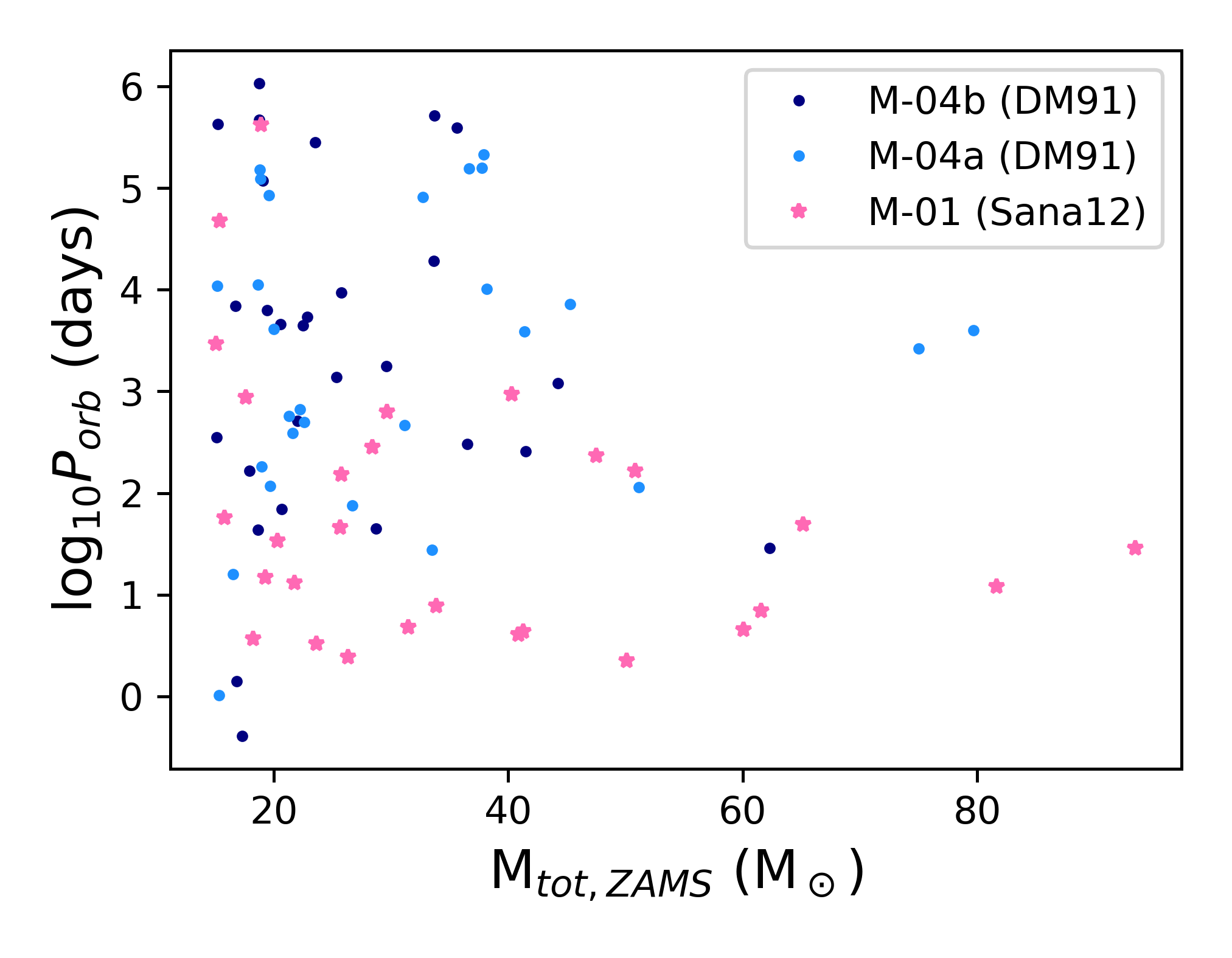}
\caption{Initial binary orbital period against total ZAMS mass for all massive primordial binaries that form DBHs.
The different points correspond to models with different initial binary orbital period distributions: 
model M-01 which uses the Sana12 distribution (magenta points) and 
model M-04 which uses the DM91 distribution (blue points).}
\label{fig:SanaDM}
\end{figure}

Now we analyse the possible effects of the initial orbital period (and thus semi-major axis) distribution of the binaries on the cluster evolution through comparison of model M-01 with M-04a (and M-04b). 
While M-01 utilizes the Sana12 initial orbital period ($P_\mathrm{orb}$) distribution the M-04a and M04b primordial binaries are chosen from the DM91 distribution. 
Note that M-04a and M-04b differ only in the random number seed
used (which affects the sequencing of distributions for setting the initial masses, positions, velocities, orbital periods and eccentricities) and otherwise have the same initial parameters. 
Crucially, models M-01, M-04a and M-04b have the same metallicity ($Z = 0.01$) and half-mass radius ($R_{\rm h} = 3.08\,$pc). 

The $P_\mathrm{orb}$ distribution described in \citet{Sana:2012} is a broken power-law distribution while that of \citet{Duquennoy:1991} is a log-normal distribution. 
DM91 allows for a comparatively larger fraction of very short, $P_\mathrm{orb} < 1\,$d, primordial binaries ($\approx$ 0.03) than Sana12 ($\approx$ 0.01) for the \citet{Kroupa:2000iv} IMF and our mass range of 0.1--100\,M$_\odot$. 
However, the Sana12 distribution preferentially chooses lower orbital periods for massive binaries (those with at least one star more massive than 15 M$_\odot$) which are prime progenitors of compact objects \citep{deMink:2015yea}. 
For the primordial binaries with a total ZAMS mass of $>15$\,M$_\odot$, the clear preference of shorter $P_\mathrm{orb}$ is observed in Fig.~\ref{fig:SanaDM}. 
Model M-01 has 10 primordial binaries of total mass over 15\,M$_\odot$ and $P_\mathrm{orb} < 10\,$d, while M-04a and M-04b have only one and two such systems, respectively. 

The global properties of the M-01 and M-04a clusters can be compared in the upper panels of 
Figs.~\ref{fig:Mclus_Grid}, ~\ref{fig:MclusIni_Zgrid1} and ~\ref{fig:R_Grid}. 
We see no clear distinguishable features between these models during most of their evolutionary phases. 
However, towards the end of their lifetimes, post $8\,$Gyr, the mass-loss rate of M-01 increases relative to M-04a, resulting in an accelerated path towards dissolution for M-01 (see the rapid decrease of half mass radius for M-01 post $9\,$Gyr in Fig.~\ref{fig:R_Grid}). Indeed, while M-04a does not drop below 500 stars remaining until $11.1\,$Gyr, this occurs earlier at $9.9\,$Gyr for M-01 as already noted. 
In terms of half-mass relaxation times elapsed this corresponds to 10 for M-01 and 11 for M-04a (see Fig.~\ref{fig:ttrh}). 
This divergence in behaviour only appears towards the end of the cluster evolution when the clusters have lost around 80\% of their mass. 
To some extent, we can account for this by the random effects due to small number statistics. During the evolution of a cluster chance encounters can lead to the formation of one or a few hard DBHs. These stable binaries tend to act as thermal engines at the core, accelerating the expansion and eventual evaporation of the cluster. The median DBH binding energies
for M-04a and M-04b are only about 0.9 and 0.8 times, respectively, of that of M-01, computed over the lifetime of the clusters. 
Similarly, the median latus-rectum
of the M-04a and M-04b DBHs  
are about 1.5 and 1.8 times respectively of those of M-01. 
While these are not huge differences they do show how differences in the characteristics of the binary populations can develop over time. 
This in turn can influence the outcomes for the cluster model, as we see 
with M-01 having the tighter DBHs and leading to a faster demise compared to M-04a and M-04b.

If we instead compare M-04a and M-04b (see lower panels of Figs.~\ref{fig:Mclus_Grid}, ~\ref{fig:R_Grid} and ~\ref{fig:ttrh},
we see that the two clusters of exactly the same initial parameters but with different random number seeds show differences between the evolution of their parameters that are of similar or greater magnitude than observed when comparing M-04a with M-01. 
For example, the evolution of the cluster mass starts to show a difference at an age of $6\,$Gyr which is about $2\,$Gyr earlier. 
In fact, the difference between the dissolution times of M-04a and M-04b is larger than between M-04a and M-01, showing that the initial $P_\mathrm{orb}$ distribution is not having a dominant effect on the global properties of clusters of our sample size. 
Though the DM91 clusters have slightly more very short $P_\mathrm{orb}$ primordial binaries, these are predominantly comprised of low-mass stars and do not play a very significant role in the cluster dynamics. 
On the other hand, the additional very short $P_\mathrm{orb}$ massive binaries arising from the Sana12 distribution still only amounts to a handful of systems and does not extraordinarily affect the core behaviour in the long run to in turn affect the global evolution of the cluster.

\subsection{Initial half-mass radius}
\label{sec:Ini_Rh}

\begin{figure}\centering
\includegraphics[width=0.37\textwidth]{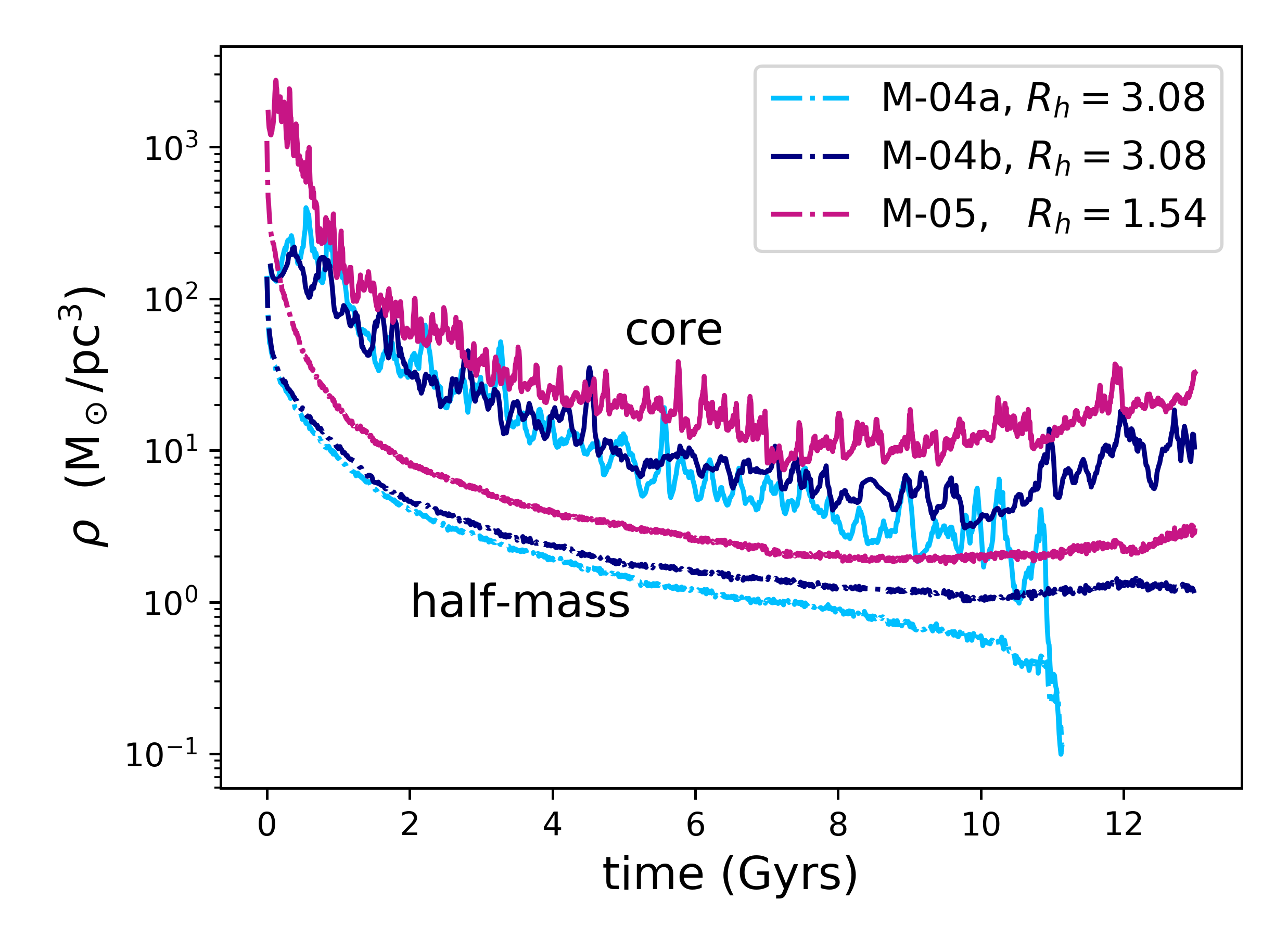}
\caption{Cluster stellar mass density ($\rho$) as a function of time for models varying the initial cluster half mass radius.}
\label{fig:rho_T}
\end{figure}

We explore the effect of initial cluster density on the global properties and evolution through models M-04a (or M-04b) and M05, with initial half mass radii ($R_\mathrm{h}$) of 3.08\,pc and 1.54\,pc, respectively. 
The factor of two difference in initial size translates to a factor of eight difference in initial density. 
It also reflects the variation in the size distribution of star clusters \citep[e.g.][]{vanLoon:2005, Misgeld:2011, Madrid:2012}
and uncertainty over whether clusters should be born compact and then expand or should already be filling their tidal radii when star formation has finished \citep{Baumgardt:2007}. 
Model M-04a is filling its tidal radius by 50\% initially and M-05 by 25\% so this provides an opportunity to compare the effect of the filling factor on the cluster evolution, which is expected to be noticeable \citep{Baumgardt:2010}.

From the lower panel of Fig.~\ref{fig:Mclus_Grid} we see that all three models have very similar initial mass loss rates. 
However, over time differences develop and the 
evolutionary timescale for M-04a ends up almost $2\,$Gyr shorter than M-05. 
Models M-04b and M-05 track each other in mass for longer, although at 13\,Gyr when both models are stopped we find that M-05 retains about 8\% of its initial mass compared to 5\% for M-04b. 

The half-mass radius $R_\mathrm{h}$ for M-05, begins at half the value of M-04a and as we see from Fig.~\ref{fig:R_Grid} (lower panel) remains lower throughout the evolution. 
The peak $R_\mathrm{h}$ reached by M-05 is about 8.4\,pc in approximately 7\,Gyr while for model M-04a it is around 10.4\,pc and occurs on a similar timescale. 
Our peak $R_\mathrm{h}$ values are within the range found by \citet{Madrid:2012} who also explored variations with Galactocentric orbit and cluster mass.
In terms of density we find that M-05 maintains its higher stellar density compared to M-04a (and M-04b) throughout but that the difference in their densities decreases over time.  

If we look at the behaviour in the central regions and take $R_\mathrm{01}$ as a representative radius (given that it is less noisy than the $N$-body core radius) we see from Fig.~\ref{fig:R_Grid} that although M-05 starts at a lower value (by construction), after about $800\,$Myr of evolution the models show nearly equal $R_\mathrm{01}$. 
Averaged across the period between 5-8\, Gyr, where the $R_\mathrm{01}$ evolution has plateaued somewhat for all models, we find that M-05 has  $R_\mathrm{01}\sim0.8$\,pc compared to $R_\mathrm{01}\sim1.0$\,pc for M-04a. 
So the difference in the size of the inner regions for the initial models has clearly decreased over time owing to increased expansion within M-05. 
This accelerated expansion of the core of M-05 indicates rapid heating resulting from enhanced dynamical activity.
However, it is also observed that this expansion rate does not continue to be sustained as we saw for the cores of the metal-poor clusters M-02 and M-03 (compare the lower and upper panels of Fig.~\ref{fig:R_Grid}). 
It appears that although the frequency of DBH interactions in the central regions affects the cluster thermal energy, the heat up is dominated by the increased BH masses. 

A lower half-mass radius and thus higher density ensures a higher initial half-mass escape velocity of $\approx10$\,km/s for M-05 compared to $\approx7$\,km/s for M-04a. This higher $V_\mathrm{esc}$ aids in retaining BHs in M-05 which leads to enhanced dynamical activity (as noted above and also see section~\ref{sec:double_black_holes}).
The difference in escape velocity between the two clusters decreases over the first couple of Gyr to about 0.5\,km/s, with values of 4\,km/s and 3.5\,km/s for M-05 and M-04a respectively, and this difference is maintained across the remainder of their evolution.

The evolution of R$_\mathrm{h}$ also plays an important role in determining the half-mass relaxation timescale of the cluster at any point in time. 
Model M-05, with the smaller initial R$_\mathrm{h}$, has a smaller maximum $t_{\mathrm{rh}}$ of 1.3 Gyr compared to M-04a and M-04b with 1.76 and 1.66 Gyr, respectively. 
Accordingly we find that the number of half-mass relaxation times elapsed for M-05 is 18 in its lifetime compared to 15 for M-04b and 11 for M-04a (Fig.~\ref{fig:ttrh}, lower panel). 
This shows that the change in initial half-mass radius (and stellar density) is having a greater impact on the dynamical ages of the clusters than on the dissolution time, noting that M-04b and M-05 have similar dissolution times. 
However, given that differences between M-04a and M-04b rival those between M-04a and M-05 is it also true that statistical uncertainties are a factor, particularly in the later phases of cluster evolution.

\subsection{Additional Models}
\label{sec:add_mods}

As mentioned earlier, we have evolved an additional set of models that we utilise in part for comparison purposes as required (for example, as has already been done in Sec.~\ref{sec:metallicity}). 
For these models we were not necessarily concerned with running them to completion as in most cases the early behaviour sufficed for what was needed. 
The models in question are M-06 through to M-10. 
They are summarised in Table~\ref{table:NBODY_models} and here we give a brief overview of how they evolved. 

Model M-06 started with a smaller number of stars (initial $N_\mathrm{tot} = 11\,000$) than our main set of models and as expected that led to a shorter evolution time-scale, with the cluster essentially dissolved after $2.5\,$Gyr. 
This was the equivalent of seven half-mass relaxation times having elapsed. 
In contrast, model M-07 which had 25\% of its initial $75\,000$ stars remaining when it was halted at an age of $9.6\,$Gyr, evolved for the equivalent of six half-mass relaxation times. 
The maximum $R_\mathrm{h}$ reached by M-06 was $8.4\,$pc after 1.2 Gyr and by M-07 was $9.5\,$pc after 6.5 Gyr. 

Model M-08 is identical to M-05 aside from the random number seed. 
It is evolved to $2.7\,$Gyr, rather than the $13\,$Gyr of M-05 but as we see from Figures~\ref{fig:MclusIni_Zgrid1} and \ref{fig:esc} the early evolution is very similar. 
For model M-09 with a primordial binary fraction of 10\% compared to 5\% for the main models we were only interested in the effect on the DBH population in the early stages of cluster evolution (see section~\ref{sec:double_black_holes}). 
As such this model was only evolved for $400\,$Myr. 
Finally, model M-10 is a low-metallicity higher-density model evolved to an age of $2.5\,$Gyr when just under 50\% of the initial stars remained. 
This model had already reached its $R_\mathrm{h}$ peak of $9.8\,$pc after 2.5\,Gyr. 

\section{Compact Object Population Statistics}
\label{sec:black_hole_populations}

\begin{figure*}\centering
\includegraphics[width=0.32\textwidth]{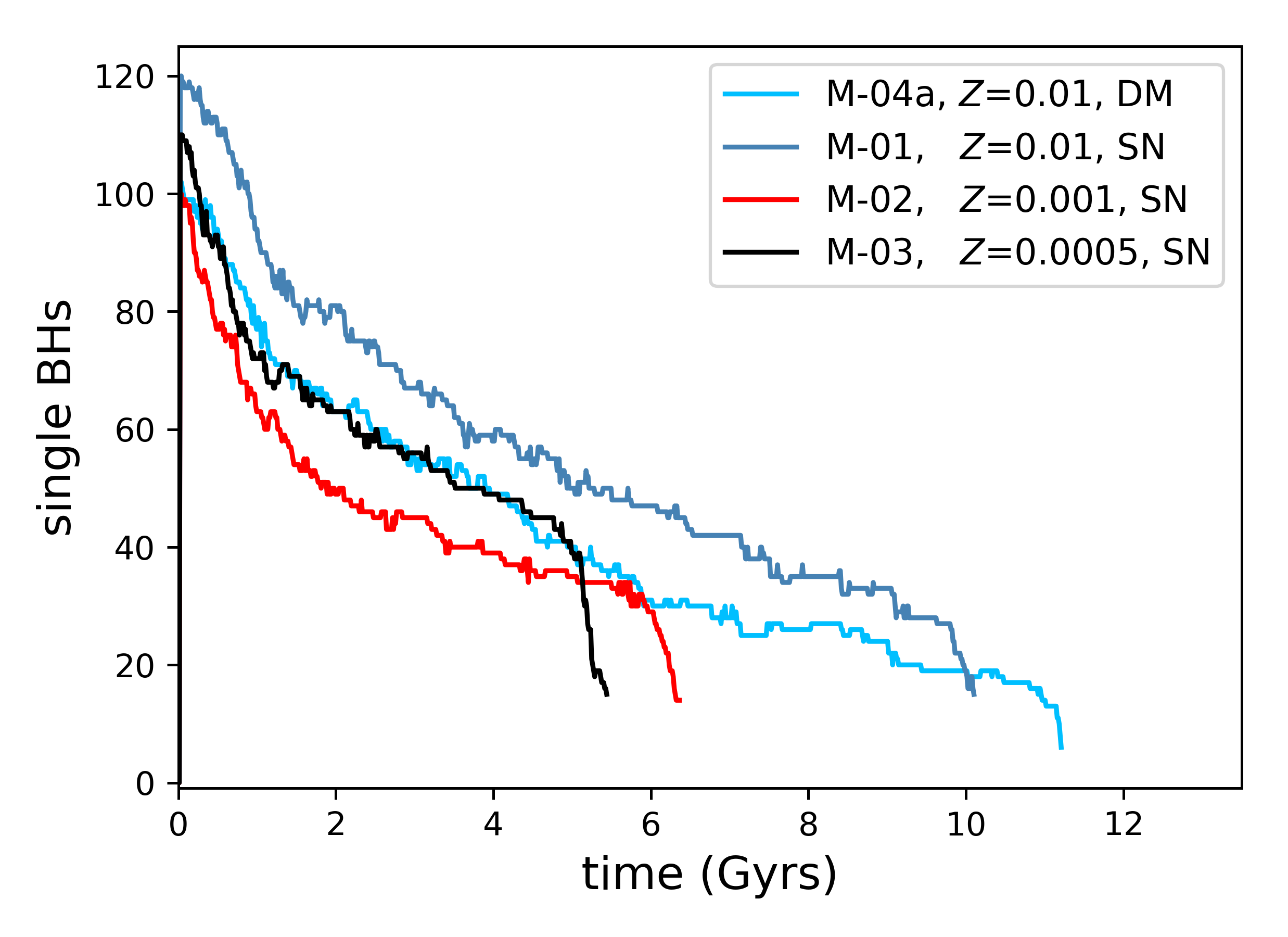}
\includegraphics[width=0.32\textwidth]{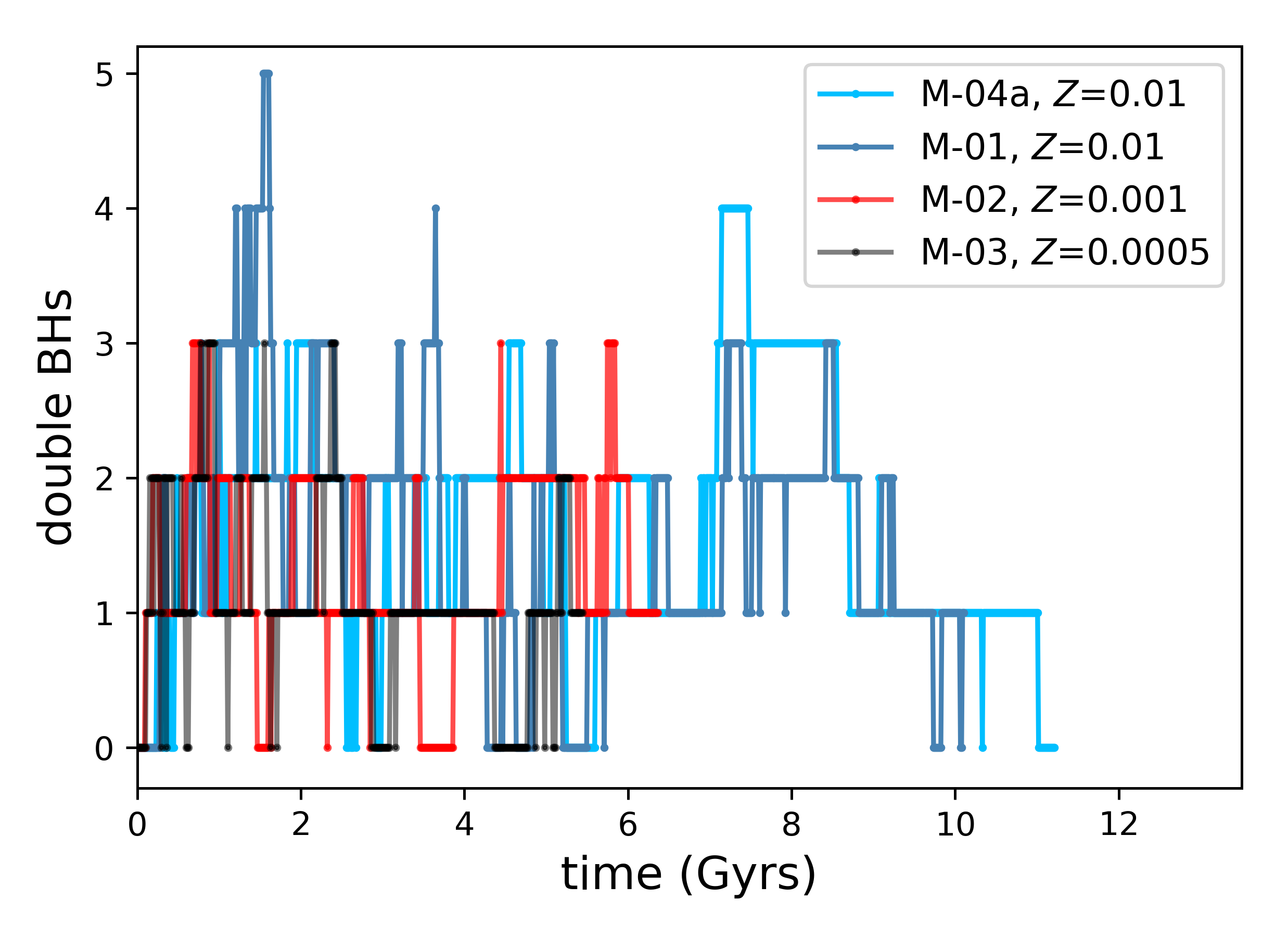}
\includegraphics[width=0.32\textwidth]{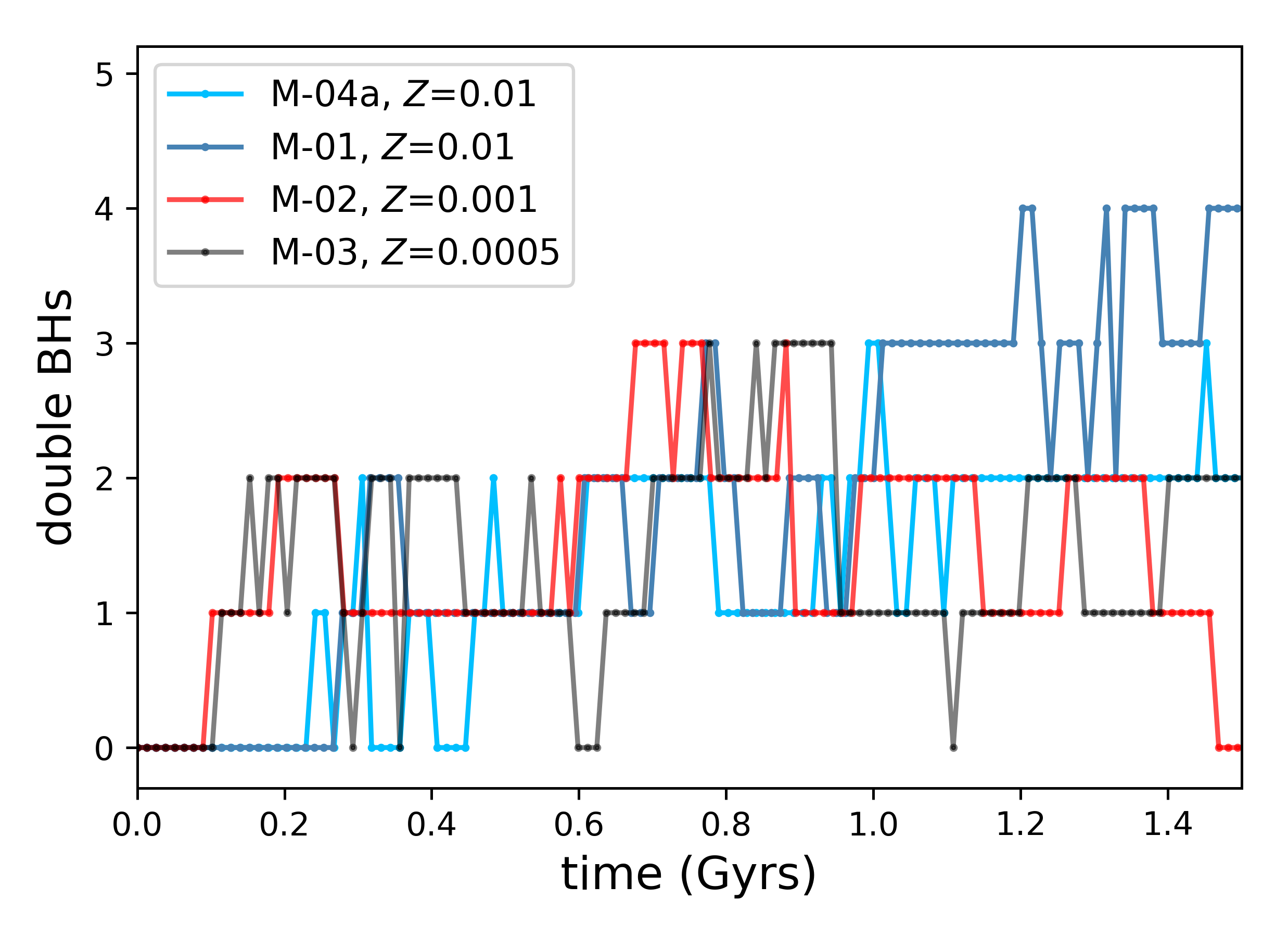}
\includegraphics[width=0.32\textwidth]{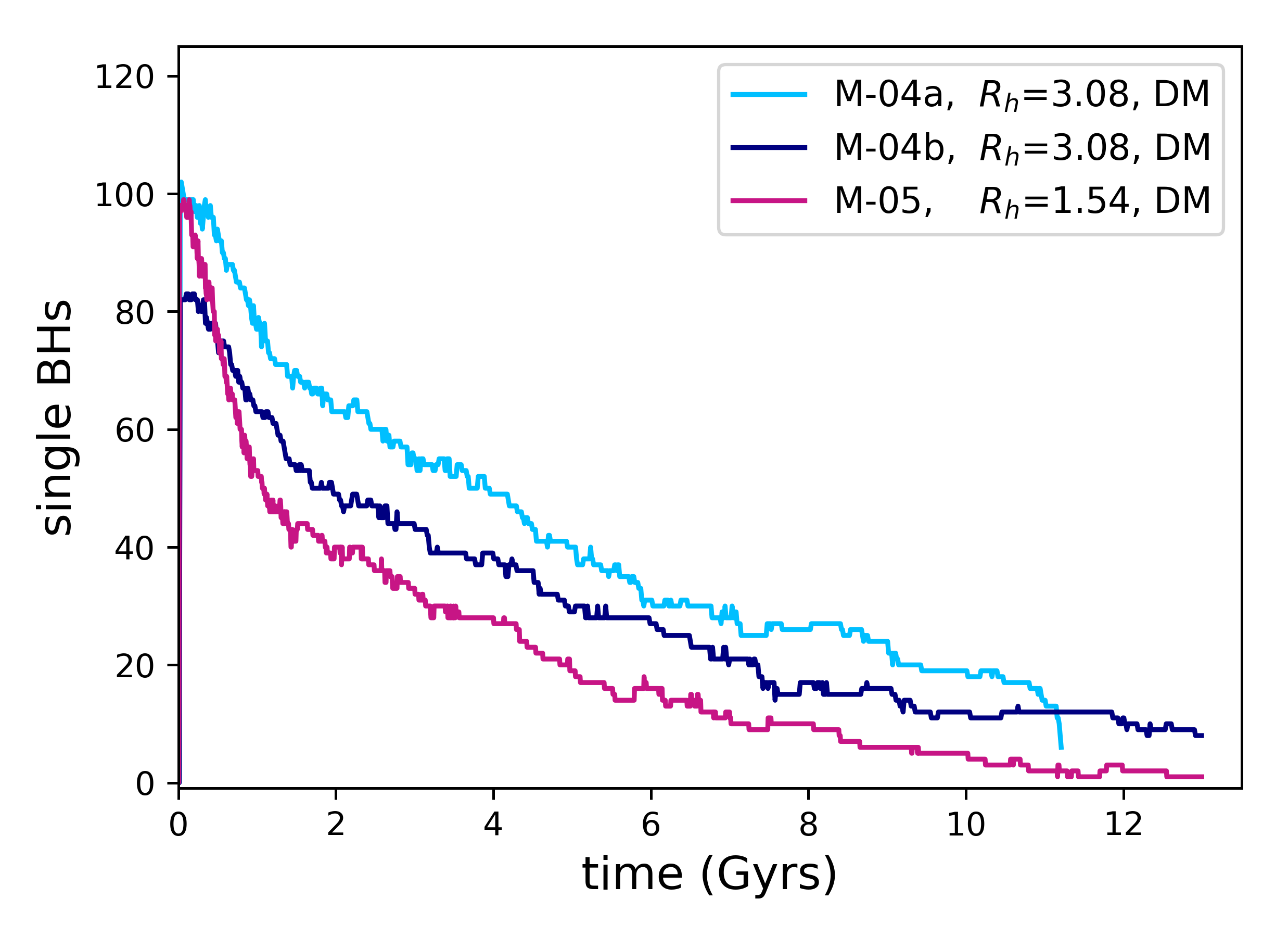}
\includegraphics[width=0.32\textwidth]{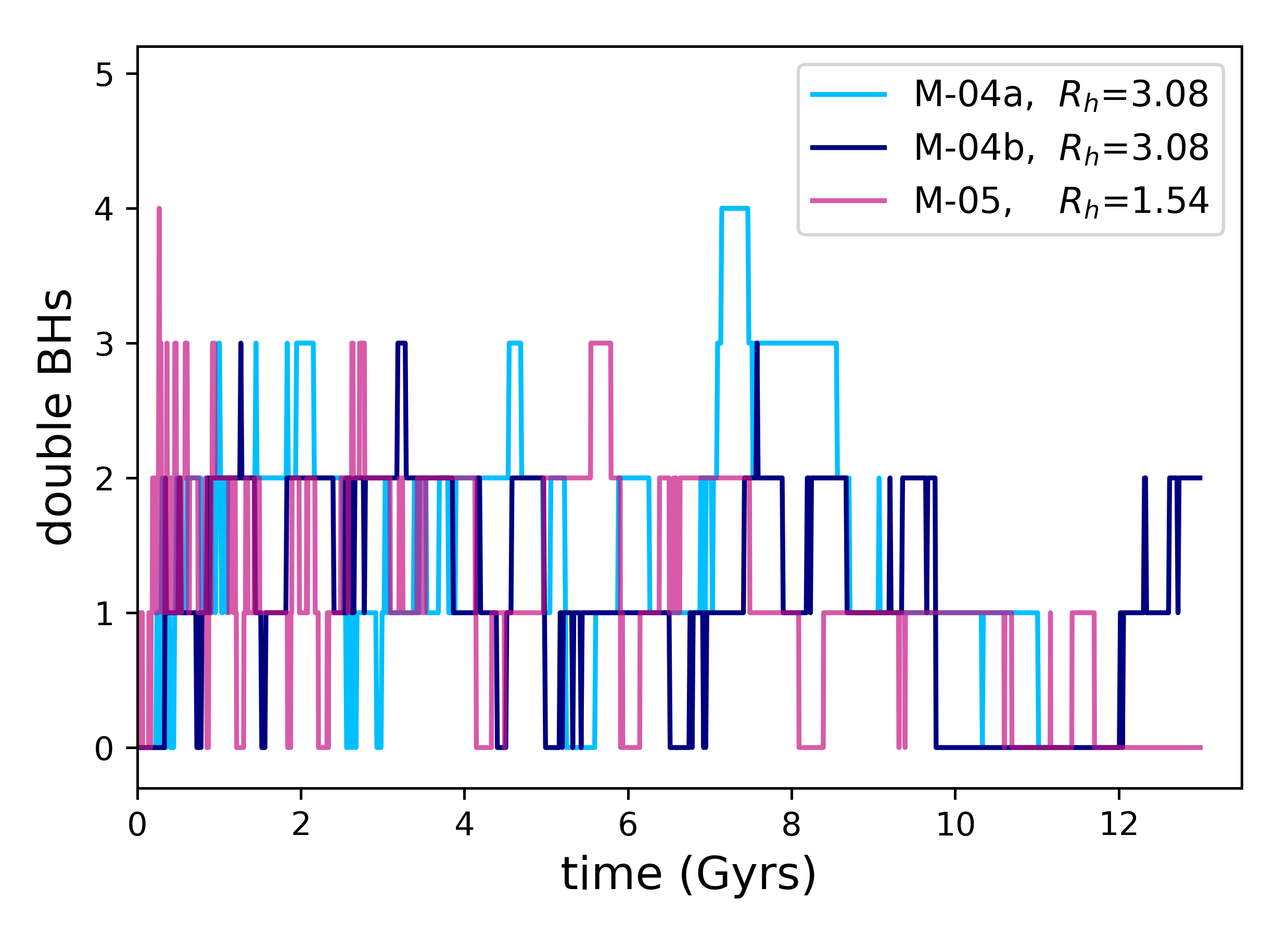}
\includegraphics[width=0.32\textwidth]{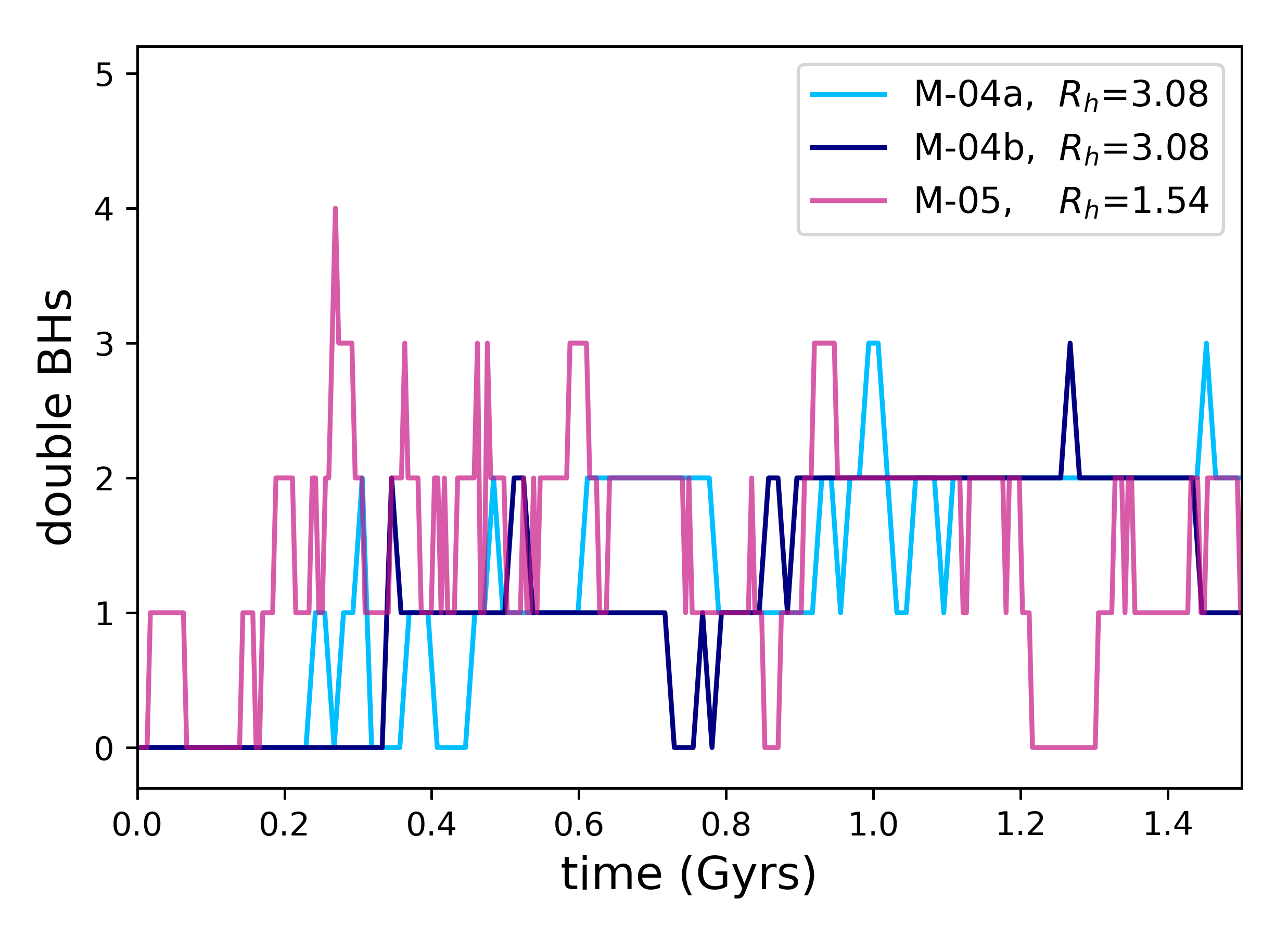}
\caption{Population numbers of BHs present in star clusters as a function of time. The upper panel represents a grid of cluster models with varying metallicities and the lower panel is for a grid with different half mass radii. The time evolution of retained single BHs (left panel) and double BHs (middle panel) over the entire cluster lifetime are shown. A more detailed zoom-in of the double BHs for the first 1.5\,Gyr of cluster evolution is also shown (right panel). }
\label{fig:BH_pop}
\end{figure*}
\begin{figure}\centering
\includegraphics[width=0.35\textwidth]{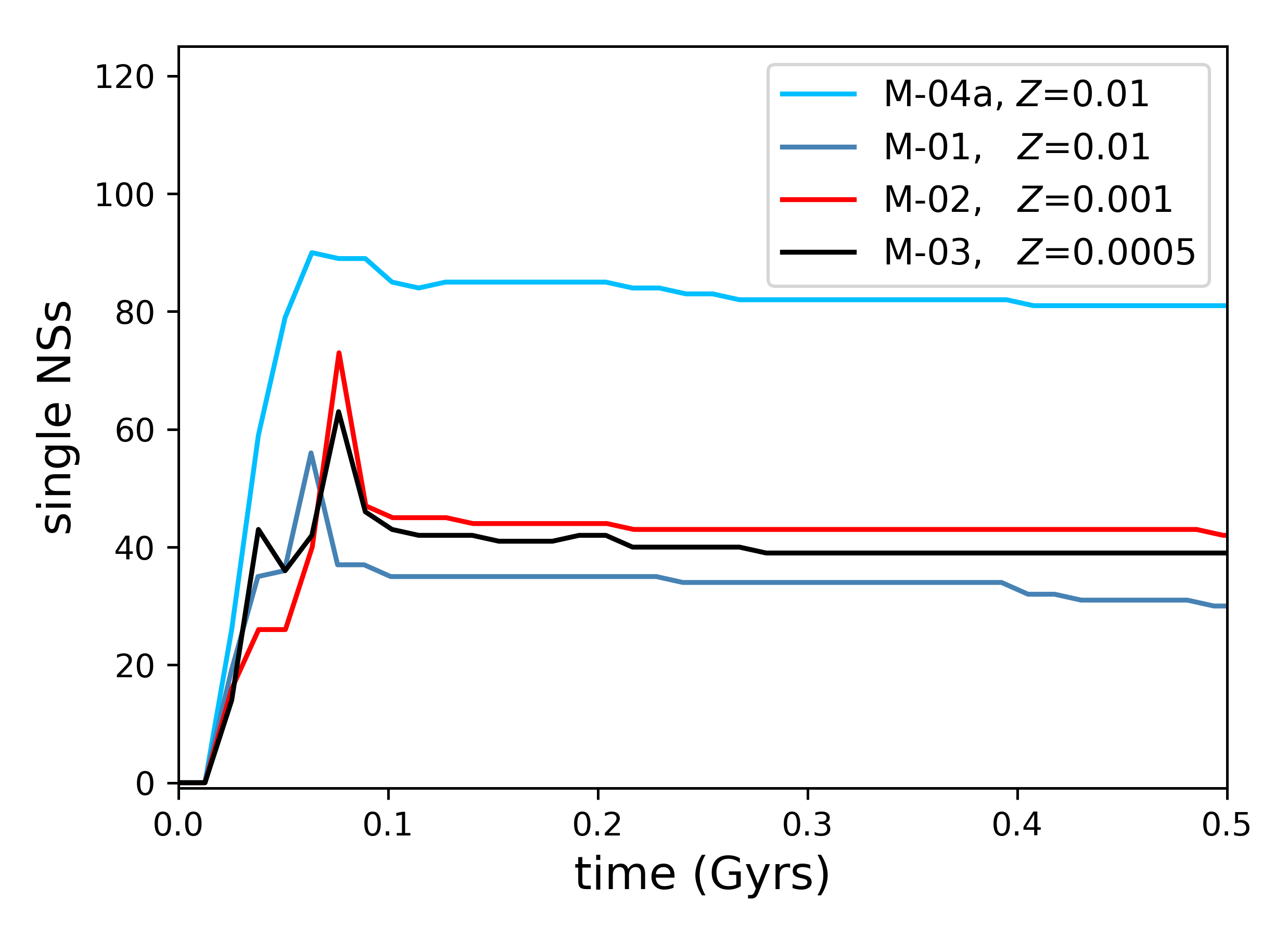}
\includegraphics[width=0.35\textwidth]{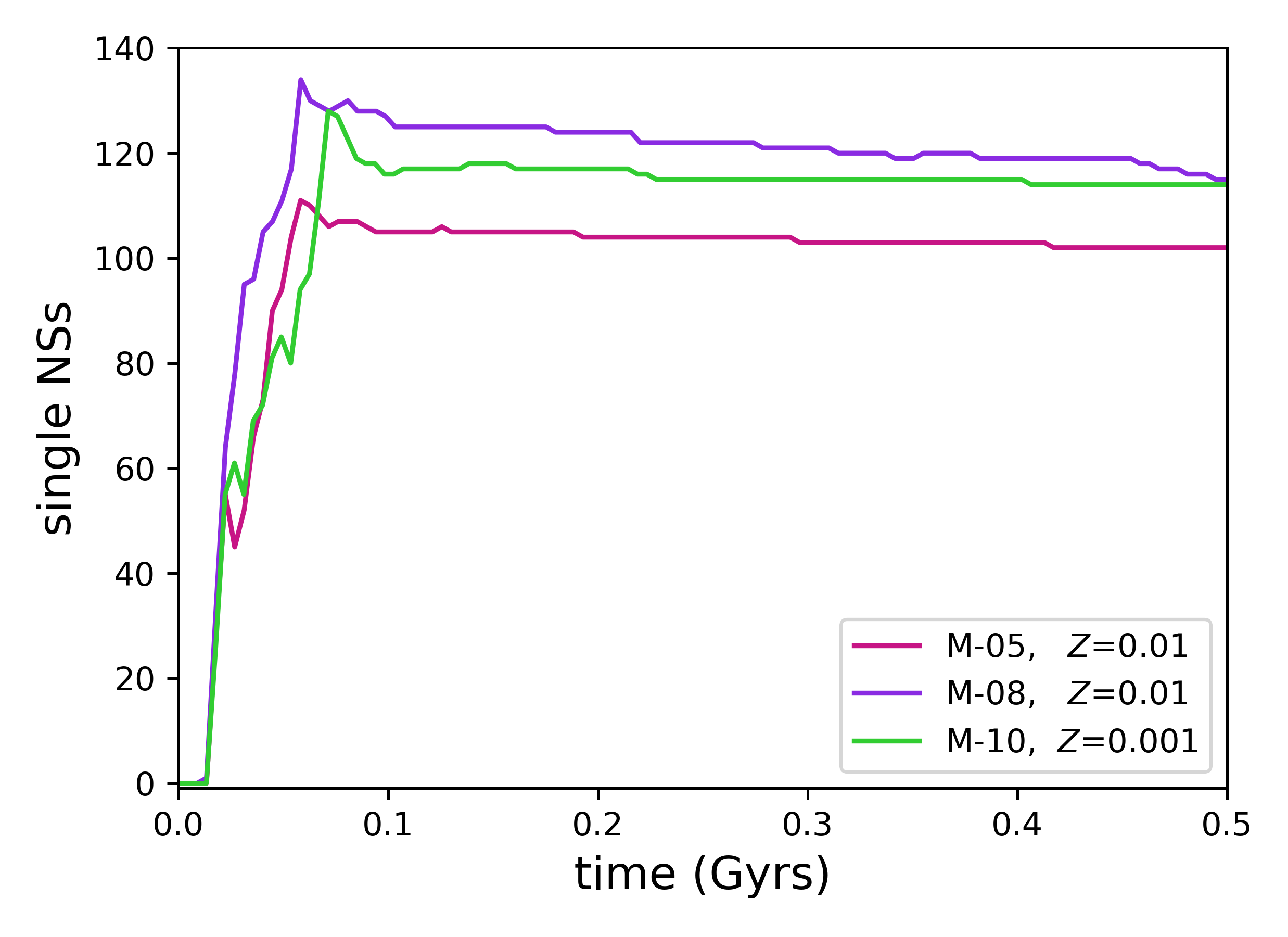}
\includegraphics[width=0.35\textwidth]{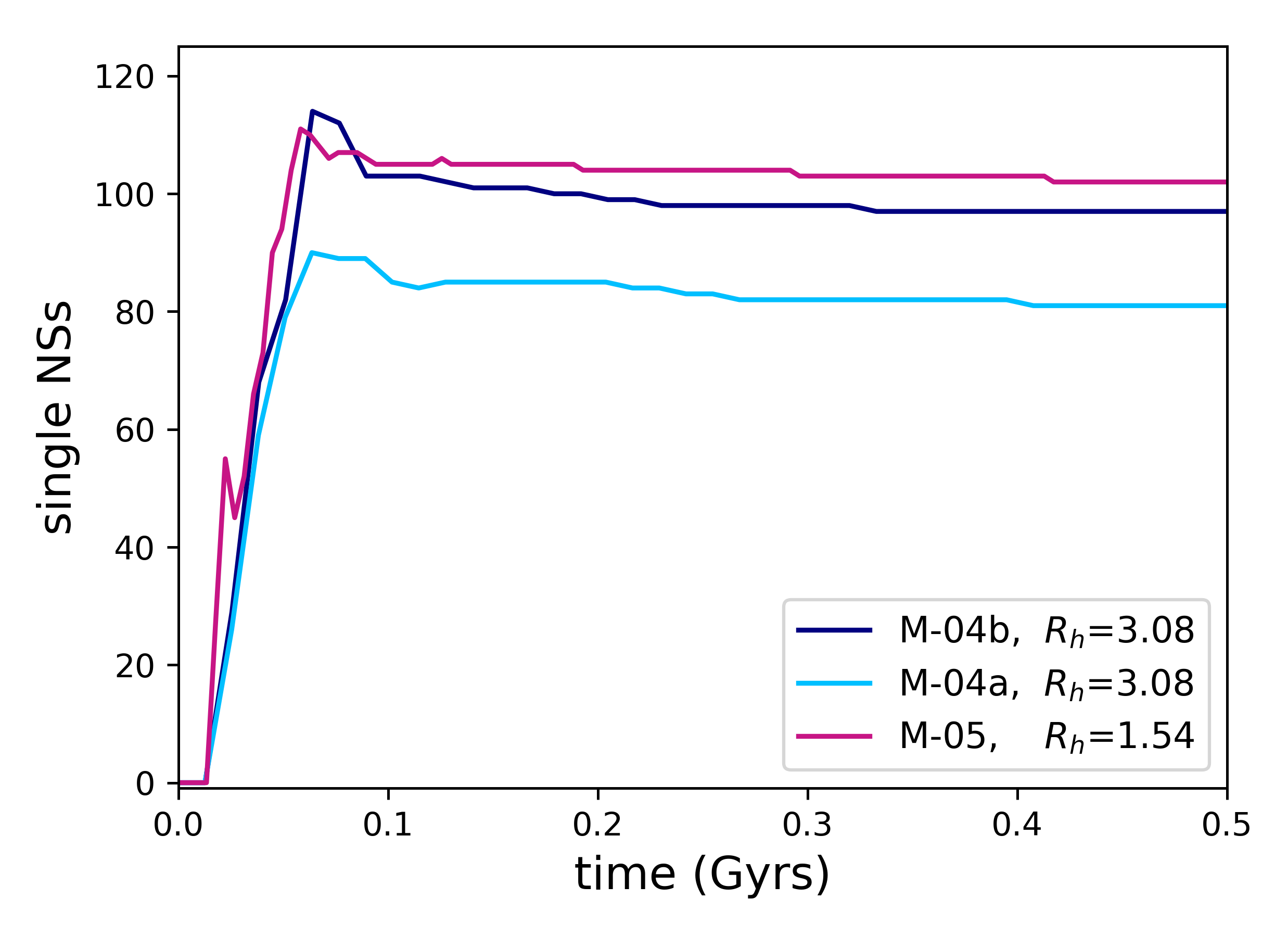}
\caption{Number of neutron stars retained within a star cluster as a function of time, for the first 500\,Myr. The impact of the initial cluster metallicity for two grids of models with initial $R_\mathrm{h} = 3.08$\,pc (top panel) and $R_\mathrm{h} = 1.54$\,pc (middle panel) on the number of single NSs is shown, while the bottom panel shows the same for a grid of models with different initial $R_\mathrm{h}$ but same $Z = 0.01$.
}
\label{fig:NS_pop2}
\end{figure}
\begin{figure}\centering
\includegraphics[width=0.35\textwidth]{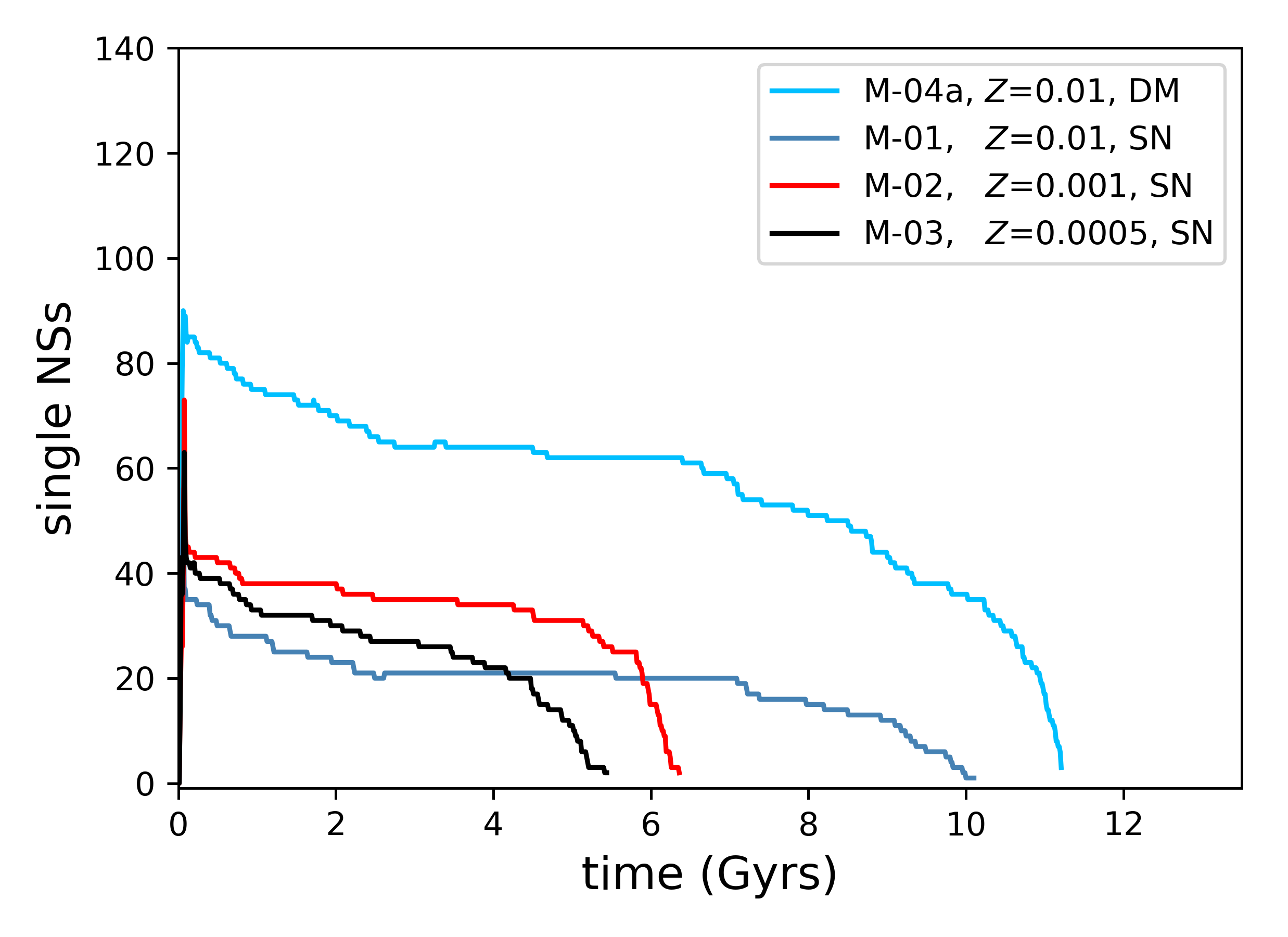}
\includegraphics[width=0.35\textwidth]{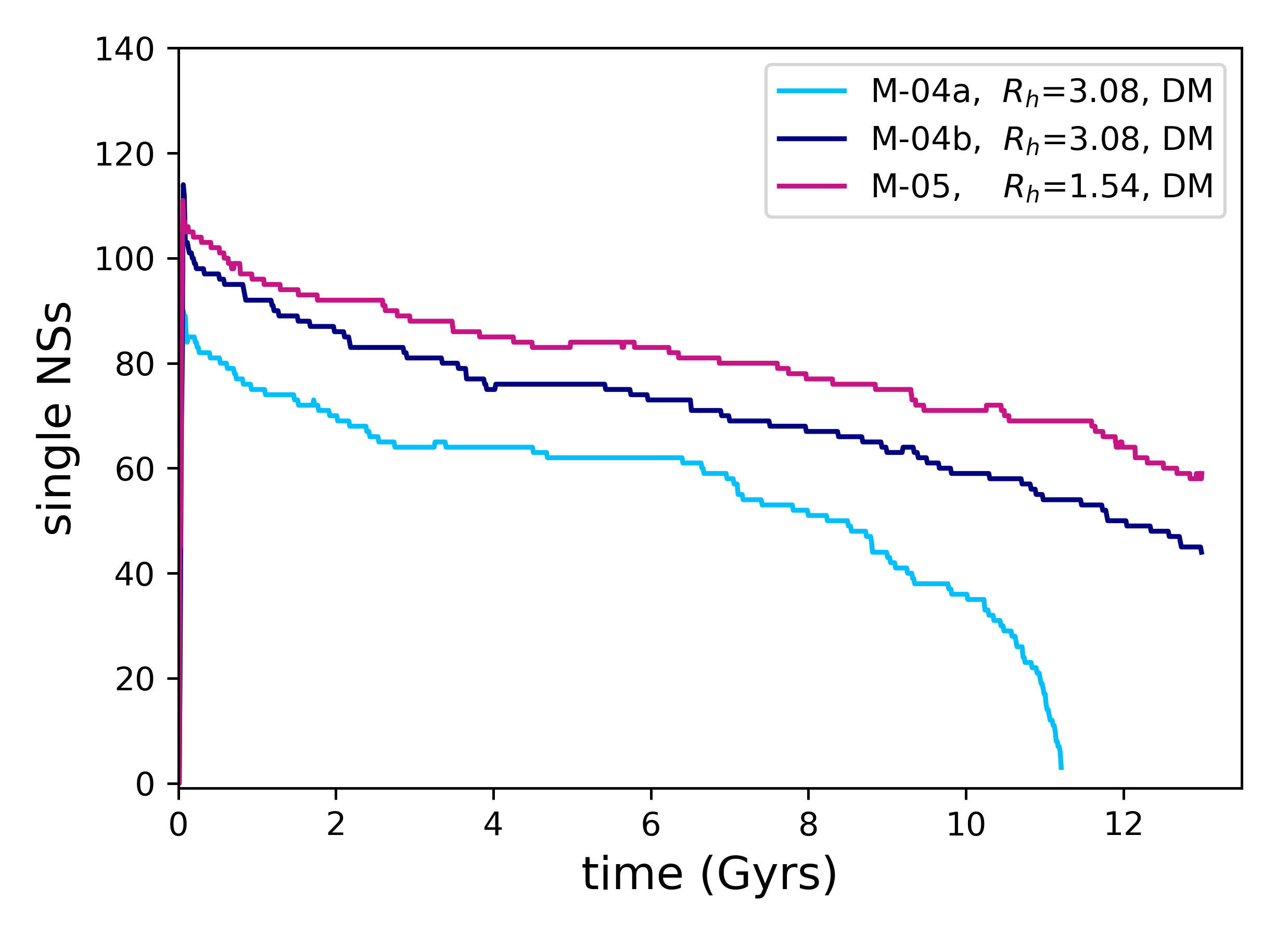}
\caption{Number of neutron stars retained within a star cluster as a function of time, over the full lifetime of the clusters. The upper panel shows a grid of models with different metallicities while the lower panel shows models with different initial half mass radii. }
\label{fig:NS_pop1}
\end{figure}
\begin{figure}\centering
\includegraphics[width=0.47\textwidth]{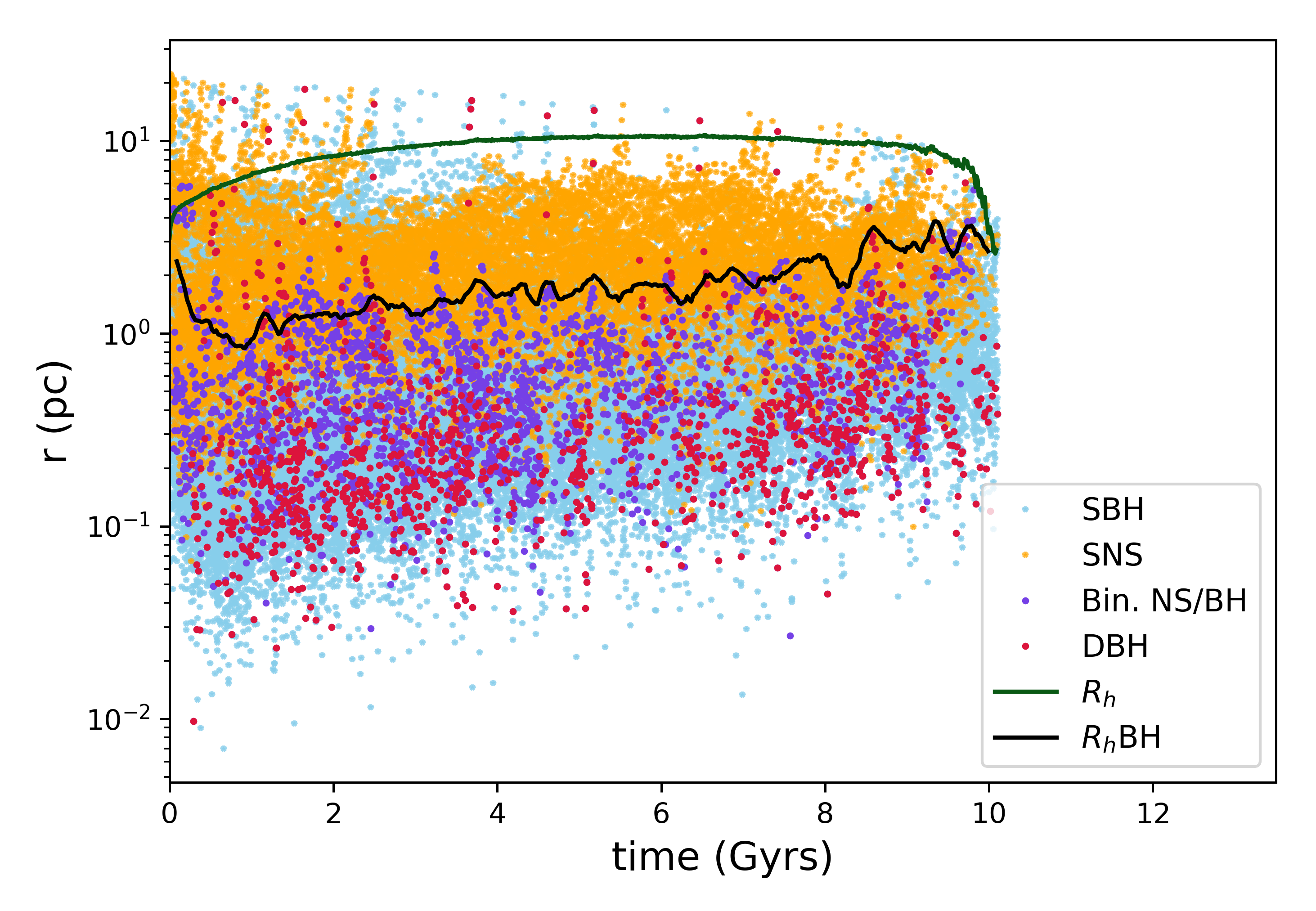}
\caption{Characteristic radii of neutron star and BH populations within the star cluster for model M-01. The green $R_\mathrm{h}$ line represents the cluster half-mass radius and the black $R_\mathrm{h,BH}$ line shows the BH half-mass radius. The positions of the double BHs (DBH in magenta), BHs and NSs in binaries with objects other than BHs (Bin. NS/BH in purple), single NSs (SNS in orange) and single BHs (SBH in blue) are identified as points.}
\label{fig:NS_BH_pos}
\end{figure}

In this section we examine in more detail the populations of NSs and BHs in our clusters.
The retention of these objects inside the cluster is primarily dependent on the supernovae, dynamical and merger remnant recoil kicks. 
While we assume for all our models a uniform NS kick distribution, the BH kicks are scaled down by the fall-back masses (see section~\ref{subsec:winds_and_sne} for details). 
Hence, the mass distribution of the BH-progenitors plays an important role in the retention fraction of BHs (and DBHs). Depending on their masses and orbital parameters, ejected DBHs can also merge outside the cluster. 
In the following sections, we will refer to the BH properties and mergers inside the cluster as `in-situ' and outside the cluster as `ex-situ'. 

Fig.~\ref{fig:BH_pop} shows how the BH and DBH populations evolve for our main set of models. The first point that we notice is that there is no conclusive trend  of the number of BHs produced as a function of metallicity. 
Though the lower $Z$ clusters M-02 and M-03 do produce a relatively lower number of single BHs (100 and 110, respectively, across the first 100\,Myr) than metal-rich M-01 (120), the intermediate metallicity M-02 has the lowest BH number of the three models. 
We also see that the random seed, through its effect on the relative number of massive ZAMS stars drawn from the stellar IMF, is having a greater impact in determining the BH production for the clusters of our size and $Z$ range. 
This effect of statistical noise on BH creation is clearly apparent by comparing M-04a and M-04b, as their difference of about 20 BHs produced is greater than between each pair of the different metallicity models. 
In fact, metal-rich models produce both the highest (M-01, 120) and lowest (M-04b, 80) number of BHs. 
The 100 BHs produced in model M-02 in the first 100\,Myr have a total mass of 1880\,M$_\odot$, evolved from their total ZAMS mass of 3410\,M$_\odot$. 
For M-03, 3760\,M$_\odot$ of ZAMS mass evolves to 110 BHs with a total mass of 2075\,M$_\odot$. 
To compare, for M-01, 120 BHs of net mass 1150\,M$_\odot$ are formed from 4310\,M$_\odot$ ZAMS mass. 
Hence, not only do M-02 and M-03 have nearly 1.6 and 1.8 times more mass in BHs than M-01 but also lose less mass in stellar evolution, converting 55\% of the BH progenitors' ZAMS mass to BHs compared to 27\% for M-01. 
The median BH mass at 100\,Myr (before any escape from the cluster) for M-01 is 9.7\,M$_\odot$, while for M-02 and M-03 it is about 20.5\,M$_\odot$ for both. 

The retention of BHs in-situ is influenced by the cluster metallicity through BH natal kicks and the dynamical activities in the core through generation of thermal energy.
While higher stellar winds in metal-rich clusters results in lower fallback masses and hence larger natal kicks, higher dynamical interactions in low $Z$ clusters cause more heat up (see section.~\ref{sec:DBH_dynamics} for more details). 
The latter becomes more dominant after the formation of the BHs in $\mathcal{O}$(Myr), causing metal-poor clusters to lose more BHs. 
In a time bin of 950--1050\,Myr, the average masses of single BHs for M-01, M-02 and M-03 are about 9.1M$_\odot$,15.6M$_\odot$ and 15.6M$_\odot$, while the BHs in DBH systems have mean individual masses of 10.1\,M$_\odot$, 18.8\,M$_\odot$ and 25.5\,M$_\odot$ respectively. 
The metal-poor clusters thus have nearly 1.8--2.5 times more mass in DBHs. 

A similar thermal effect is observed in the lower initial $R_\mathrm{h}$ cluster model M-05, which initially produces $\approx 100$ BHs. 
Due to the high incidence of dynamical activity in M-05, these BHs then escape from the cluster much more rapidly than in the lower density clusters M-04a and M-04b, leading to the lower density clusters having a larger population of BHs after the first Gyr (lower left panel of Fig.~\ref{fig:BH_pop}). 

As we see from the middle panel of Fig.~\ref{fig:BH_pop}, the number of BH-BH pairs in-situ for all our cluster models oscillates between 0--5 at any given time. 
While the metal-rich clusters M-01 and M-04a have a higher maxima of 5 DBHs than the lower-$Z$ M-02 and M-03 (whose BH-BH count does not exceed three at any point) due to a higher retention fraction of BHs over time, we note that noise can become significant in small number statistics (upper middle panel of Fig.~\ref{fig:BH_pop}). 
The comparatively rapid evolution of M-02 and M-03 in the first 200\,Myr through dynamics is apparent in the top right panel of Fig.~\ref{fig:BH_pop}, showing the quicker formation of DBHs than for M-01. 
The higher density cluster M-05 shows similar behaviour of prompt formation of BH binaries (bottom right panel of Fig.~\ref{fig:BH_pop}) relative to M-04a and M-04b. 
Even though the young M-05 cluster appears to have more BH-BH interactions happening than M-02 and M-03, the latter two show higher thermal radiation and hence more rapid mass-loss and eventual evaporation. 
This all seems to point to a trend where cluster heat up is determined primarily by BH masses while the dynamical interactions are governed mainly by cluster density.  

We also notice that all our models either do not form primordial DBHs, and if they do form, the binaries are very 
short lived. After about a hundred Myrs, the model with the most primordial DBH binaries remaining is M-09 with 4 and the bulk of the models have either 0 or 1 remaining. 
The natal kicks and enhanced participation of BHs in dynamical activities at the onset disrupts primordial binaries that would have otherwise gone on to form DBHs.

The number of single NSs in our set of cluster models varies from 0--135 
for our primary grid of models.
Most of the primordial binaries that produce these NSs are disrupted by supernova kicks, and NSs in binaries remain too rare (less than two) to include in our analysis. 
The effect of statistical noise on the number of NSs is apparent through comparing models M-04a and M-04b with different random seeds with M-04b producing nearly 30 more NSs than M-04a 
which is about 30\% more (Fig.~\ref{fig:NS_pop1}, lower panel). 
Conversely, the difference in the initial half-mass radius does not effect the total number of NSs produced in our simulations (Fig.~\ref{fig:NS_pop1}, lower panel). 
Also, there appears to be no strong correlation between metallicity and the number of NSs produced in our range of models (upper panel(s) of Fig.~\ref{fig:NS_pop1} and ~\ref{fig:NS_pop2}). 
However, the DM91 initial semi-major axis distribution, produces on average more NSs (90--135, the lower and upper limits from models M-04a and M-08) than Sana12 models (none exceeds 90 single NSs). 
Although the Sana12 models have close massive (total mass approximately $>20$M\,$_\odot$, $\log_{10}(P_\mathrm{orb} / \mathrm{day}) \lessapprox 1$) binaries, DM91 models have more binaries in tighter orbits (with $\log_{10}(P_\mathrm{orb} / \mathrm{day}) \lessapprox 0$) 
that are less massive (total mass $<20$M\,$_\odot$), as can be seen from Fig.~\ref{fig:SanaDM}.

The reduction of the number of NSs as a result of natal kicks can be observed in the lower panel of Fig.~\ref{fig:NS_pop2} by the sharp peak at around $100\,$Myr denoting the sudden increase in numbers from star formation/evolution followed by a rapid decrease as many are ejected. 
Fig.~\ref{fig:NS_pop1} then shows the longer-term evolution of the NS population and we see a subsequent 
steady decline of the NSs numbers as they are 
lost through interactions with other cluster members and/or tidally stripped. 

For reference we note that after 1\,Gyr, M-01 has 28 single NSs and one in a binary with an oxygen-neon white dwarf, 
whereas 
M-02 has 38 single NSs (with one NS-carbon-oxygen white dwarf binary) and M-03 has 33 single NSs.

Further utilising 1\,Gyr as a census reference point we find that there are no 
primordial DBHs across models M-01,M-02 and M-03, 
in contrast to the two NS binaries (with white dwarfs as mentioned before) in M-01 and M-02 at the same age that are primordial in nature. 
As non-primordial (i.e. dynamical) binaries\footnote{While compact object binaries born through isolated binary evolution (i.e. without any exchange interactions with other stars or systems) are referred to as primordial binaries, the double compact object formed through interchanging stellar companions are often termed dynamical systems \citep{Hong:2018bqs}.} 
are indicative of 
close gravitational interaction between the compact objects, 
a lack of primordial DBHs and the primordial nature of the few binaries containing a NS that are retained 
demonstrates the dynamically active nature of the BH population 
and on the other hand the non-participation of NSs in exchange interactions. 
Hence, for our suite of models the cluster dynamics is primarily and predominantly determined by the central BH population. 
The BH activity at the cluster core mostly prohibits active dynamical participation of the less massive NSs.

\section{Double Black Hole Properties}
\label{sec:double_black_holes}

We now examine in further detail the populations of double black hole (DBH) binaries in our clusters. 
As demonstrated already, these binaries are of particular importance for the overall evolution of the cluster, and merging DBHs can also be observed directly with gravitational waves. 
For all our models, from very early on in the cluster evolution (within the first $400\,$Myr), all primordial DBHs have been disrupted and any  existing DBHs are of a dynamical nature. 
Thus, the DBHs presented in this section are implied to be dynamically-formed pairs unless stated otherwise. 
In this section we will study the interactive and orbital properties of in-situ DBHs (sec.~\ref{sec:DBH_dynamics}, ~\ref{sec:tdelay}) as well as discuss the ejected ex-situ binary BHs (sec.~\ref{sec:ex_situ}).

\subsection{In-situ DBH: Dynamics}
\label{sec:DBH_dynamics}

\begin{table*}
\label{table:Promiscuity_Mortality_UniquePair }

    \resizebox{0.85\textwidth}{!}{\begin{minipage}{\textwidth}
    \begin{tabular}{lrrrrrrrrrrr}
    \hline
     Model & $\mathcal{U}$ & 3$t_\mathrm{rh}$ & $\mathcal{P}_\mathrm{md}$(3$t_\mathrm{rh}$) & $\mathcal{P}_\mathrm{md}$(1.5) & $\mathcal{P}_\mathrm{md}$($\tau$) &  $\mathcal{P}_\mathrm{mx}$($\tau$) &
    log$_\mathrm{10}\mathcal{S}_\mathrm{md}$(3$t_\mathrm{rh}$) & log$_\mathrm{10}\mathcal{S}_\mathrm{md}$(1.5) &
    log$_\mathrm{10}\mathcal{S}_\mathrm{md}$($\tau$)& log$_\mathrm{10}\mathcal{S}_\mathrm{mx}$($\tau$) & log$_\mathrm{10}\mathcal{S}_\mathrm{mn}$($\tau$)  \\
    - & - & Gyr & - &  - &  - &  - &  Myr &   Myr &  Myr  &  Myr  &  Myr\\

    \hline
M-01 & 139 & 4.7 &2&1.5& 3 & 12 &1.704&1.579& 1.801 & 2.941 & 1.102 \\
M-02 & 64 & 5.8 &2&2& 2 & 12 &1.708&1.583& 1.757 & 3.025 & 1.106 \\
M-03 & 107 & 5.1 &2&2& 3 & 12 &1.707&1.406& 1.582 & 2.905 &  1.105\\
M-04a & 86 & 4.1 &2&1.5& 2 & 9 &1.406&1.406& 1.582 & 3.358 & 1.105 \\
M-04b & 91 & 4.5 &2&2& 3 & 11 &1.709&1.584& 1.952 & 3.619 &  1.107 \\
M-05 & 157 &2.5 &3&2& 3 & 15 &1.129&0.953& 1.351 & 3.286 & 0.652 \\
\hline
M-06 & 18 &1.4& 1&2& 2 & 4 &1.118&1.118& 1.119 & 3.030 & 0.818 \\
M-07 & 209 &4.0&3&2& 3 & 16 &0.997&0.872& 0.997 & 3.276 & 0.395 \\
M-08 & 115 &2.3& 3&2& 3 & 9 &1.254&1.254& 1.255 & 2.786 & 0.653 \\
M-09 & 46 &&&& 2 & 8 &&& 0.953 & 2.129 & 0.652 \\
M-10 & 80 &&&2& 2 & 8 &&1.127& 2.141 & 2.793 & 2.097\\

    \hline
    \end{tabular}
    \end{minipage}}
    \caption{UniquePair ($\mathcal{U}$), Promiscuity ($\mathcal{P}$) and Survivability ($\mathcal{S}$) of all star cluster models presented in this paper (see section~\ref{sec:DBH_dynamics} for more details). The third column shows the equivalent of three half-mass relaxation times (3$t_\mathrm{rh}$) in physical time for each model. The following four columns show $\mathcal{P}$ medians (md) calculated at 3$t_\mathrm{rh}$, 1.5\,Gyr, entire cluster lifetime ($\tau$) and $\mathcal{P}$ maxima (mx) for $\tau$ respectively. The next five columns depict the median values of $\mathcal{S}$ in logarithm after 3$t_\mathrm{rh}$, 1.5\,Gyr, $\tau$ and the maxima and minima of $\mathcal{S}$ in logarithm for $\tau$.
    }
    \label{tab:Prom_imm_unqpr}
\end{table*}

To quantify the DBH dynamical activity in our cluster models, we coin three quantities - ``UniquePair'', ``Promiscuity'' and ``Survivability''.

The number of unique BH pairings that each BH has (i.e. the number of different individual DBH combinations) integrated over i) all BHs and ii) the entire lifetime of the cluster, is termed as UniquePair ($\mathcal{U}$). 
Any binary that is disrupted but is later re-coupled is counted as a new system.
The Promiscuity ($\mathcal{P}$) is defined as the number of unique BH partners per individual BH.
In this case, any rejoining of a BH with its previous companion after a hiatus does not increase its Promiscuity count. 
Binary systems can be broken-up through interactions within binary-single, binary-binary or even higher order chain subsystems. 
For instance, the types of interaction that can contribute to a higher Promiscuity value include an incoming fly-by BH either snatching or replacing another BH in a binary,  
and two binaries interchanging partners in an exchange interaction.
The lifespan of each unique DBH pairing (i.e. the lifetime of each distinct binary) is identified as the Survivability ($\mathcal{S}$) timescale in this paper. 

Promiscuity, UniquePair and Survivability can be used to estimate the extent of the BH dynamical interactions in the cluster. 
In particular, using the three measures in conjunction with each other can help to paint a more detailed story of the cluster dynamics. 
Furthermore, studying the saddle point of the 3-dimensional space of Promiscuity, UniquePair and Survivability that creates the maximum number of possible mergers for a given cluster can lead us in analysing clusters with an ideal DBH merger environment.
While we define UniquePair for the total evolutionary time-scale of the cluster ($\tau$), we compute both Promiscuity and Survivability at three different characteristic times.
Firstly, after a physical time of 1.5 Gyr has elapsed, secondly, after three half-mass relaxation times have elapsed ($3t_\mathrm{rh}$) and finally at the end of the cluster lifetime, $\tau$. 
Analysing Promiscuity and Survivability at separate time snapshots can help to shine more light on the DBH dynamical evolution of the cluster. 
However, unless specified, we will refer to Promiscuity and Survivability at $\tau$, the full duration of the cluster. 
A larger Promiscuity and smaller Survivability would hint at a heightened level of BH-BH interactions, while a higher value of Survivability would suggest less dynamical activity. 
Another consideration is that while an increased level of BH-BH encounters may increase the chances of DBH mergers, too many frequent encounters also has the possibility of breaking up pairs which may have otherwise spiralled in to coalesce. 
Table~\ref{tab:Prom_imm_unqpr} shows the UniquePair, Promiscuity and Survivability distributions of each cluster model presented in the paper.

\subsubsection{Promiscuity}
\label{sec:Promiscuity}

\begin{figure}\centering
\includegraphics[width=0.4\textwidth]{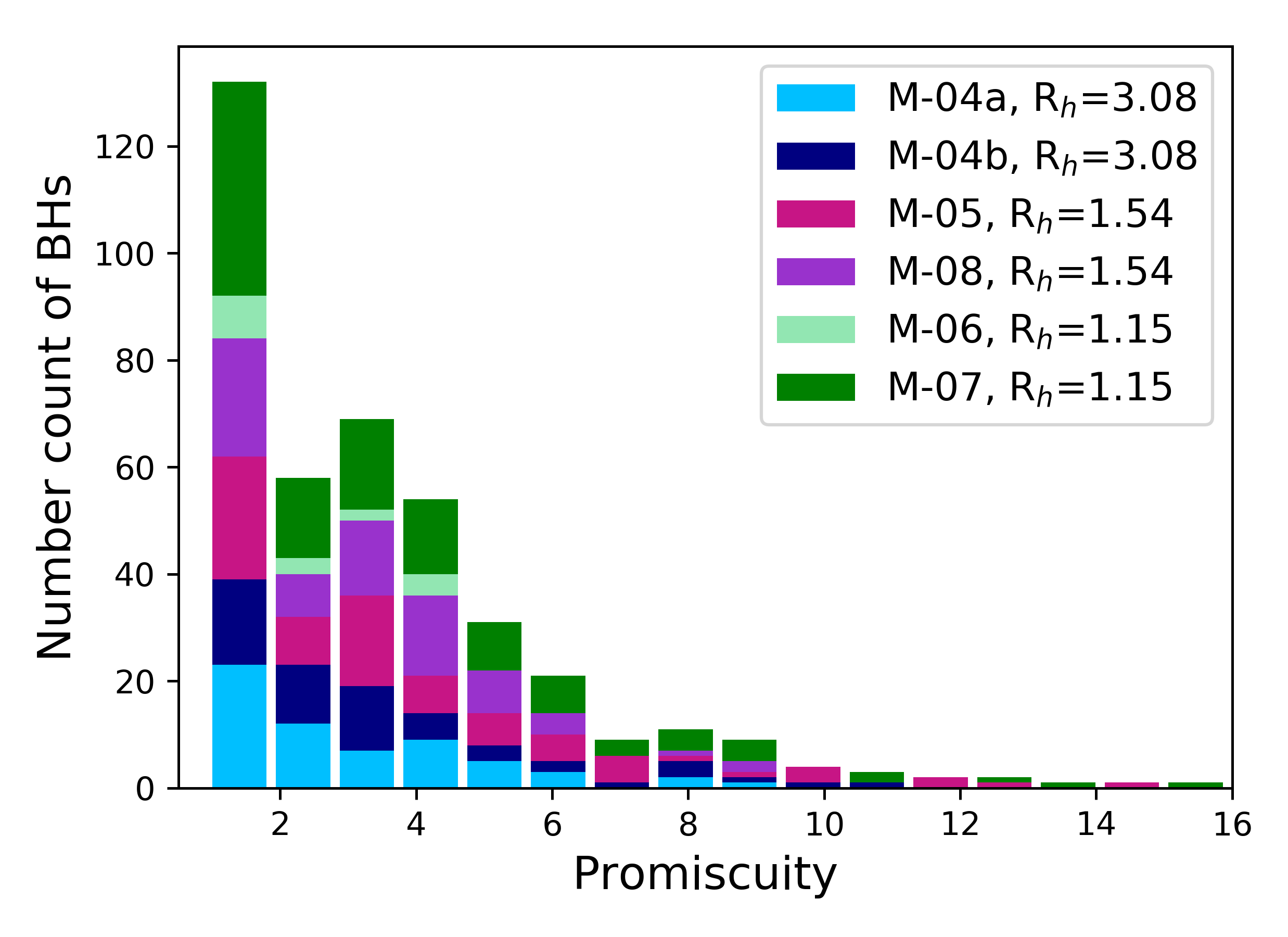}
\includegraphics[width=0.4\textwidth]{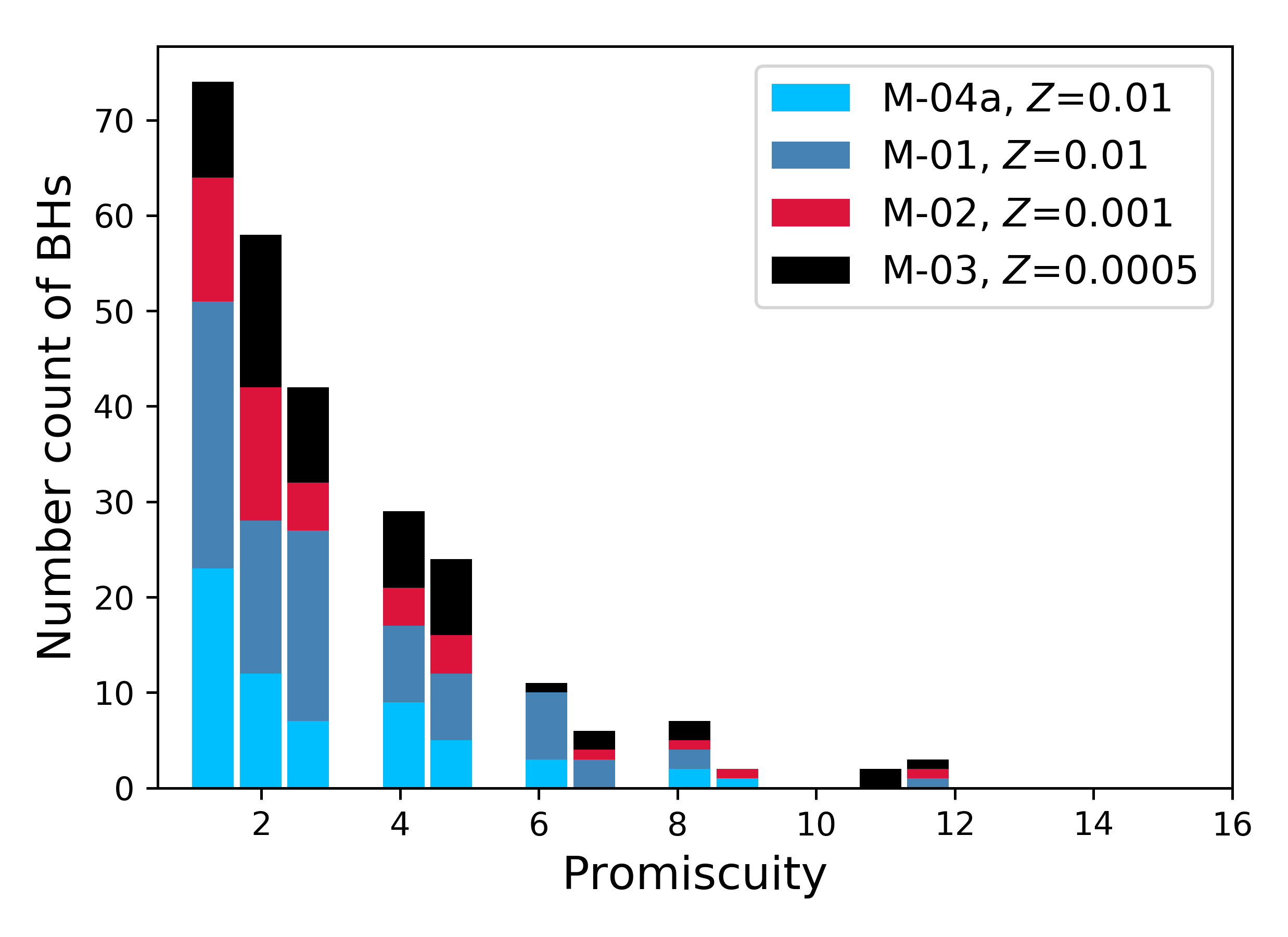}
\caption{The Promiscuity distributions of different cluster models for their entire evolution represented through stacked histograms. The upper panel shows models with varying initial half mass radii and the lower panel with different metallicities.
}
\label{fig:Promiscuity}
\end{figure}

The initial half mass radius of the cluster plays a key role in determining the DBH Promiscuity of a cluster. 
A lower initial $R_\mathrm{h}$ leads to a denser cluster core which, 
as we see from Figs.~\ref{fig:R_M01} and~\ref{fig:NS_BH_pos}, is where most BHs reside and hence leads to a higher level of Promiscuity. 
The top panel of Fig.~\ref{fig:Promiscuity} shows the Promiscuity distribution of individual BHs for a grid of clusters with initial R$_\mathrm{half} = 1.15, 1.54,$ and 3.08\,pc. 
While the  majority of BHs have only one BH companion, clusters of lower initial $R_\mathrm{h}$ (M-07, M-06, M-08 and M-05) show an increased trend towards higher Promiscuity than those with larger $R_\mathrm{h}$ (M-04a and M-04b). 
It is further noted that although M-07 is dynamically younger (as well as younger in physical age) than model M-05, the two models have the same Promiscuity median of 3 and M-07 has a slightly larger maxima at 16 than 15 for M-05. 
An even younger cluster is M-08 which, although sharing the same initial properties as M-05 (aside from the random seed) shows the exact same Promiscuity medians for times $3t_\mathrm{rh}$, 1.5\,Gyr and $\tau$. 
However, the maximum Promiscuity for M-05 is 1.6$\times$ higher than for M-08, 
likely reflecting that the BHs have more time to interact in M-05. 
If we tally Promiscuity with the UniquePair value in Table~\ref{tab:Prom_imm_unqpr}, we observe that M-07 has its UniquePair value nearly 1.5 times higher than the average UniquePair of M-05 and M-08. 
The higher $R_\mathrm{h}$ models M-04a  
and M-04b  
have their UniquePair values even lower than that of M-07. 

This is despite the fact that they are more evolved in time: M-07 reaches an age of $9.6\,$Gyr with 25\% of the initial mass remaining while M-04a and M-04b are evolved to close to dissolution (at $11.2$ and $13\,$Gyr, respectively). 
Furthermore, the metallicity of M-07 is half that of the other models shown in the top panel of Fig.~\ref{fig:Promiscuity} ($Z = 0.005$) and M-07 has more mass (7.1 times that of M-06 and 1.4 times the other models).  
The metal-poor environment and the higher initial mass of M-07 both assist in enhancing the dynamical activity within this model. 
The effect of initial cluster mass becomes even more apparent when M-06, with 
the lowest initial mass, 
shows reduced DBH dynamical activity through an UniquePair value of only 18 and Promiscuity maxima at 4. 
However, the Promiscuity median remains 2, equal to that of M-04a and close to the other models, showing that this, individually, is not such a useful indicator. 
We instead emphasize the importance of the third variable Survivability to realise a more complete picture of cluster core dynamics, as will  be discussed in the following section. 

The initial semi-major axis distribution appears to affect the DBH activity of the cluster as illustrated in the middle panel of Fig.~\ref{fig:Promiscuity}. 
M-01, using the Sana12 initial semi-major axis distribution, shows a slightly higher Promiscuity maxima (and also about 1.6 times larger UniquePair value) than M-04a and M-04b with the DM91 initial semi major axis distribution.
With a closer initial separation for more massive stars than DM91 (see Fig.~\ref{fig:SanaDM}), the Sana12 distribution hence seems to assist in increased cluster core activity. 

Lower metallicity seems to slightly increase the Promiscuity of the cluster BHs as is shown in the middle panel of Fig.~\ref{fig:Promiscuity}, where the low-$Z$ cluster models are biased further towards higher Promiscuity values, even though they have the same maxima. 
However, the higher-$Z$ cluster M-01 has a higher UniquePair value than M-02 and M-03. 
In metal-rich environments with elevated stellar winds, lower-mass BHs are formed. 
M-01 creates less massive but slightly more numerous BHs (Fig.~\ref{fig:BH_pop}) than M-02 and M-03, giving rise to an increased number of unique individual DBH binaries. 
However, the more massive BHs of the low metallicity clusters are more likely to reside in the denser environment at the very centre of the cluster thus experiencing stronger gravitational perturbation and thus have a slightly higher interaction rate as shown in the Promiscuity histogram.

To compare the effect of the time evolution of the clusters, we compute the Promiscuity medians at the age of 1.5\,Gyr and after 3$t_\mathrm{rh}$ have elapsed. 
We observe that at the same dynamical age (3$t_\mathrm{rh}$), the denser clusters M-05 and M-08 have a higher Promiscuity median (than less dense M-04a, M-04b, M-01, M-02, M-03) but at 1.5\,Gyr this difference is less apparent. 
Comparing the Promiscuity medians at 3$t_\mathrm{rh}$ and $\tau$ for all six primary models (that evolve to completion, by our definition), we observe that for most the median increases as we go from 3$t_\mathrm{rh}$ to $\tau$.
However, for M-02 and M-04a the two values remain the same (their UniquePair values are also lower than the other primary models) showing that their slightly lower number of BHs (see Fig.~\ref{fig:BH_pop}) slightly hampers the Promiscuity. 
M-02 still reaches a Promiscuity maxima of 12, showing that even with the low UniquePair count, the low-$Z$, more massive BH environment creates ample exchange interactions.

\subsubsection{Survivability}
\label{sec:Survivability}

\begin{figure}\centering
\includegraphics[width=0.4\textwidth]{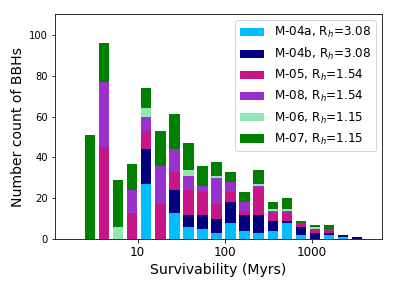}
\includegraphics[width=0.4\textwidth]{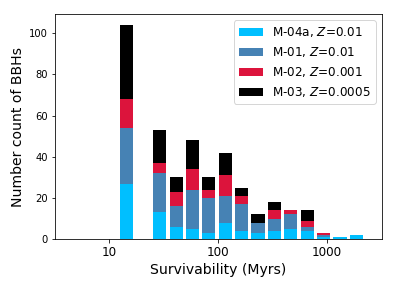}
\caption{Stacked histogram of the Survivability parameter for grids of cluster models for the entirety of their evolution with different initial half mass radii (top panel) and with different metallicities (bottom panel).
}
\label{fig:Survivability}
\end{figure}

An increased number of BH-BH exchange interactions leads to short-lived binary pairings. 
Hence, the Survivability coefficient of a cluster model is ordinarily expected to be inversely related to the Promiscuity count. 
For instance, M-07 which shows the largest UniquePair and Promiscuity maxima also has the lowest Survivability medians (Table.~\ref{tab:Prom_imm_unqpr}). 

A lower initial half mass radius for clusters of similar initial mass ensures a denser environment and hence increased rate of close encounters. 
This can decrease the lifetime of each DBH system and lower the calculated Survivability. 
The effect is illustrated in the top panel of Fig.~\ref{fig:Survivability}, with smaller $R_\mathrm{h}$ clusters M-05 and M-08 showing lower Survivability minima than M-04a and M-04b of the same metallicity and initial number of stars. 
M-06 is a small cluster of initial mass  $< 20\%$ of M-01. 
In spite of its small number of stars it shows the signature of high rate of BH-BH interaction through having smaller Survivability values. 
The higher Survivability medians at both 3$t_\mathrm{rh}$ and 1.5 Gyr of M-04a, M-04b compared to M-05, M-08, M-06 and M-07 
(see Table~\ref{tab:Prom_imm_unqpr}) 
shows the similar trend of anti-correlation with their initial $R_\mathrm{h}$.  
From this we infer that the dynamical timescales are not significantly effected by the initial density and hence the relative magnitudes of the DBH signatures at the same physical and dynamical timescales for these clusters do not change.  

The importance of the initial semi-major axis distribution is apparent through the lower panel of Fig.~\ref{fig:Survivability}, where we can compare M-01 to M-04a and observe the bias of M-04a towards larger Survivability values. 
The presence of long-lived DBH binaries in M-04a and M-04b can also be seen in the upper panel of Fig.~\ref{fig:Survivability} where the distribution tails towards Survivability values higher than $1\,000\,$Myr. 
The higher Survivability maxima of M-04a (and M-04b) relative to M-01 is further evident in Table.~\ref{tab:Prom_imm_unqpr}.

Although the trend with metallicity is not as strongly apparent as for models with varying initial $R_\mathrm{h}$,  
the lower panel of Fig.~\ref{fig:Survivability} does show a steeper slope of Survivability distributions for the metal-poor models M-02 and M-03 compared to M-01 with higher metallicity. 
M-01, M-02 and M-03, arranged by descending $Z$, have consecutively lower Survivability medians (albeit slightly); signifying an increased incidence of dynamical interactions. 
Comparing the Survivability medians at 1.5\,Gyr, we still observe the metal-poor clusters to have consecutively shorter DBH lifespans. 
To account for differences in dynamical timescales of clusters of different $Z$, we can compare the Survivability medians at 3$t_\mathrm{rh}$ and see that M-01, M-02, M-03 all have very similar values. 
That the difference in BH dynamical interactions for clusters of different $Z$ is not significant is further illustrated by comparing the Survivability of M-05, M-08 and M-10. 
Since M-10 is only evolved to about 1 $t_\mathrm{rh}$, we compare Survivability at 1.5\,Gyr from Table.~\ref{tab:Prom_imm_unqpr} and observe their similarity. 
In fact, M-08 actually has a slightly lower value than M-10 even though the latter is comparatively metal-poor. 
This emphasizes that while cluster heat-up and hence the dynamical time-scale is dominated by the BH mass spectrum (different $Z$ clusters producing BHs of different mass ranges), the BH-BH interaction is mainly influenced by cluster density.

A way to compare Survivability is through the inverse of encounter rate $\mathcal{R} \sim \rho$/$(a\sigma)$ as introduced by \citet{Hills_Day:1976}, where $\rho$ is the local density, $\sigma$ is the velocity dispersion and $a$ is the semi-major axis of the binary. For dynamical DBHs of equal masses $m_\mathrm{BH}$;  $a\sim m_\mathrm{BH}/\sigma^2$, making $\mathcal{R} \sim \rho\sigma/m_\mathrm{BH}$. At the same dynamical age, metal poor clusters with more massive DBHs thus should have a lower $\mathcal{R}$, reflected by a slightly higher  log$_\mathrm{10}\mathcal{S}_\mathrm{md}$(3$t_\mathrm{rh}$). However, as the density and velocity dispersion also determine  $\mathcal{R}$, this trend is not observed for the Survivability calculated at similar physical timescales or at the end of cluster evolution (clusters of different metallicity have significantly different lifetimes in our models).

\subsection{Chirp Mass and Mass Ratio: In-situ}
\label{sec:mchirp_q}

We compare the in-cluster DBH masses through the chirp mass and mass ratio distributions. 
The chirp mass ($M_\mathrm{chirp}$) is defined as \begin{equation}
    \label{eqn:chirpmass}
    M_\mathrm{chirp}=(m_1 m_2)^{3/5}(m_1 + m_2)^{-1/5} ,
\end{equation}
and the mass ratio $q=m_{2}/m_{1}$, where $m_1 > m_2$ are the masses of the two BHs in a binary. 
$M_\mathrm{chirp}$ is the dominant term determined directly from the gravitational waves signal \citep{Abbott:2016bqf}. Observationally, the individual masses $m_1$ and $m_2$ are derived from $M_\mathrm{chirp}$ 
and 
$q$ (which has more uncertainty), leading to larger error bars in $m_1$ and $m_2$ than $M_\mathrm{chirp}$. 
Thus, instead of showing the mass distribution of the individual BHs ($m_1$ and $m_2$) in DBH binaries, we express our mass distribution in terms of $M_\mathrm{chirp}$ and $q$. 
Since most of our models have no surviving or very short-lived primordial DBHs, the discussion and plots of DBHs in the following sections are effectively of dynamical DBHs, unless mentioned otherwise.

\subsubsection{Chirp Mass}
\label{sec:mchirp}

\begin{figure}
\includegraphics[width=0.35\textwidth]{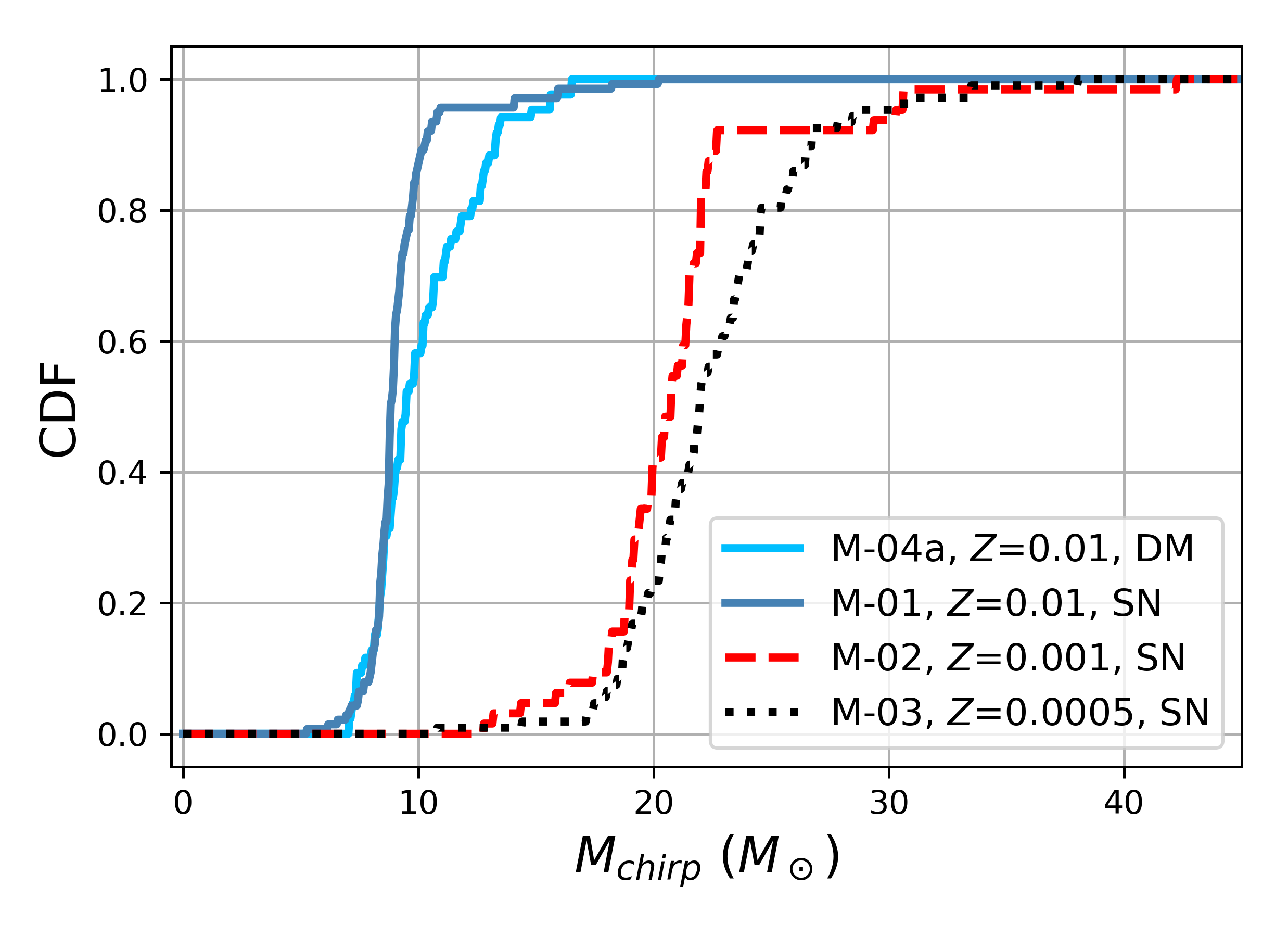}
\includegraphics[width=0.35\textwidth]{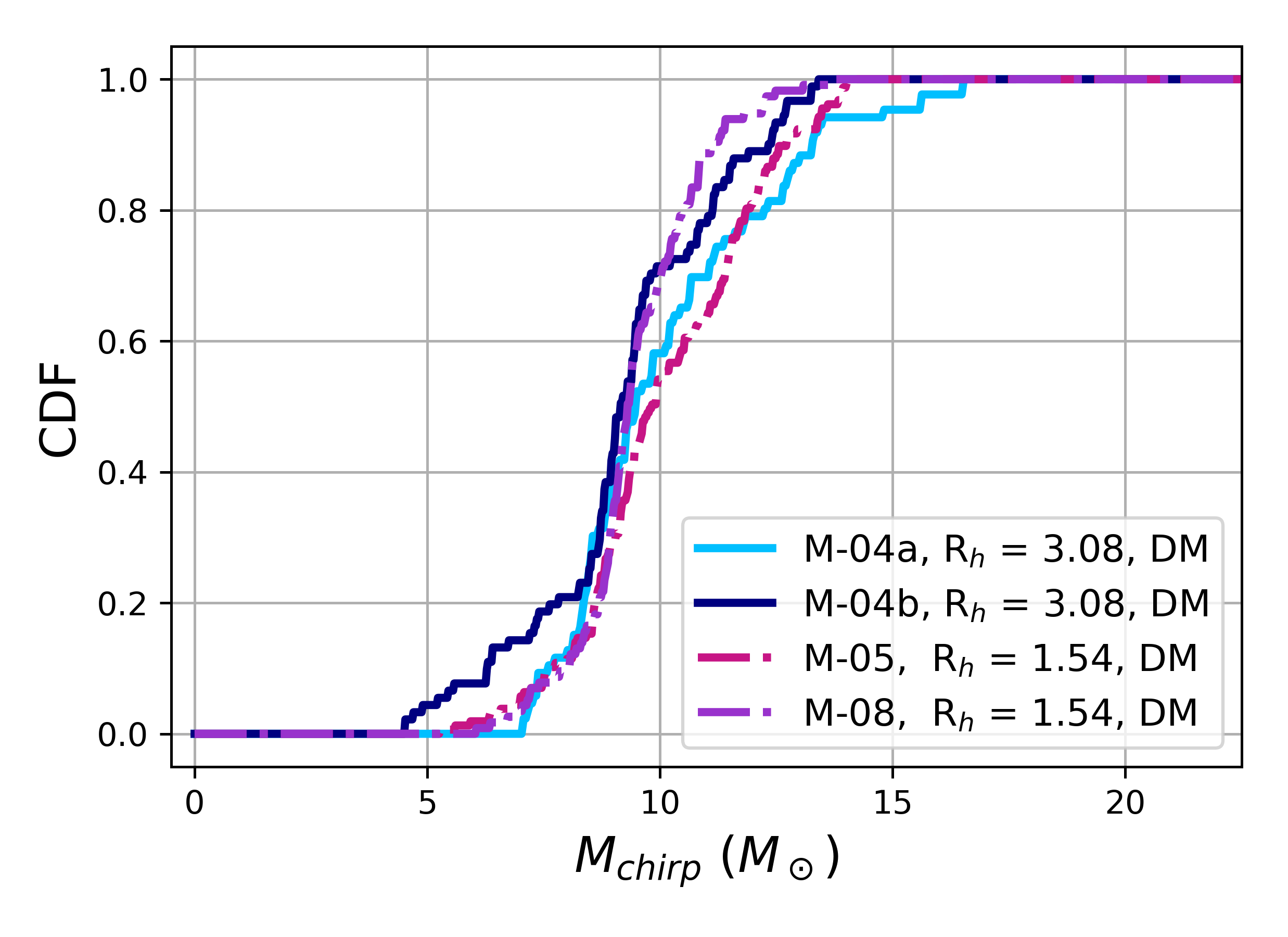}
\includegraphics[width=0.35\textwidth]{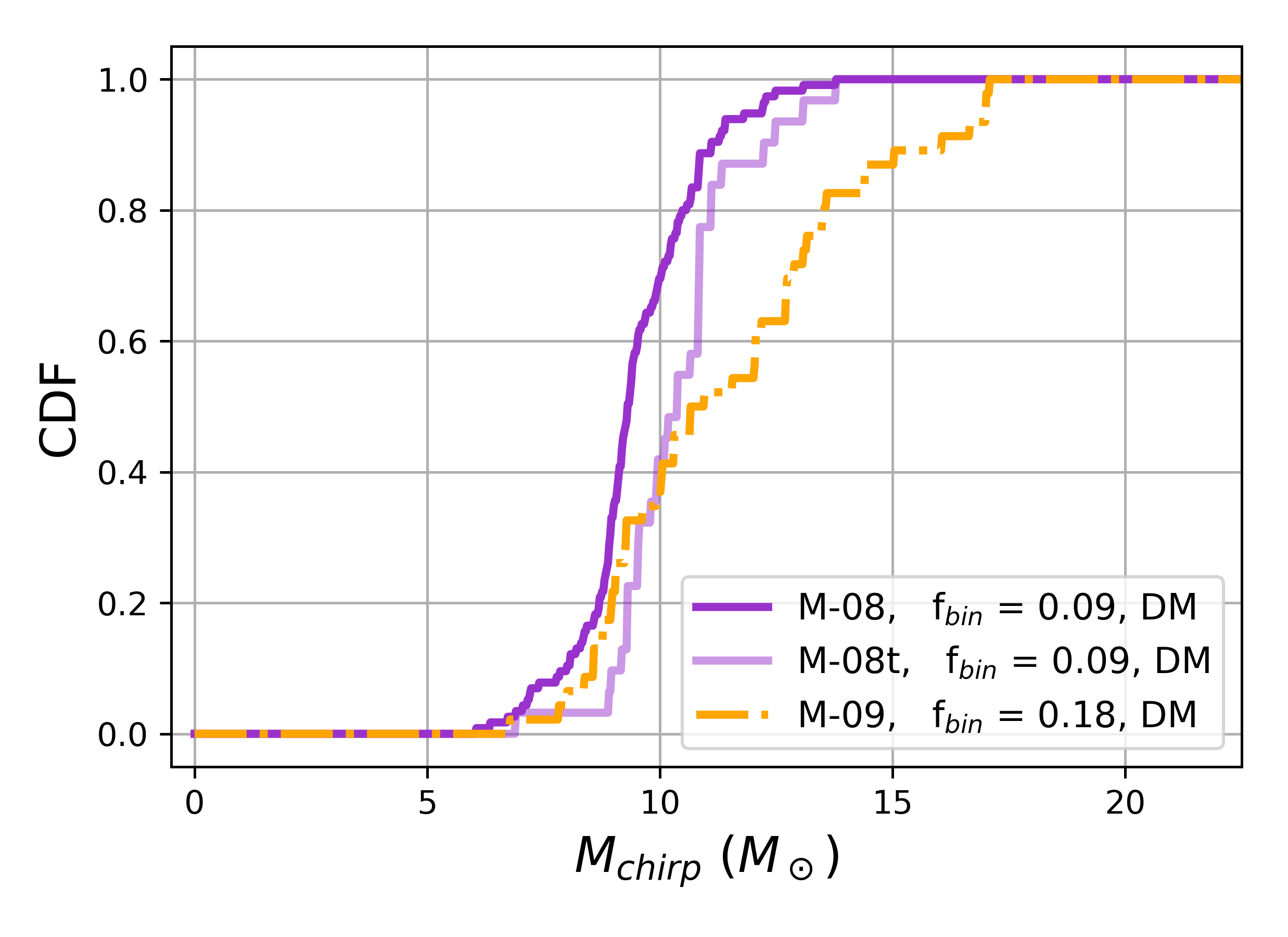}
\caption{Distributions of DBH chirp masses. The top panel shows the variation with metallicity, the middle panel shows the effect of the initial cluster half-mass radius whilst the bottom panel shows the effect of varying the initial binary fraction.}
\label{fig:Mchirp}
\end{figure}

Cluster metallicity plays the key role in determining the BH masses and hence the DBH chirp mass ($M_\mathrm{chirp}$) distribution. 
Lower $Z$ allows the formation of heavier BHs, due to decreased stellar winds \citep{Hurley:2000pk, Belczynski:2010}. 
The more massive BH progenitors also have typically larger fall-back masses, increasing the BH mass and decreasing the BH natal kick, thus retaining more of these heavier BHs in the cluster. 
The median BH mass in a binary with another BH for M-01 ($Z=0.01$) is $\approx$10.1 M$_\odot$, for M-02 ($Z=0.001$) is $\approx$24.3 M$_\odot$ and for M-03 ($Z=0.0005$) is $\approx$26.5 M$_\odot$. 
The top panel of Fig.~\ref{fig:Mchirp} shows the $M_\mathrm{chirp}$ distributions for clusters with different metallicities through their respective Cumulative Density Functions (CDFs).
M-01 has
a median $M_\mathrm{chirp}$ of 8.9\, M$_\odot$,
while M-02 and M-03 have median $M_\mathrm{chirp}$ of 20.7\,M$_\odot$ and 22.0\,M$_\odot$ respectively.

The initial $R_\mathrm{h}$ of a cluster model does not significantly alter the $M_\mathrm{chirp}$ distribution within our range of variation of $R_\mathrm{h} = 1.54, 3.08$ pc (see Fig.\ref{fig:Mchirp}, middle panel). 
The median $M_\mathrm{chirp}$ of M-04a and M-04b are 9.4\,M$_\odot$ and 9.2\,M$_\odot$ respectively,
while for the lower 
$R_\mathrm{h}$ models M-05 and M-08 the medians are $9.8$\,M$_\odot$ and $9.2$\,M$_\odot$ respectively. 
It is also to be noted that though M-05 and M-08 are at different physical and dynamical ages the mass distributions are not significantly affected by this (see the next paragraph for further details). 

Models M-08 and M-09 have the same initial total number of stars, metallicity, initial semi-major axis distribution (DM91) and initial half mass radii but different binary fraction: 
M-09 having double the number of primordial binaries than M-08. 
Since the final physical ages of M-08 and M-09 are different, we define M-08t, 
which is model M-08 at the same maximum physical age reached by M-09 (about 400\,Myr) to compare the effect of initial binary fraction (f$_\mathrm{bin}$=0.09 and 0.18 for M-08 and M-09, respectively) on the DBH mass distribution. 
As shown in the lower panel of Fig.~\ref{fig:Mchirp}, the higher initial f$_\mathrm{bin}$ cluster M-09 has a significantly more massive $M_\mathrm{chirp}$ distribution: 
80\% of the M-09 
DBHs have $M_\mathrm{chirp}<14$\,M$_\odot$ compared to 80\% of M-08 (and M-08t) DBHs with $M_\mathrm{chirp}<9$ (and $<11$) \,M$_\odot$. 
Although accounting for the difference in cluster ages (through changing M-08 to M-08t) slightly shifted the $M_\mathrm{chirp}$ CDF (lower panel, Fig.~\ref{fig:Mchirp}), the qualitative inference of our results does not change (as discussed in the previous paragraph). 
Interestingly, the M-08t and M-09 CDFs are similar up until the 50$^\mathrm{th}$ percentile, after which differences appear. It can be argued that M-09 produces more hard BH binaries as a result of having a higher initial binary fraction. Since massive binaries have higher binding energy compared to binaries with lower masses (orbital separation remaining constant) and Heggie's law shows that hard binaries become harder, the more massive DBHs of M-09 contribute to its higher $M_\mathrm{chirp}$ CDF tail. The low mass, softer binaries of M-09 are easily disrupted and the $M_\mathrm{chirp}$ distributions of M-08t and M-09 appear similar at the lower mass end. 
The median $M_\mathrm{chirp}$ values for M-08, M-08t and M-09 are 9.3\,M$_\odot$, 10.4\,M$_\odot$ and 10.8\,M$_\odot$ respectively.

The initial semi-major axis distribution only slightly affects the net $M_\mathrm{chirp}$ distribution of the DBHs. 
M-01 with Sana12 has a $M_\mathrm{chirp}$ median of 8.9\,$M_\odot$ compared to M-04a and M-04b (with DM91) with 9.4\,$M_\odot$ and 9.2\,$M_\odot$ respectively, 
showing the difference between M-04a and M-04b is close to that between M-01 and M-04a. 
While the median values are similar, M-01 has the $M_\mathrm{chirp}$ maxima at 20.2 M$_\mathrm{\odot}$, compared to 
M-04a with 16.5 M$_\mathrm{\odot}$ and 
M-04b with 13.4 M$_\mathrm{\odot}$.  
Hence, we comment that the initial semi-major axis distribution of Sana12 appears to allow the formation of slightly more massive DBHs, while keeping the median value similar to DM91 but the data is insufficient to conclude so with certainty. 
\subsubsection{Mass Ratio}
\label{sec:q}
The distribution of the mass ratios $q$ of the DBHs in all our models varies from about 0.3 (with the exception of M-09, discussed later) to 1.0, with the median in the range of 0.8--0.9, depending on the cluster model. 
\begin{figure}\centering
\includegraphics[width=0.37\textwidth]{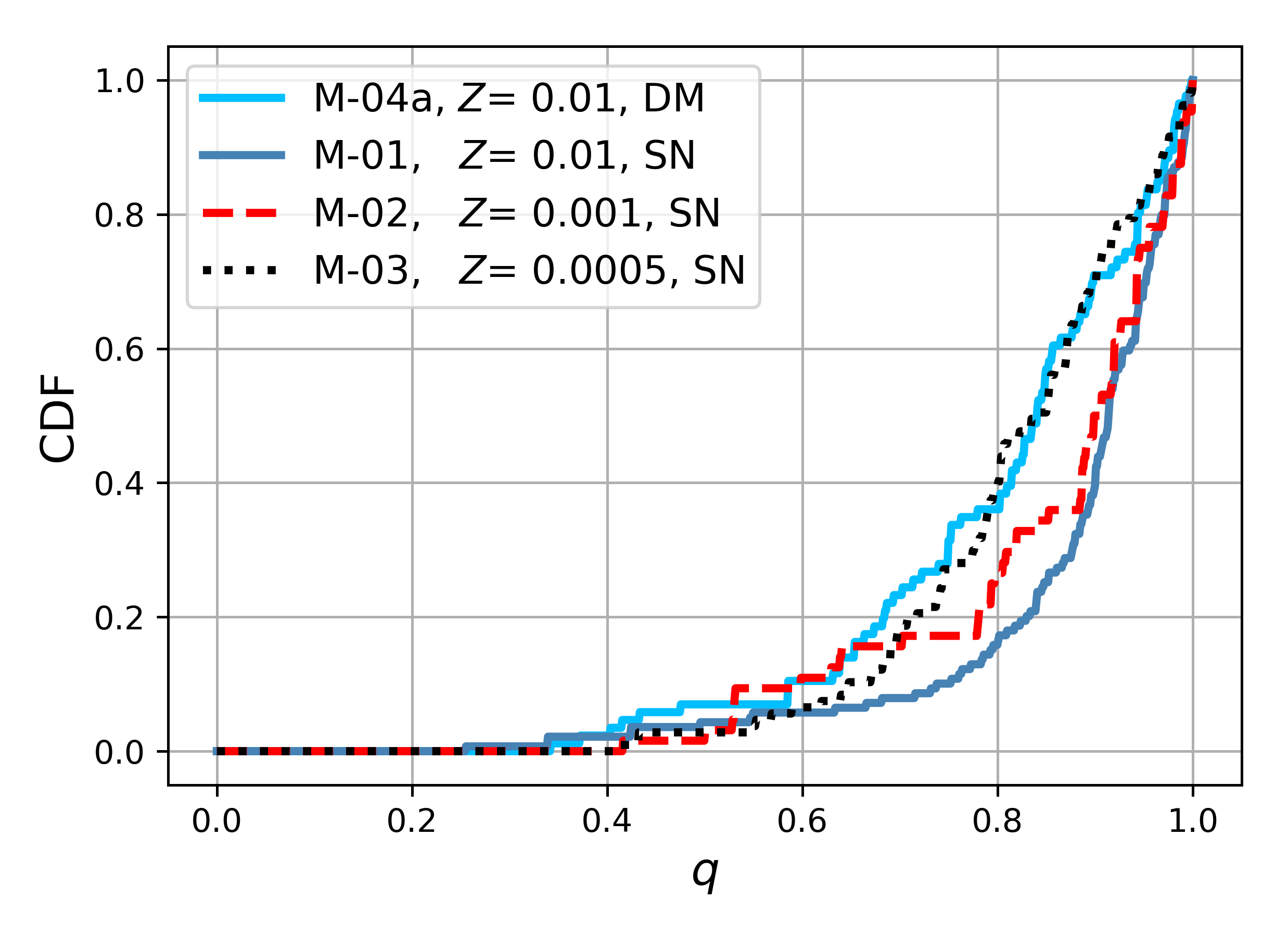}
\includegraphics[width=0.37\textwidth]{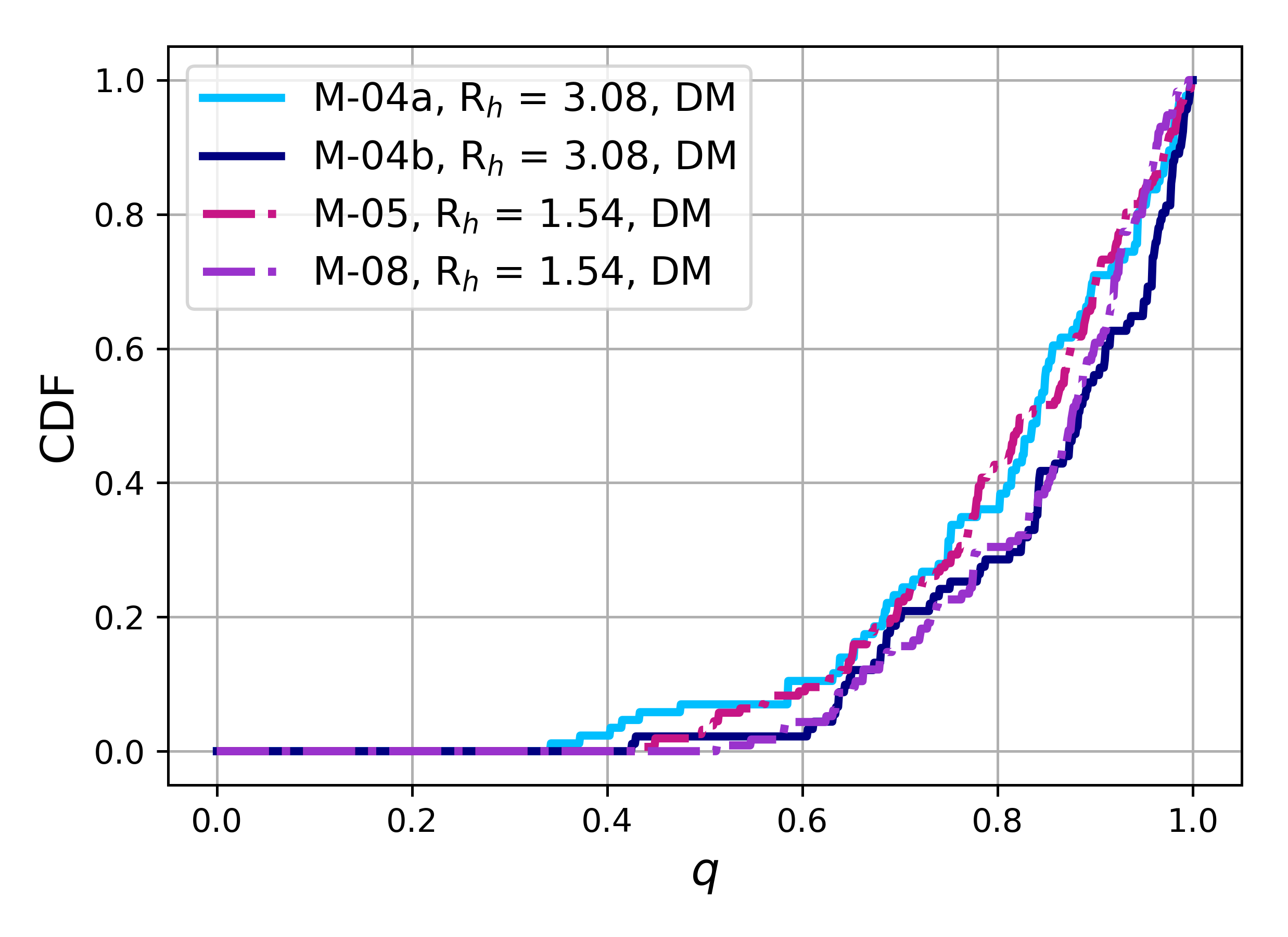}
\includegraphics[width=0.39\textwidth]{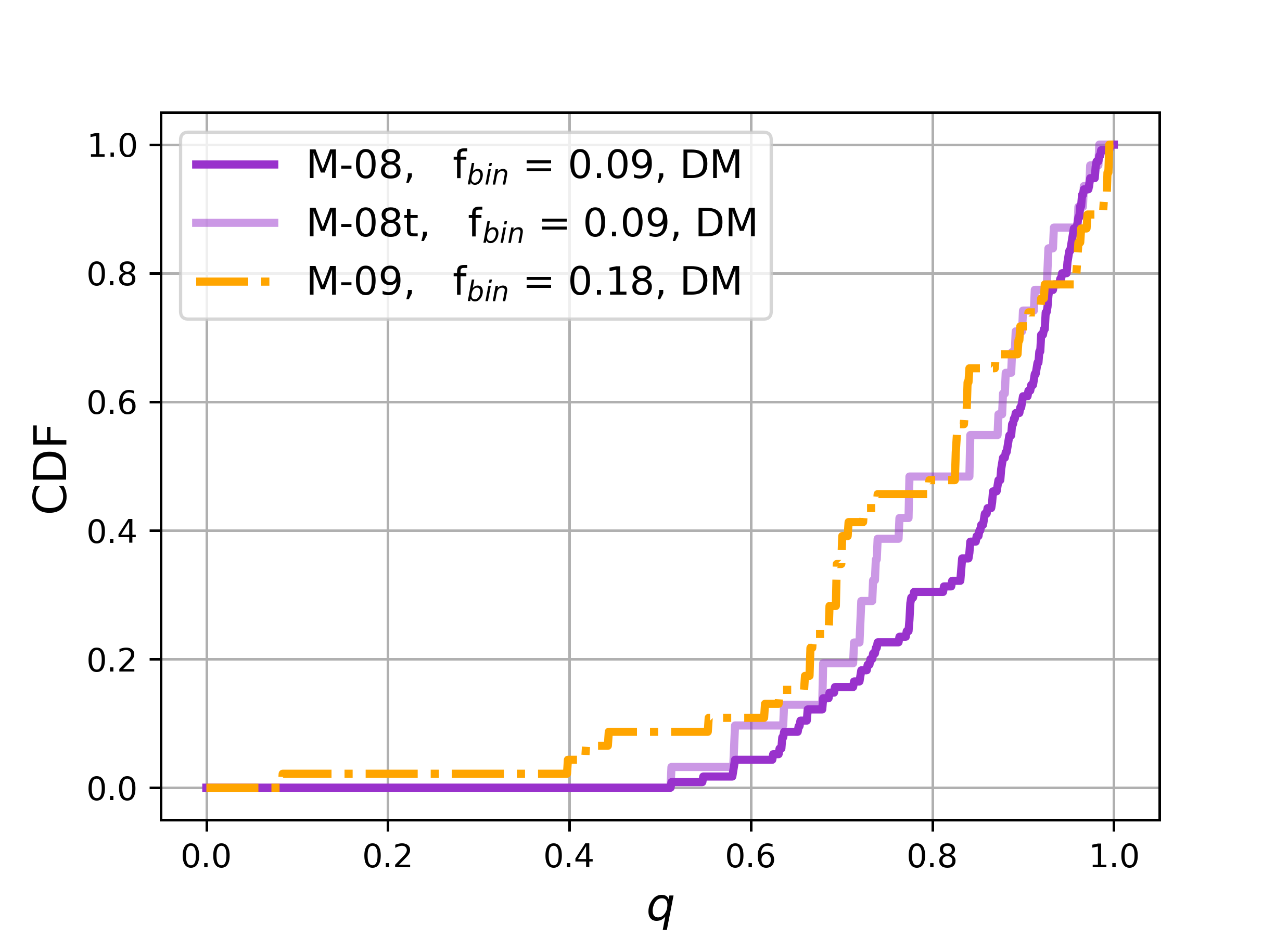}
\caption{Cumulative distribution of DBH mass ratios $q$. The three panels show the same models as in Fig.~\ref{fig:Mchirp}.}
\label{fig:q}
\end{figure}
Fig~\ref{fig:q} shows the CDFs of $q$ for our different cluster models. 
For the initial binaries in our models we choose $q$ from a uniform distribution so there is no bias towards equal-mass partners. 
However, we expect that over time, binary evolution and dynamical interactions will affect the DBH $q$ distribution with a tendency towards equal mass binaries. 

The mass ratio distribution of BH binaries is more biased towards lower $q$ for low metallicity models M-02 and M-03 than for metal-rich model M-01 (Fig.~\ref{fig:q}, top panel). 
The primary reason for this difference is because at lower $Z$, the BH mass distribution becomes broader, with more chances of asymmetric masses pairing together.  
In terms of the median $q$ values, for models M-01, M-02 and M-03 these are 0.91, 0.90 and 0.83, respectively.

While a change in the $R_\mathrm{h}$ of the clusters does not seem to significantly impact the DBH $q$ distribution (Fig.~\ref{fig:q}, middle panel), the initial semi-major axis distribution appears to play a small role. 
The median $q$ values for models with the DM91 semi-major axis distributions and $Z=0.01$ (M-04a, M-04b) are 0.84 and 0.88, respectively, whereas the median $q$ for the equally metal-rich model but with the Sana12 mass distribution (M-01) is 0.91 (as noted above). 
In fact, the metal-rich DM91 models are quite similar in behaviour to the metal-poor Sana12 model, as is apparent from the respective median values and comparing M-03 to M-04a in Fig.~\ref{fig:q} (top panel). 
Deviations from the initial uniform $q$ distribution towards $q=1$ is a result of more prolonged dynamical activity and more mass segregation.

A higher initial binary fraction ($f_\mathrm{bin}$) increases the mass range of the DBHs (as was apparent from the $M_\mathrm{chirp}$ distribution in the bottom panel of Fig.~\ref{fig:Mchirp}). 
From the lower panel of Fig.~\ref{fig:q}, we see that the $q$ distributions of M-09 and M-08t are not as remarkably different as  for $M_\mathrm{chirp}$ but we still find that M-09 ($f_\mathrm{bin} = 0.18$) shows more asymmetric mass-ratio DBHs and hence a lower $q$ distribution than M-08t ($f_\mathrm{bin} = 0.09$). 
It is interesting to compare M-08 to the younger version M-08t 
where we see a clear demonstration of dynamical interactions (mostly exchange interactions) leading to a preference for systems of similar mass over time. 

\subsection{Orbital Properties: In-situ}
\label{sec:orbital_properties}

Unlike isolated binary evolution where the orbital period ($P_\mathrm{orb}$) and eccentricity ($e$) can provide clues about the mass transfer and supernovae kick history of a binary, 
in the dynamical environment of a star cluster such orbital information can be lost or scrambled. 
This is especially true in the central regions of the clusters where BHs reside 
and can result from 
the formation and breaking of binaries or chain systems. 
Strong gravitational perturbation from nearby stars can also modify the orbital properties of the DBHs such that the $P_\mathrm{orb} - e$ distribution instead becomes a marker for the degree of cluster dynamical activity.  
While a distribution of binary orbital parameters skewed towards high eccentricity values is expected to be a clear imprint of elevated dynamical activity, 
a distribution that favours longer orbital periods could be caused by two opposing effects - i) a low density (less close encounters) environment that leaves wide binaries intact or ii) too many encounters that either break-up or accelerate the evolution towards contact of close pairs.  
Thus even if the Promiscuity, Survivability and $e$ distributions are known and point towards increased dynamical activity, a $P_\mathrm{orb}$ distribution favouring long-periods may also emerge from interactions breaking soft binaries (contrary of hard binaries evolving towards shorter orbital periods).
Using this combination of indicators to find the optimal 
cluster configuration amongst the initial parameter hyperspace is a longer-term project and beyond the scope of this paper. 
Here we simply present $P_\mathrm{orb}$ and $e$ distributions for our models to add to the portfolio of information as we form a picture of what may be possible in the future with an expanded set of models. 
\subsubsection{Orbital Period}
\label{sec:Porb}
\begin{figure}\centering
\includegraphics[width=0.37\textwidth]{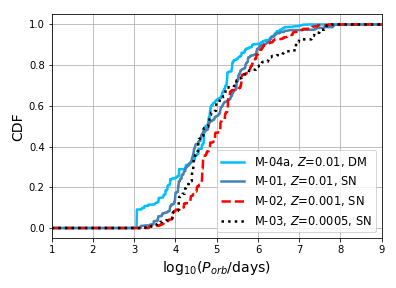}
\includegraphics[width=0.37\textwidth]{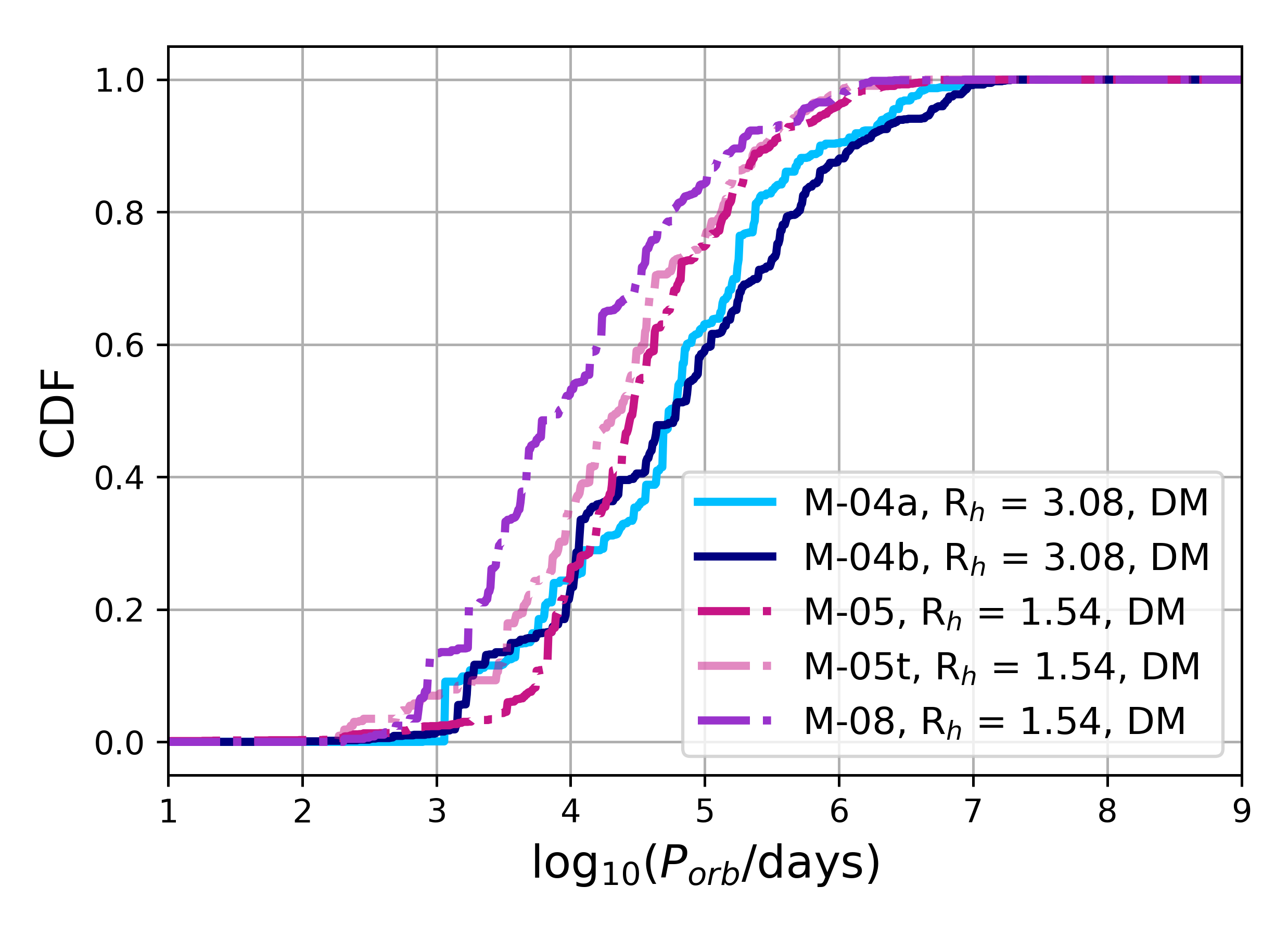}
\caption{Cumulative distribution of $\log_{10} P_\mathrm{orb}$. 
The top panel compares models with different initial half mass radii, whilst the bottom panel compares models with different initial metallicities.
}
\label{fig:Porb}
\end{figure}

From the upper panel of Fig.~\ref{fig:Porb} showing the distribution of DBH orbital periods for clusters of different metallicity, we conclude that for our models, covering the range of $Z= 0.01, 0.001, 0.0005$, there is no conclusive evidence of the cluster metallicity affecting the $P_\mathrm{orb}$ distribution of the DBHs.  
Similarly, the initial semi-major axis distribution does not appear to show any major effect on the DBH $P_\mathrm{orb}$ distribution (c.f. Figure~\ref{fig:Porb}). 
If we compare M-01 with the Sana distribution and M-04a and M-04b with the DM91 initial distribution (across the upper and lower panels of Figure~\ref{fig:Porb}) we see they have typically similar $P_\mathrm{orb}$ values. 

The initial $R_\mathrm{h}$ of a cluster has a more noticeable effect on the cumulative DBH $P_\mathrm{orb}$ 
as can be seen in the lower panel of Fig.~\ref{fig:Porb}. 
The slopes of the CDFs for the simulations with larger $R_\mathrm{h}$ (M-04a and M-04b) are less steep, showing a slight preference towards larger $P_\mathrm{orb}$ relative to M-05 from the main model set which has an initial $R_\mathrm{h}$ reduced by a factor of two. 
As a smaller $R_\mathrm{h}$ facilitates more close encounters due to a higher density, 
the corresponding depletion of wide (and thus loosely bound) DBH systems matches the behaviour that we would expect to observe. 
Model M-08 has the same initial conditions as M-05 aside from a different random number seed but has a clear preference for even shorter $P_\mathrm{orb}$ systems than M-05, even though both have the same initial $R_\mathrm{h}$. 
A key difference is that M-08 is only evolved for $2.8\,$Gyr and is thus a younger cluster.  
To determine if this is a significant factor we 
compare M-05 and M-08 at the same physical time 
by creating M-05t which is M-05 restricted to considering only systems within the same time-frame as for M-08 ($2.8\,$Gyr). 
We observe in Fig.~\ref{fig:Porb} 
that M-05t does have a slightly shorter orbital period distribution than M-05, indicating that an age effect may be at play where given more time a cluster is more effective at depleting short-period systems. However, the most significant differences exist between M-05t and M-08 showing that differences in the evolution pathways of these models arising from random fluctuations 
(in the same way as discussed for M-04a and M-04b in Sec.~\ref{sec:porb}) is the main factor. 
As a final comparison, models M-04a and M-04b ($R_\mathrm{h}$ = 3.08) have approximately 60\% of the DBHs with log$_{10}P_\mathrm{orb}<5$ 
while M-05 and M-08 ($R_\mathrm{h}$ = 1.54) have about 80\% and 90\% respectively.

\subsubsection{Eccentricity}
\label{sec:e}
\begin{figure}\centering
\includegraphics[width=0.37\textwidth]{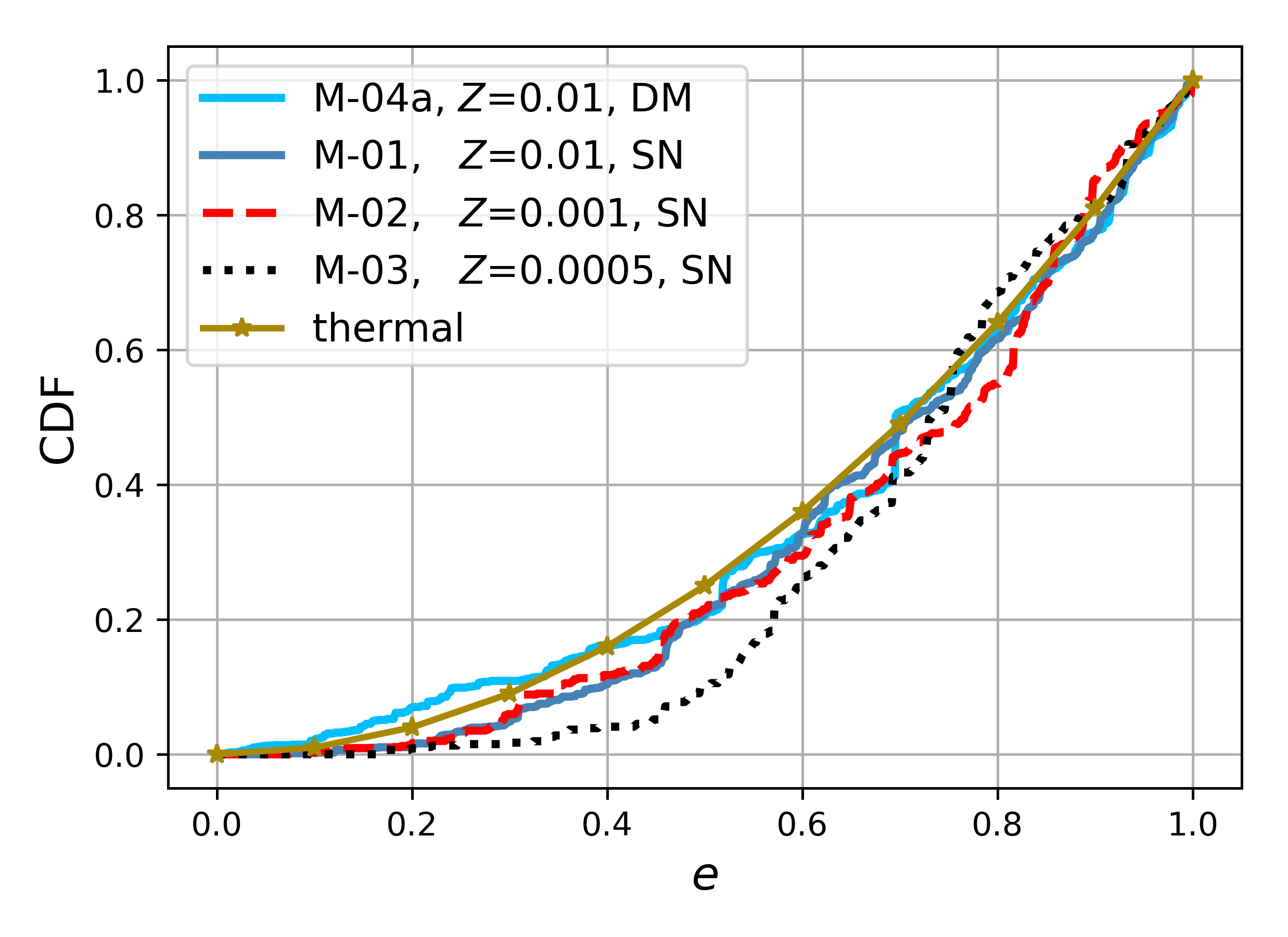}
\includegraphics[width=0.37\textwidth]{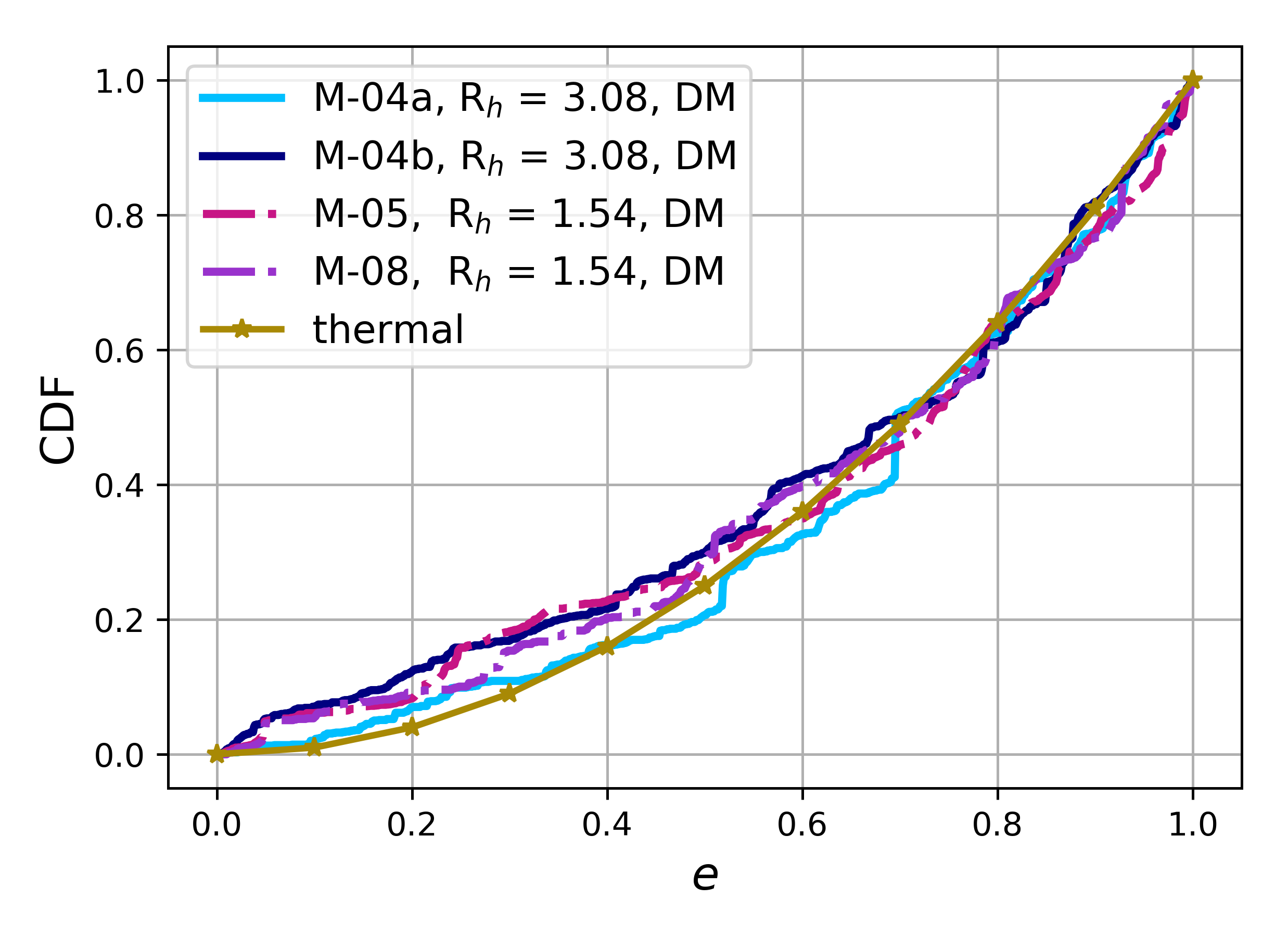}
\caption{Cumulative distribution of orbital eccentricities of DBHs. 
The panels show the same models as Figure~\ref{fig:Porb}. 
The solid yellow line shows the theoretical thermal eccentricity distribution (CDF($e) = e^2$) for comparison.}
\label{fig:e}
\end{figure}
The eccentricity distributions of the DBHs for our models are shown in Fig.~\ref{fig:e} alongside a thermal (Boltzmann) distribution $f(e)de=2ede$ \citep{Jeans:1919,Heggie:1975} which they appear to follow in general. 
Under the assumption of a population solely composed of dynamically interacting binaries, \citet{Jeans:1919} showed that the binaries achieve thermal equilibrium over multiple energy exchange interactions. 
Although, in star clusters the presence of singles and triples (or even higher order chain systems) can cause deviation from the idealistic scenario. Though the average binding energy is solely a function of the semi-major axis, the binding energy varies periodically through the change in the orbital separation in eccentric orbits.
Orbital $e$ is mathematically connected to binary orbital separation (through latus-rectum), which in turn makes the binding energy at each point of an eccentric orbit an implicit function of $e$. 
The deviation from \citet{Jeans:1919}'s law was explained by \citet{Heggie:1975}, who showed that due to binary-single interactions, a hard binary (binding energy $>$ kinetic energy of the passer-by third body) will become more tightly bound and a soft binary (binding energy $<$ kinetic energy of the passer-by third body) will become more loose (and eventually break). 

As for orbital period 
the effect of the initial cluster metallicity on the DBH $e$ distribution appears to be inconclusive for our grid of models (upper panel of Fig.~\ref{fig:e}). 
The only notable observation is a slight bias towards higher eccentricity  ($e > 0.6$) for the lower-$Z$ cluster M-03 with $Z=0.0005$ (compared to models M-01 and M-02 with $Z=0.01$ and $Z=0.001$, respectively). This perhaps is caused by the higher rate of DBH activity for M-03 with stronger gravitational perturbations from massive 
BHs or exchange interactions creating slightly more eccentric binary orbits.
Indeed, M-03 is the only model to show any significant departure from the thermal eccentricity distribution. 

There appears to be no variance in the orbital eccentricity distribution across clusters with different initial $R_\mathrm{h}$ (lower panel Fig.~\ref{fig:e}). 
M-04b, M-04a, M-05 and M-08 all with $Z=0.01$, and the same initial number of systems but different $R_\mathrm{h}$, all show that approximately 60\% of the binaries have $e>0.6$.
For the models with different initial semi-major axis distributions we see some differences, for example M-04a (DM91) has more low eccentricity orbits compared to M-01 (Sana12), but nothing that couldn't be attributed to statistical noise. M-01 and M-04a have 70\% compared to M-04b with 60\% of their respective DBH systems with $e>0.6$.
\subsection{Delay Time: In-situ}
\label{sec:tdelay}
The delay time ($t_\mathrm{delay}$) is defined as the time span from the formation of a compact object pair through to its merger (potential or actual). 
We have calculated $t_\mathrm{delay}$ as formulated by \citet{Peters:1964}, assuming two point-mass particles in a binary spiralling in towards each other solely due to the loss of gravitational energy. 
The main factor in determining the $t_\mathrm{delay}$ of a binary is the initial orbital separation ($a$), followed by the masses of the two bodies ($m_\mathrm{1}$ and $m_\mathrm{2}$) and the orbital eccentricity ($e$).
\citet{Peters:1964} show that the time taken by the two bodies, BHs in our case, in a binary to merge is given by
\begin{equation}
    \mathcal{T}=\frac{12c_\mathrm{0}^4}{19\beta}\int_{0}^{e_\mathrm{0}}\frac{e^{29/19}[1+(121/304)e^2]^{1181/2299}}{(1-e^2)^{3/2}} , 
\end{equation}
where
\begin{equation}
    \beta=\frac{64G^2m_1m_2(m_1+m_2)}{5c^5} , 
\end{equation}
and
\begin{equation}
    c_0=\frac{a_0(1-e_{0}^2)}{e_{0}^{12/19}}\left[1+\frac{121e_{0}^2}{304}\right]^{ -870/2299} \, . 
\end{equation}
Here $e_0$ and $a_0$ are respectively the initial eccentricity and initial semi-major axis of the elliptical orbit. 
For a circularized binary, the merger time is solved to be $\mathcal{T}(e=0) = a_0^4/4\beta$. 
For isolated binary mergers $t_\mathrm{delay} = \mathcal{T}$ always. 
For a dynamical environment, 
orbital perturbations 
from nearby stars 
can change $a$ or $e$ of the initial binary and hence $t_\mathrm{delay}$ needs to be computed in steps rather than as a continuous integral.  
It is also possible that 
the initial BH-BH pair can get disrupted, 
for example in an exchange interaction, 
and a new dynamical BH binary can emerge from the interaction, making $t_\mathrm{delay}$ different from the initial $\mathcal{T}$ computed for the initial pair. 
Thus $t_\mathrm{delay}$ needs to be re-calculated with the masses and the orbital parameters of the new BH-BH system. 
To avoid confusion between isolated primordial mergers and dynamical mergers in dense environments (as in this paper), we define $t_\mathrm{delay}$ as the time it will take for two compact objects to merge after forming a bound binary system 
(as opposed to the time after the second supernova that marks the formation of the double compact object in an isolated systems). 

\begin{figure}\centering
\includegraphics[width=0.37\textwidth]{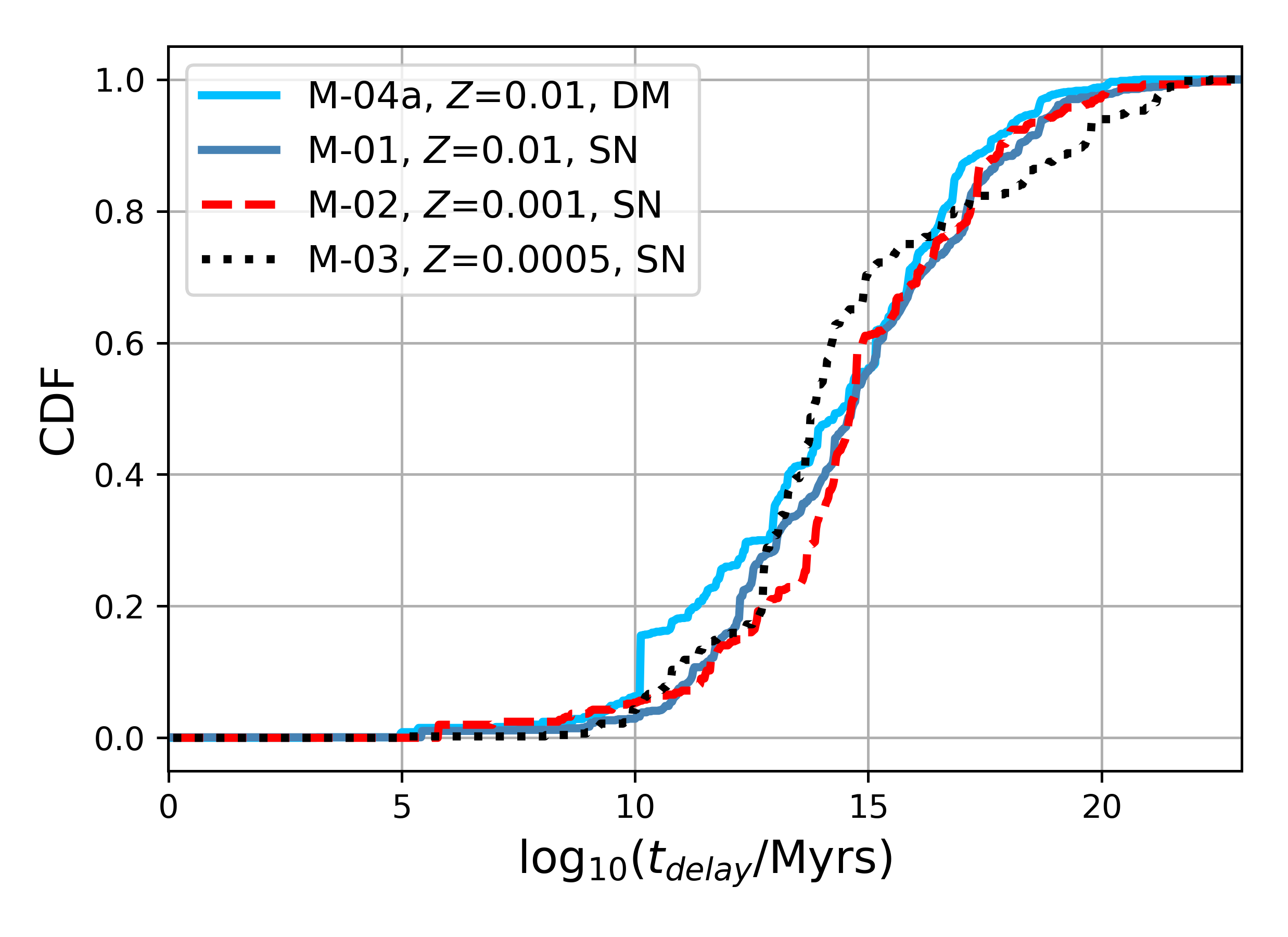}
\includegraphics[width=0.37\textwidth]{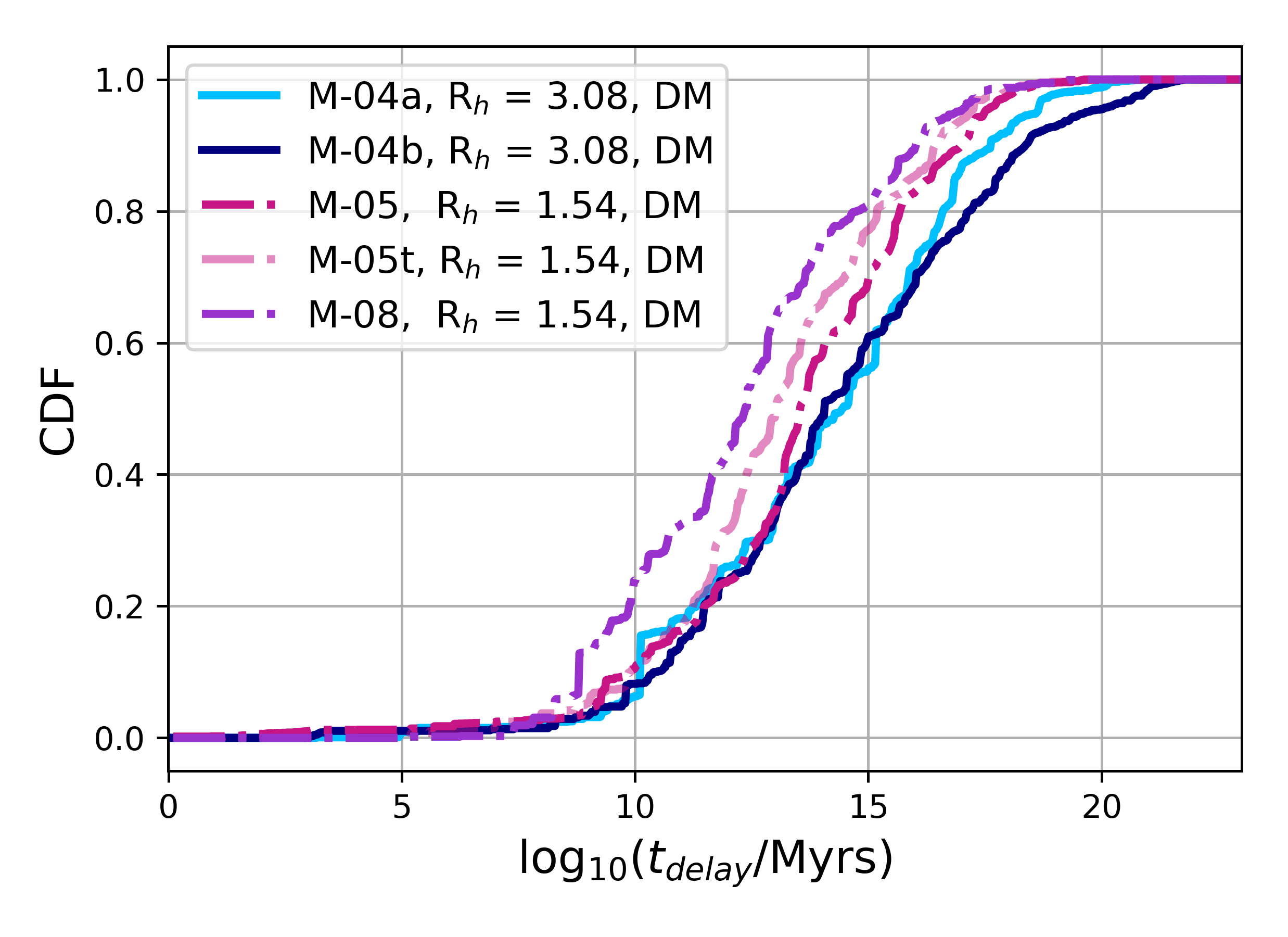}
\caption{Cumulative distributions of the delay time between DBH formation and merger for our models.}
\label{fig:tdelay}
\end{figure}

For nearly circular binaries the time delay remains proportional to the fourth power of separation ($t_\mathrm{delay} \propto a^4$). 
Decreasing the binary mass increases the $t_\mathrm{delay}$ of the system while a higher orbital $e$ (with all other factors remaining constant) reduces the merger time. 
Given that the variation of the DBH $e$ distribution over different cluster global parameters has been fairly small for our array of models (see Sec.~\ref{sec:e}), we expect ordinarily that $a$ will have the dominant impact on $t_\mathrm{delay}$. 
However, if the variation in $a$ remains small, a considerable change of DBH masses will effect the $t_\mathrm{delay}$ distribution. 
To 
study this 
we calculate the $t_\mathrm{delay}$ at every major output time-step (i.e. snapshot interval, which is $\sim 10\,$Myr for our models) for each DBH present in the model at that time, since the $a$ and $e$ of the orbit can change on this timescale owing to perturbations from neighbouring systems. 

The distributions of $t_\mathrm{delay}$ for our models are shown in Fig.~\ref{fig:tdelay}. 
For metallicity we did not see any significant variation in the distribution of $P_\mathrm{orb}$ in Fig.~\ref{fig:Porb} and thus we would not expect variations in $a$ to be a factor. 
Instead, any differences in the delay times for clusters of different metallicity are likely caused by the mass difference of BHs in these clusters, 
where the average DBH mass for M-01 is 19.7 M$_\odot$, 
while for M-02 and M-03 it is 44.6 M$_\odot$ and 47.5 M$_\odot$,  respectively. 
In contrast the median $\log_{10} (a / R_\odot)$ of the DBHs in M-01, M-02 and M-03 is 4.2, 4.5 and 4.3, respectively. 
The median $t_\mathrm{delay}$ numbers do show a slight decrease in merger time for lower metallicity: 
median log$_\mathrm{10} (t_\mathrm{delay} / \mathrm{Myr})$  
is 14.7 for M-01 ($Z = 0.01$), 14.6 for M-02 ($Z = 0.001$) and 13.8 for M-03 ($Z = 0.0005$). 
However, overall there is no clear pattern for metallicity influencing the delay times. 
It is a similar story for the initial semi-major axis distribution (comparing M-01 and M-04a in the upper panel Fig.~\ref{fig:tdelay}) 
where our models do not significantly affect the $t_\mathrm{delay}$ distribution. 

The behavior of $t_\mathrm{delay}$ across clusters of different initial $R_\mathrm{h}$ (lower panel of Fig.~\ref{fig:tdelay}) closely follows that observed for $P_\mathrm{orb}$ in Fig.~\ref{fig:Porb}. 
This is to be expected and as such the corresponding analysis in Sec.~\ref{sec:Porb} holds here as well. 
The main result is that denser clusters are effective in shortening the delay times of the DBH binaries.

\subsection{Ex-situ: ejected DBHs}
\label{sec:ex_situ}

As described earlier, 
for NSs the high natal kick will usually eject the object out of the cluster. 
In the case of BHs, where the supernova kicks are scaled down by fallback mass and the BHs reside mainly within the core of the cluster, the escape rate is relatively lower than for NSs. 
Nevertheless, it is still possible for BHs and even DBHs to escape from our model clusters. 
Unlike the in-situ binaries, escaped BH pairs will not be subsequently interrupted by other BHs as they continue to evolve in the Galactic field. 
This can be an advantage because 
while it is true that gravitational perturbations from other BHs or DBHs can facilitate mergers, 
too many close interactions may also hinder potential mergers 
(see Sec.~\ref{sec:in_cluster_mergers} for an example). 
Thus if a DBH from a cluster escapes with a tightly bound or highly eccentric orbit, instead of potential unwanted hindrances from other BHs, it can be left to merge in peace (MIP). 
While the importance of the orbital separation $a$ in determining the $t_\mathrm{delay}$ is explained in Sec.~\ref{sec:tdelay}, we also highlight the role of eccentricity $e$ in DBH mergers. 
Unlike in-cluster DBHs where $a$ and $e$ of the binary can be changed due to external influence, the orbital separation of escaped BH pairs will gradually reduce solely due to emission of gravitational radiation. 
Hence all the ex-situ mergers, when observed by ground-based gravitational-wave detectors, will show circular orbits at a gravitational-wave frequency of 10Hz, whereas there remains a possibility of in-situ mergers having $e>0.1$ at the same frequency (see Sec.~\ref{sec:introduction}).

To illustrate the importance of ex-situ DBHs, we select M-07 with the highest UniquePair, largest Promiscuity maxima and lowest Survivability median (see Table~\ref{tab:Prom_imm_unqpr}), 
which are all signatures of high dynamical activity in the cluster.
The median mass of the BHs in a DBH inside the cluster for M-07 is $m\approx 17.3$ M$_\odot$, the median $q \approx 0.94$, median log$_{10}(a / R_\odot) \approx 4$ and the median $e\approx0.94$. For the escaped DBHs of M-07, these medians are  $m\approx 17.7$ M$_\odot$, $q \approx 0.96$, log$_{10}(a / R_\odot) \approx 3$ and $e\approx0.91$.
We thus select a typical case of a 18.1\,M$_\odot$ and 17.3\,M$_\odot$ DBH, with an initial orbital separation of log$_{10}(a / R_\odot) = 3$. 

It is important to note the nature of the $t_\mathrm{delay}$ vs $e$ curve and the sensitivity to the initial conditions of our typical M-07 binary: an eccentricity of 0.99 will result in a merger in a Hubble time compared to 0.98 which would not.
Of course, a smaller $a$ and more massive BHs can allow less eccentric binaries to merge. 
For model M-07, which is the most massive that we evolved, we have 23 DBHs escaping from the cluster over the span of 9.5\,Gyr, out of which 3 merge in a Hubble time. 
For comparison, model M-01 has 15 DBHs ejected during its lifetime of about 10\,Gyr, none of which merge in a Hubble time. 

The first system of the three ex-situ M-07 DBHs mergers escapes the cluster at 0.33\,Gyr with masses $m_\mathrm{1}=16.1$\,M$_\odot$ and $m_\mathrm{2}=15.4$\,M$_\odot$, and orbital properties log$_{10}(a / R_\odot) = 3.01$ 
and $e=0.994$ at the time of escape. 
The system mergers outside cluster 5.38\,Gyr from the time it leaves. 
The next DBH of interest departs the cluster at 0.94\,Gyr. 
It has masses and orbital parameters (at escape) of 
$m_\mathrm{1}=17.7$\,M$_\odot$,  
$m_\mathrm{2}=17.7$\,M$_\odot$, log$_{10}(a / R_\odot) = 2.95$ and $e=0.994$. 
This system merges 2.61\,Gyr after escape. 
The third system leaves the M-07 cluster at 1.20\,Gyr with  $m_\mathrm{1}=18.1$\,M$_\odot$,  
$m_\mathrm{2}=18.2$\,M$_\odot$, log$_{10}(a / R_\odot) = 2.88$ and $e=0.995$.  
It merges 0.63\,Gyr after leaving the cluster. 
All three systems are highly eccentric in nature. 
The third DBH with a lower $a$ and higher $e$ merges in the shortest time (out of the three).

M-07 has the highest incidence of dynamical activity of all our models. It has no in-situ merger even though it is a fairly evolved model  (Table.~\ref{table:NBODY_models}). 
This points to the possibility that the heightened incidence of in-cluster dynamical interactions creates a too agile environment for DBHs to coalesce inside cluster M-07 without interruption. 
However, the increased intensity of dynamical interactions like i) binary-binary encounters, which (apart from producing more eccentric binaries) are typically have longer timescales than binary-single encounters \citep{Zevin:2018kzq} - and thus is more effective in a more massive cluster like M-07 and ii) more exchange interactions \citep{Ziosi:2014} due to higher stellar density also results in more elliptic orbits which in turn can help ex-situ DBHs to MIP. 
Thus clusters with a high in-situ DBH merger rate may not correspond to the clusters with more ex-situ mergers.


\section{In-cluster Mergers}
\label{sec:in_cluster_mergers}

We record a total of 12 distinct in-cluster binary BH mergers occurring in models M-01, M-03, M-04a, M-04b, M-05 and M-08 (see Table~\ref{table:NBODY_models}), all of  dynamical origin, such that the merging pair of BHs were formed through multiple chain systems and exchange interactions. 

M-05 has four in-situ BH-BH coalescences, the highest in our suite of models, followed by M-08 with three such mergers.  
Both these models have low initial half mass radii ($R_\mathrm{h}=1.54$\,pc) and are metal rich ($Z=0.01$). 
M-08 is not fully evolved (with about half its mass remaining), unlike M-05 which has nearly evaporated (Table~\ref{table:NBODY_models}). 
In spite of this, M-08 has a higher merger rate than all models other than M-05. 
Even though M-08 only evolves to $2.8\,$Gyr it has completed three half-mass relaxation times after $2.3\,$Gyr (see Table.~\ref{tab:Prom_imm_unqpr}) and all of the mergers have occurred within this timeframe. 
M-05 completes 18 half-mass relaxation times across 13\,Gyr of evolution but only increases the total number of in-situ DBH mergers by one from M-08. 
Indeed the last DBH merger in M-05 occurs around 3.1\,Gyr (with the earliest at around 150\,Myr) which is just after it has completed three half-mass relaxation times (also see Table.~\ref{tab:Prom_imm_unqpr}). 
This suggests that most in-cluster mergers are completed (or at least set in motion) early in the dynamical lifetime of a cluster and that even 
for long-lived clusters there is less need to focus on the later evolution stages when looking for DBH in-cluster mergers. 
Furthermore, the abundance of DBH mergers for M-05 and M-08 shows the importance of initial cluster density in increasing the number of in-situ mergers. This is also supported by the UniquePair, Promiscuity, Survivability quotients discussed in sections~\ref{sec:Promiscuity} and \ref{sec:Survivability}. 

Interestingly, M-07 with the the lowest initial $R_\mathrm{h}=1.15$\,pc, comparatively metal-poor $Z=0.005$ and the most massive of our set of models, although dynamically mature (Table~\ref{table:NBODY_models}), does not have any in-cluster DBH merger. 
This might appear counter-intuitive as M-07 proves itself to be the most dynamically active out of all our models (see sections~\ref{sec:Promiscuity} and \ref{sec:Survivability}, Table.~\ref{tab:Prom_imm_unqpr}, Fig.~\ref{fig:Promiscuity} and \ref{fig:Survivability}). 
However, as we noted in the previous section, extreme dynamical activity may also hamper compact binary mergers 
(note also that compared to the three ex-situ mergers for M-07 we find none for M-05 and M-08). Dynamical activity actually prohibiting mergers is also observed by \citet{DiCarlo:2019fcq,Zevin:2018kzq,DiCarlo:2020MNRAS}, who discuss that lower mass binaries are actually stopped from in-spiral mergers due to dynamics.
The difference between in-cluster BH merger probability between M-05/M-08 and M-07 shows 
that further investigation of 
the `cluster initial density versus number of in-cluster DBH mergers' parameter space 
is warranted (as well as the total number of stars, BH masses and BH mass spectrum),  
with the possibility that mergers increase with density up to a certain point beyond which in-situ mergers are inhibited. 
It is important to compare this in terms of cluster density than half-mass radius as the density also varies depending on cluster size. For instance, even though M-07 and M-06 have the same initial $R_\mathrm{h}=1.15$\,pc; their initial half-mass densities are 3.79$\times10^3$\,M$_\odot$/pc$^3$ and 0.54$\times10^3$\,M$_\odot$/pc$^3$ respectively. 
The initial half-mass cluster density of M-05 and M-08 is 1.15$\times10^3$\,M$_\odot$/pc$^3$. 
However, we realise the need for further investigation of this behaviour with a larger array of models. 

Cluster metallicity does not appear to affect the number of DBH mergers with one in-situ merger for both M-01 ($Z=0.01$) and M-03 ($Z=0.0005$), and none for M-02 ($Z=0.001$). Although $Z$ does not affect the number of DBH mergers, the masses of the merging BHs is indeed dictated by $Z$ (see section.~\ref{subsubsec:M-03}).
M-04a and M-04b have two and one mergers respectively, highlighting statistical noise can slightly alter the total number of mergers in otherwise similar conditions. 
The initial semi-major axis of the binary population does not appear to affect the number of DBH mergers for the range of models that we have performed. 

We observe a higher fraction of in-situ DBH mergers than ex-situ mergers. 
This trend is noted in {\tt{NBODY}} cluster simulations as shown by \citet{Anagnostou:2020eff} using models from \citet{deVita:2019}, as well as by \citet{Banerjee:2021}, unlike Monte-Carlo simulations  \citep[e.g.,][]{RodriguezChatterjee:2016}. 
We find that the ejected systems that merge (basically, the ejected systems of M-07) are biased towards higher eccentricity, unlike the ejected systems of all other models which roughly follow the thermal distribution. \citet{Anagnostou:2020eff} also finds ejected mergers having steeper eccentricity, as expected from \citet{Peters:1964} calculations showing that higher ellipticity aids in more efficient emission of gravitational waves (also discussed in section~\ref{sec:ex_situ}). 
It can be argued that the smaller clusters of our dataset---as well as \citet{Banerjee:2021}---do not allow enough  interactions to eject hard (smaller semi-major axis), eccentric binaries and instead eject softer binaries more readily than they should be compared to more massive clusters. 
The models from \citet{deVita:2019}, which show a similar trend, also consist of 200,000 initial systems, as do the models in \citet{RodriguezChatterjee:2016}; and \citet{Anagnostou:2020eff} argues the lower density of the latter's model causes the disparity of in-situ vs ex-situ mergers. 
The addition of PN terms increases the fraction of in-cluster mergers, as shown by \citet{Rodriguez:2017pec}, unlike \citet{RodriguezChatterjee:2016}. 
The higher in-situ DBH mergers in {\tt{NBODY}} models can hence arise from multiple factors including their smaller size and the inclusion of PN terms.

We select the details of four in-situ DBH coalescences from models M-03 (one), M-04a (one) and M-05 (two). 
We make the selections based on some unique properties of these mergers (as described in the following subsections)
For each event, we use the terms `Star1' and `Star2' to refer to the two individual stars that merge. 
The metallicity of the stars are $Z=0.0005$ in M-03 and $Z = 0.01$ in both M-04a and M-05.  
We discuss the properties of the merger remnants in section~\ref{subsec:merger_remnant_properties} and the BH merger rates in section~\ref{subsec:BBH_merger_rates}.

\subsection{M-03 merger}
\label{subsubsec:M-03}

This example starts with two single stars of ZAMS masses 70.2\,M$_\odot$ and 43.1\,M$_\odot$ that evolve to become BHs within the first $6\,$Myr of cluster evolution. 
The BH masses are 28.1 and 19.4\,M$_\odot$ and we will refer to these as Star1 and Star2, respectively. 
Star1  forms its first compact object binary around 100\,Myr with a 18.9\,M$_\odot$ BH. 
Over the next couple of Gyr, Star1 forms binaries with seven different BHs, with masses ranging from 25.4--28.3\,M$_\odot$, as well as with a couple of low-mass (< 1\,M$_\odot$) main-sequence stars. 
Out of the array of BHs that pair up with Star1, 
six (out of the seven) DBH systems have $0.6<e<0.9$ 
while the other is less eccentric ($0.2<e<0.3$) but in a wide (5.2<log$_{10}(a / R_\odot)<5.3$) orbit with a 25.4M$_\odot$ BH from $1.2-1.3$\,Gyr. 

The 19.4\,M$_\odot$ Star2 BH mostly lingers around as single star, forming a short-lived binary with a 24.8\,M$_\odot$ BH at 1.6\,Gyr. 
Around 1.9\,Gyr, Star2 disrupts a DBH system containing Star1 with a 25.8\,M$_\odot$ BH and forms a binary with Star1. 
The orbital parameters of the binary are $e=0.4$ and log$_{10}(a / R_\odot)=4.2$. 
The newly formed Star1-Star2 DBH system soon becomes a chain system with another BH binary formed by the previous 24.8\,M$_\odot$ BH (that was a partner with Star2) and its new 22.4\,M$_\odot$ BH companion. 
The gravitational perturbation of the added binary in the chain system causes Star1 and Star2 to merge at 2.8\,Gyr, forming a remnant BH of mass 47.5\,M$_\odot$. 
The chain system is disrupted as Star1 and Star2 merge. 
This is the most massive merger remnant produced in our suite of models highlighting the importance of low metallicity clusters in understanding LIGO detections of higher mass BHs. 
The 47.5\,M$_\odot$ BH forms a binary with 26.2\,M$_\odot$ at $2.9\,$Gyr and it escapes the cluster a few Myr later.
\subsection{M-04a merger}
\label{subsubsec:M-04a}
For this merger we start with Star1 as a 54.1\,M$_\odot$ main-sequence star that evolves to become a 12.5\,M$_\odot$ BH in about 6\,Myr,
while Star2 originates from a 38.8\,M$_\odot$ main-sequence star to become an 11.1\,M$_\odot$ BH on a similar timescale. 
By an age of 1.6 Gyr they have both sunk into the cluster core. 
About 0.1 Gyr later they form a short lived binary
with eccentricity of 0.7 with each other, but this get disrupted by another nearby DBH system. 
The BHs then become involved in a succession of chain system interactions (varying between configurations of three to five bodies) with BHs of similar mass before experiencing a hyperbolic collision within the subsystem and merging. 
\subsection{M-05 mergers}
\label{subsubsec:M-05}
\begin{figure}
\includegraphics[width=0.47\textwidth]{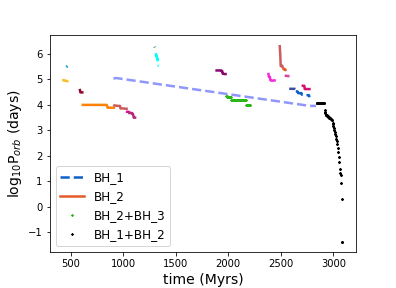}
\caption{Evolution in time of the orbital period $P_\mathrm{orb}$ of BH-BH binaries in model M-05.
All BH\_1 encounters are shown in dotted lines and all BH\_2 encounters shown in bold lines. 
The individual colours of lines signify each unique encounter with other BHs. 
The eccentric BH\_2 and BH\_3 merger is identified by the green stars while the black stars represent the final BH\_1 and BH\_2 inspiral merger (see section~\ref{subsubsec:M-05} for more details).}
\label{fig:merger1}
\end{figure}

In this example, Star1 begins its life as a 26.2 M$_\odot$ ZAMS single star and evolves to become a 10.2 M$_\odot$ BH in the span of 8\,Myr. It proceeds to form binaries with seven unique stars (excluding Star2, the final merger companion) including five BHs, a low-mass ZAMS star and a star on the first giant branch. 
These interactions occur in the time-frame from BH formation up until an age of $2.8\,$Gyr when Star1 first encounters Star2. 

Star2 evolves from a 41.0 M$_\odot$ ZAMS single star to become an 11.4 M$_\odot$ BH over the span of 6\,Myr. 
This BH partners with 14 different stars in 19 distinct systems before it encounters the Star1 BH. 
Notable amongst these pairings is a binary formed with a 1.8 M$_\odot$ main sequence star in which the two stars eventually become close enough together for the BH to accrete some of the companion and the BH mass increases to 12.4 M$_\odot$ in the process. 
Also of important consequence is a binary formed with another BH amongst a short-lived subsystem where the two BHs undergo an eccentric collision and the Star2 BH increases to 19.8 M$_\odot$. 

Star1 and Star2 have had three common BH companions of mass 10.1 M$_\odot$, 9.8 M$_\odot$ and 10.2 M$_\odot$, where they each paired with these BHs but at different times. 
Subsequently, Star1 and Star2 formed a triple with the third of those BHs and from that triple (at a cluster age of $3\,$Gyr) emerged the Star1 and Star2 binary that goes onto result in the BH-BH merger via inspiral. 
The initial orbital period and eccentricity of this Star1 and Star2 binary BH (with masses 10.1 M$_\odot$ and 19.8 M$_\odot$) were 11749 days and 0.863 respectively. 
For these conditions, if the binary was kept isolated, we estimate  $t_\mathrm{delay}=4.4\times10^{11}$\,Myr, which means it would not have merged within a Hubble time  \citep{Peters:1964}.  
However, the orbit becomes perturbed under the influence of a third body, causing it to shrink. 
The period changes quickly with brief pauses at 93 days and 29 days (where the eccentricity is 0.29). 
The binary then goes into an in-spiral coalescence about $t_\mathrm{delay}=0.1$\,Myr after it first formed 
to merge and form a 30.0 M$_\odot$ BH (at  cluster physical age of about 3.1\,Gyr). 
Then, 9.7 Myr after the merger, the remnant BH escapes the cluster. 

All the encounters of Star1 and Star2 as BHs (hence BH\_1 and BH\_2) with all other BHs mentioned in the description above are shown in Fig.~\ref{fig:merger1} in terms of their orbital period evolution over time. The comparatively short time-span eccentric merger of BH\_2 with another BH (BH\_3) at around 2.2\,Gyr is plotted, as well as the in-spiral merger of BH\_1 and BH\_2 showing a pre-merger orbital period of less than a day.

\subsection{Merger remnant properties}
\label{subsec:merger_remnant_properties}

For each of our BH mergers, we estimate the final mass, spin and kick of the remnant BH. 
Since these binaries have undergone multiple dynamical interactions, we assume that the orientations of their spin vectors have been randomised \citep{RodriguezSpin:2016}.
The final BH mass is typically around 5\% lower than the total merging binary mass due to energy lost in gravitational waves \citep{Varma:2018aht,LIGOScientific:2018mvr}, whilst the final BH spin is typically around 0.7 (where a spin of 0 corresponds to a non-spinning BH and a spin of 1 is maximally spinning)
due to the orbital angular momentum of the binary at the last stable orbit \citep{Varma:2018aht,LIGOScientific:2018mvr}.
The gravitational-wave recoil kick\footnote{In the future, gravitational-wave recoil kicks may become direct observables \citep{Gerosa:2016vip,Varma:2020nbm,Abbott:2020mjq}.} however depends sensitively on both the binary mass-ratio, and the spins of the merging BHs. 
For an equal-mass or extreme mass-ratio non-spinning binary, the recoil kick will be zero, whilst at intermediate mass-ratios the kick peaks at around 150\,km\,s$^{-1}$ for a mass ratio of around 1/3. 
Incorporating generically spinning BHs, recoil velocities up to thousands of km\,s$^{-1}$ are possible for certain configurations \citep{Herrmann:2007,Campanelli:2007ew,Gonzalez:2007hi}, sufficient to eject BHs from even the most massive star clusters.
At present our models do not make robust predictions for the spins of BHs (for an early attempt at including models of BH spins in NBODY calculations, see \citealp{Banerjee:2020bwk}).
In the absence of reliable models for BH spins, we calculate the properties of the merger remnant under two different spin models to demonstrate the uncertainty. 
We show that our conclusions are robust to these uncertainties.
We determine the merger remnant properties using the \textsc{surfinBH} \citep{Varma:2018aht} software package, which uses fits to results from numerical relativity simulations. 
Specificallly, we use the \textsc{NRSur7dq2} approximant \citep{Blackman:2017pcm}, which is appropriate for comparable mass binaries with misaligned spins. 
Using the kick velocity, we then calculate the probability of each remnant BH being retained within the cluster -- the retention probability -- by calculating the fraction of the kick distribution below the escape velocity in the core of the cluster at the time of the merger.

We first assume that both component BHs are non-spinning, as is consistent with most gravitational-wave observations \citep[e.g.][]{LIGOScientific:2018mvr,Abbott_GWTC2:2020niy,Farr:2017uvj} and predicted by stellar models incorporating strong core-envelope coupling \citep[e.g.][]{Qin:2018nuz,Fuller:2019sxi}.
In this model, we find that the final BH from the M-03 merger (section~\ref{subsubsec:M-03}) results in a remnant BH of mass 45.3\,M$_\odot$, a dimensionless spin of $0.66$ and a recoil kick of $90.7$\,km\,s$^{-1}$. The M-04a merger (section~\ref{subsubsec:M-04a}) yields a BH remnant with a mass of $22.6$\,M$_\odot$, a dimensionless spin of $0.69$ and a recoil kick of $20$\,km\,s$^{-1}$. 
The M-05 merger (section~\ref{subsubsec:M-05}) remnant has a mass of $28.8$\,M$_\odot$, a dimensionless spin of $0.62$ and a recoil kick of $130$\,km\,s$^{-1}$.
The escape velocities in the cores of the clusters these mergers occurred in were only a few km\,s$^{-1}$.
Thus, under this spin assumption, we expect that each of these remnant BHs would be ejected from their host clusters. 

Our second spin model is motivated by stellar models with weak core-envelope coupling \citep{Belczynski:2017spin}, which can lead to moderately or highly spinning BHs. 
In this model, we independently draw each of the dimensionless spins of the component BHs from a uniform distribution\footnote{This model also matches the prior distribution used to perform parameter inference on gravitational-wave signals by the LIGO/Virgo collaborations \citep[e.g.][]{Abbott_GWTC2:2020niy,Romero-Shaw:2020owr}.} between 0 and 1. 
In this spin model, the final parameters of each remnant BH are not unique, but depend upon the six parameters describing the spin vectors of the component BHs. 
We find that for the M-03 DBH merger under this spin assumption, the remnant BH mass is $45.3^{+0.4}_{-0.7}$\,M$_\odot$, the final BH spin is $0.67^{+0.14}_{-0.15}$ and the kick velocity is $1110^{+1180}_{-760}$\,km\,s$^{-1}$ (all quantities quoted are the median and upper and lower 90\% bounds).
For the M-05 merger remnant the final BH mass is $28.8^{+0.3}_{-0.5}$\,M$_\odot$, the final BH spin is $0.65^{+0.19}_{-0.16}$ and the kick velocity is $370^{+820}_{-230}$\,km\,s$^{-1}$. 
For the M-04a merger remnant, the final BH mass is  $22.6^{+0.2}_{-0.3}$\,M$_\odot$, the final BH spin is $0.69^{+0.12}_{-0.12}$ and the kick velocity is $550^{+1120}_{-410}$\,km\,s$^{-1}$.
Hence we conclude that for star clusters with parameters similar to those we study here, it is likely that most DBH merger remnants are ejected from the cluster.

In this section we have computed the final properties of BH merger remnants using a state-of-the-art model \citep{Varma:2018aht}. 
In the future, it would be desirable to calculate these properties on-the-fly, so that, in the case the remnant is ejected from the cluster, it can be removed from the simulation, and the remainder of the simulation can be performed self consistently \citep[see][]{Banerjee:2020bwk}. 
However, we note that the M-05 merger remnant did leave the cluster soon after the event owing to the recoil velocity obtained in dynamical encounters. 

\subsection{Binary black hole merger efficiencies}
\label{subsec:BBH_merger_rates}

We have evolved six base models (each with an initial mass of 3.5$\times10^4$\,M$_\odot$) until either only 500 stars remain or until a cluster age of 13\,Gyr. 
Using only this base set (M-01, M-02, M-03, M-04a, M-04b and M-05 in Table~\ref{table:NBODY_models}), we obtain a DBH in-situ merger efficiency of $4.2\times10^{-5}$\,mergers/M$_\odot$. 
Adding our five additional models (M-06, M-07, M-08, M-09 and M-10 in Table~\ref{table:NBODY_models}) which are evolved to various early-intermediate stages, 
we obtain a DBH in-situ merger efficiency of $3.2\times10^{-5}$\,mergers/M$_\odot$. 
Adding the three ex-situ mergers from M-07 (for a total of 15 mergers) 
results in a DBH merger efficiency of $4.1\times10^{-5}$\,mergers/M$_\odot$ from our all eleven models. 

We now compare our DBH merger efficiencies to previously published results.
Using Table\,1 of
\citet{Banerjee:2017}, who evolved 12 star cluster models with similar initial masses to our clusters ($\sim10^4$M$_\odot$), we calculate an in-situ DBH merger efficiency of 2.9$\times10^{-5}$\,mergers/M$_\odot$. 
Their total merger efficiency (in-situ and ex-situ) is 4.6$\times10^{-5}$\,mergers/M$_\odot$. 
From \citet{Banerjee:2020bwk}, who performed a larger suite of 65 simulations (63 models $\sim10^4$M$_\odot$ and 2 models $\sim10^5$M$_\odot$) with varying initial settings, we obtain (from their Table\,C1) an in-situ DBH merger efficiency of  4.0$\times10^{-5}$\,mergers/M$_\odot$ and total merger efficiency of 5.0$\times10^{-5}$\,mergers/M$_\odot$. 
Taking only the 63 models with $\sim10^4$M$_\odot$ initial mass, we calculate in-situ and total DBH merger efficiencies of  3.7$\times10^{-5}$\,mergers/M$_\odot$ and 4.6$\times10^{-5}$\,mergers/M$_\odot$, respectively.  
\citet{Kumamoto:2019MNRAS} finds, using two grids of simulation (with initial masses of $\sim10^{3}$ and $\sim10^{4}$\,M$_\odot$ and initial half mass radii of $\sim$0.3 and 0.5\,pc, respectively), similar DBH total merger efficiencies of $1.7$ and $4.0 \times 10^{-5}$\,mergers/M$_\odot$ (Table\,2). 
All of these studies are in broad agreement with our obtained merger efficiencies.

The DBH merger efficiency has been shown to be a function of cluster mass, with more massive globular clusters found to have higher efficiencies. 
For example, the in-cluster and total DBH merger rates obtained from \citet{RodriguezChatterjee:2016}, who present 24 models of clusters with initial masses of $\sim 10^5$\,M$_\odot$ and initial half-mass radii of 1--2\,pc
at various metallicities (see their Table\,1) are 8.0$\times10^{-5}$\,mergers/M$_\odot$ and 15.8$\times10^{-5}$\,mergers/M$_\odot$, considerably higher than for star clusters with initial masses around $10^4$M$_\odot$. 

Translating these merger efficiencies into a predicted DBH merger rate is fraught with uncertainties. 
These are primarily due to uncertainties in the cosmological formation rate of star clusters, the initial distributions of their masses and radii, and uncertainties in the cosmic star formation rate \citep[e.g.,][]{Rodriguez:2018rmd,Neijssel:2019,Santoliquido:2020bry}.
Considering all this, along with 
our relatively small set of models (Table~\ref{table:NBODY_models}), we do not attempt to calculate a volumetric merger rate.
For cosmological integration, a larger data-set covering the full parameter space of cluster initial masses, densities, metallicities and host galaxy types is necessary. 
Other studies of clusters with similar properties to the ones we study typically find BBH merger rates of 1--100\,Gpc$^{-3}$\,yr$^{-1}$
\citep[e.g.,][]{Kumamoto:2020MNRAS,Santoliquido:2020bry,Banerjee:2021}, comparable to the empirically determined BBH merger rate \citep{Abbott_Pop_GWTC2:2020gyp}.

\section{Discussion}
\label{sec:discussion}

\begin{figure}
\includegraphics[width=0.4\textwidth]{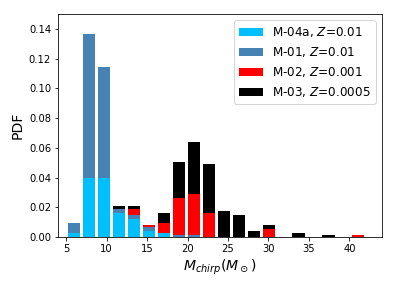}
\includegraphics[width=0.4\textwidth]{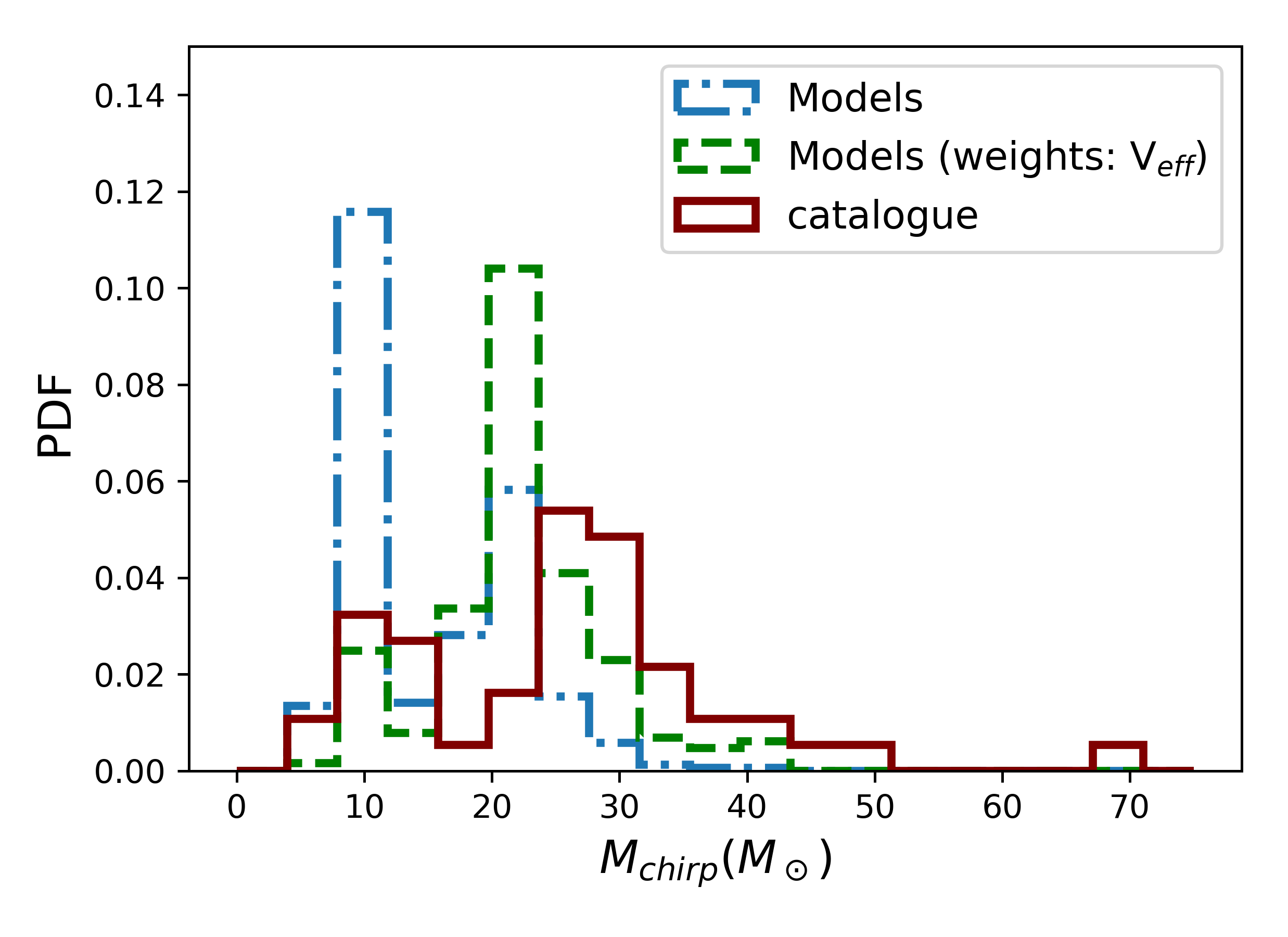}

\caption{Chirp mass distributions comparing our four low and high metallicity models (top panel) to the gravitational wave data from the first two gravitational-wave transient catalogues \citep[GWTC1+2;][]{LIGOScientific:2018mvr,Abbott_GWTC2:2020niy}, excluding the two DNSs GW170817 \citep{2017PhRvL.119p1101A} and GW190425 \citep{Abbott:2020uma} and the low significance BHNS candidate GW190426 (bottom panel). }
\label{fig:Mchirp_obs}
\end{figure}

In this paper we have investigated the effect of varying several initial parameters of a star cluster on the evolution of its global properties and its double BH population. 
In particular, we looked at the impact of varying the cluster metallicity, the initial half-mass radius, binary fraction and binary distributions through a suite of models (see Table~\ref{table:NBODY_models}).

Out of the parameters we varied, we find that the initial cluster metallicity $Z$ plays the most important role in dictating the DBH dynamics and evolution of the cluster. 
We began with a fiducial star cluster with $N = 55\,000$ stars at a metallicity $Z = 0.01$ (model M-01).
This was compared to clusters with $Z = 0.001$ and $Z = 0.0005$ (M-02 and M-03, respectively) that otherwise had the same initial conditions. 
We found that the lower-$Z$ clusters dissipate faster:   
to reach the state of only 500 M$_\odot$ remaining, which is approximately 1\% of the initial cluster mass, M-01 ($Z = 0.01$) took about 10\,Gyr while M-02 ($Z = 0.001$) and M-03 ($Z = 0.0005$) took around 6\,Gyr and 5\,Gyr, respectively, to reach the same final mass. 
This result is somewhat contradictory to \cite{Hurley:2004}, who find metal rich clusters to have the shorter lifespans by a factor of about 10\%. 
We attest this discrepancy to the key difference that we use the updated $Z$-dependant stellar wind mass-loss prescriptions from \cite{Belczynski:2010}, creating more massive BHs and thus a cluster core comprised of heavier remnants for metal-poor clusters than in the \cite{Hurley:2004} models. 
Moreover, the \cite{Hurley:2004} suite of models have no primordial binaries and start with $30\,000$ stars (0.55 times our starting $N$),  both of which are factors in determining lifetimes but should not be expected to be the primary cause of variations with metallicity.

Overall, we find that the presence of more massive double BH systems in metal-poor clusters plays the key role in setting the dynamical evolution picture and determining the life-span of these clusters.
More massive DBHs that sink to the cluster core generate more thermal energy, expanding the outer envelope of the cluster (see details in section.~\ref{sec:metallicity}) and resulting in the earlier dissolution of the low-$Z$ host cluster. 
We also conclude that metal-poor clusters, though comparatively short-lived, show slightly enhanced BH dynamical activity (see section.~\ref{sec:DBH_dynamics}).

The dissolution times of our models implies that any low mass, low metallicity clusters (with similar properties to those we study here) born more than 6\,Gyr ago would be expected to have dissolved by the present age and would not be observable in the Milky Way today.
However, globular clusters that are more massive, by at least an order of magnitude, even of low metallicity are expected to still linger on \citep[e.g. the metal-poor NGC 4372 with an estimated age of about 12.5\,Gyr, see][]{San_Roman:2015}. 
Furthermore, dynamical mergers from low-$Z$ clusters at higher redshift that have already dissolved in the past can still be observable by the current gravitational wave detectors. 

Following on from that point it is interesting to consider how the gravitational wave signatures of the DBH mergers varies with $Z$ and how that compares to the detections to date. We plot the stacked probability density function (PDF) chirp mass distribution of DBHs from two of our higher metallicity models (M-01 and M-04a, both $Z=0.01$) and two metal-poor models (M-02 and M-03, $Z=0.001$ and $0.0005$, respectively) in the upper panel of Fig.~\ref{fig:Mchirp_obs}. 
To compare our models to the observations we show in the lower panel of Fig.~\ref{fig:Mchirp_obs} the M$_\mathrm{chirp}$ distributions from i) the combined data from the four models (M-01, M-04a, M-02, M-03), 
ii) the same data weighted by the effective volume (V$_\mathrm{eff}\propto$M$_\mathrm{chirp}^{5/2}$) to account for the gravitational-wave selection effect favouring more massive sources
and iii) the current confirmed DBH LIGO/Virgo detections from GWTC-2 \citep{Abbott_GWTC2:2020niy}. 
There appears to be two peak structures at low (about 10 M$_\odot$) and higher (about 25 M$_\odot$) M$_\mathrm{chirp}$ for the observed DBH mergers, which roughly corresponds with our metal-rich and metal-poor cluster models, respectively. 
A more careful future calculation, accounting for the cluster distribution with respect to metallicity and redshift, will be able to shed more light on this mass distribution feature of the DBH observations.  

We find a total of 12 in-situ DBH mergers and three ex-situ DBH mergers for our 11 models (Table.~\ref{table:NBODY_models}). 
Across the models we notice that the initial half mass radius $R_\mathrm{h}$, through determining the initial density, plays the most important role in determining the total number of mergers in a cluster model. 
While a higher initial density produces more in-situ mergers, if the density is too high it appears that it can prohibit inspiral mergers within the cluster. 
Our models suggest a transitional density point in the range of approximately $1 - 4 \times 10^3$M$_\odot$/pc$^3$, below which in-situ mergers increase with increasing initial density and beyond which increasing density actually inhibits DBH in-situ inspiral events (see section~\ref{sec:in_cluster_mergers}). 
Furthermore, while we find that a high density can prevent DBH mergers inside the cluster, the associated active dynamical environment can instead imprint a signature of high eccentricity ($e>0.9$)  on the DBH systems ejected from the cluster.
This aids those systems in merging when outside the cluster (as shown by M-07, see section~\ref{sec:ex_situ}). 
Therefore, while a heightened rate of BH-BH interaction can either aid or hinder DBH mergers inside the cluster (section.~\ref{sec:DBH_dynamics}), highly eccentric binaries (produced by major dynamical activity) once ejected out of the cluster can have expedited mergers compared to their more circular counterparts (section.~\ref{sec:ex_situ}).

Although we find that the number of DBH mergers is unaffected by the initial metallicity, the mass of such mergers is indeed a function of $Z$, with metal-poor clusters producing more massive DBH mergers (see Section.~\ref{subsubsec:M-03}, where we discuss the production of a 47.5\,M$_\odot$ remnant in the $Z=0.0005$ model M-03). 
Combined across all of our 11 models we obtain an in-situ DBH merger efficiency of 3.2$\times10^{-5}$merger/M$_\odot$ and a total (in-situ and ex-situ) merger efficiency of $4.1\times10^{-5}$\,mergers/M$_\odot$. 

We also examined the impact of other initial properties on the evolution of star clusters. Changing the initial orbital period distribution through altering the initial semi-major axis distribution from \citet{Duquennoy:1991} to \citet{Sana:2012} reduces the number of NSs produced by the cluster (see section.~\ref{sec:black_hole_populations}) but does not significantly impact the global cluster properties, such as dissolution time, or the characteristics of the DBH populations. 
A higher initial binary fraction, however, has a consequential effect in biasing the chirp mass distribution of the DBH population towards more massive systems (section.~\ref{sec:mchirp}). 

The initial cluster density, determined by the starting half mass radius $R_\mathrm{h}$, does not significantly change the cluster global properties or evolutionary timescales for our range of models. 
On the other hand, it can have a pronounced effect on the DBH dynamics of the cluster, stimulating more exchange interactions (see section.~\ref{sec:DBH_dynamics}). 
What is interesting to note is that cluster models starting with the same initial conditions, such as the same $R_\mathrm{h}$, but with different random seeds can have non-identical evolutionary timescales due to statistical fluctuations. 
Chance encounters can create hard binaries that persistently transfer thermal energy to the cluster halo and boost the evaporation rate of the cluster. 
As a result the outcome for the cluster evolution (such as dissolution time) can hinge on the formation (or not) and subsequent effects of a few subsystems in the centre of the cluster. This statistical effect can be observed in the models as subtle differences arising in the properties of the binary populations as they evolve, 
as shown for models M-01, M-04a and M-04b in Sec.~\ref{sec:porb}.
For these models we start to see a difference in the rate of $R_\mathrm{h}$ expansion around 2\,Gyr, with greater expansion reducing the cluster escape velocity as well as allowing more stars to cross the tidal radius, which finally leads to enhanced mass-loss from the cluster and its more rapid demise.

In general the evolution of the half-mass radius of a cluster (whether expanding or contracting) depends on the varying factors of thermal expansion, self gravity, mass loss, dynamical interactions and the action of the external tidal field \citep{Fujii:2016}. 
We determine (from section.~\ref{sec:metallicity} and ~\ref{sec:DBH_dynamics}) that the mass distribution of DBHs within the cluster core plays the principal role in driving the expansion of the cluster (through generation of energy) thereby determining the cluster's lifetime, rather than the BH-BH interactions (which can have a secondary effect). This trend of the mass of the DBH population factoring into cluster heat-up and dissolution can be understood more clearly through core radius oscillations (Fig.~\ref{fig:R_Grid}) and escaping systems (Fig.~\ref{fig:esc_ejec_eva}) from the clusters of varying metallicity and initial half mass radii.

The lack of a distinct core-collapse phase for our grid of models is attributed to the sizeable ($\approx$10--20\%) primordial binary population and is in agreement with previous studies \citep[e.g.][see section.~\ref{sec:model1}]{Vesperini:1994, Chatterjee_CoreCollapse:2013}.
Observationally, the Galactic globular clusters can be sub-categorized as core-collapsed and non-core collapsed \citep{Harris:2010,McLaughlin:2005,Heggie_Hut:2003}, with about 20\% of all Milky Way globular clusters being core-collapsed \citep{Harris:2010}. While the central regions of the non-core collapsed clusters show a plateau feature in their surface brightness profile, the brightness of core-collapsed clusters continues to increase to the core. 
The importance of stellar mass BHs in restraining complete core collapse has been observed in several studies before. \citet{Morscher:2013} found the lack of evidence of the "Spitzer instability" \citep{Spitzer:1969, Kulkarni:1993}, which is the complete segregation of BHs to the cluster core and their eventual rapid evaporation. 
Instead, BH-rich clusters show no evidence of a clear core collapse phase and instead, three-body interactions such as the Kozai-Lidov mechanism involving the heavier BHs prevents complete decoupling of the cluster core and allows the lower mass BHs to intermingle with other stars, prohibiting complete mass segregation  \citep{Morscher:2013, Morscher:2015}. Previously,
modelling of BH-abundant clusters have shown the ability to halt complete core-collapse, whilst simulations of BH-poor clusters have shown a clear core-collapse phase \citep{Kremer_CoreCollapse:2019}.
Moreover, the central dense cores of massive clusters can be mis-interpreted as IMBHs, because the lack of extreme mass-segregation and less radial anisotropy can instead point towards a sizeable population of stellar-mass BHs  \citep[see for example the $\omega$ Centauri study by][]{Zocchi:2019}.

Finally, we introduced three variables --- UniquePair, Promiscuity and Survivability (defined in section.~\ref{sec:DBH_dynamics}) that allow us to quantify BH-BH interactive dynamics with ease. 
When we looked at the statistics from these variables, particularly the comparisons of Promiscuity (Fig.~\ref{fig:Promiscuity}) and Survivability (Fig.~\ref{fig:Survivability}) across the models, 
the dominance of BH dynamical activity in the denser clusters became apparent. 
In particular, these measures correlated with M-07 being the model with the highest incidence of dynamical activity, reflected in a high UniquePair number, the largest Promiscuity maximum and the lowest Survivability median (indicating that DBHs have short lifetimes in the model).

\subsection{Future improvements}
\label{subsec:future}

As discussed in section~\ref{subsec:winds_and_sne}, our models include several important improvements to the modelling of massive stellar evolution.  
That being said, there are a number of further improvements which should be made in the future to improve the accuracy of our predictions for populations of compact object binaries \citep[see also][for further discussion]{Rodriguez_MonteCarlo_NBODY:2016,Banerjee:2020bwk,Kamlah:2021}.

The masses of the BH population 
formed through stellar evolution \citep{Hurley:2000pk, Hurley:2002BSE} 
depend on the mass loss prescription assumed. 
As discussed in section.~\ref{sec:introduction}, BHs formed from very massive stars directly through stellar evolution are not expected to exist in an upper mass gap due to (P)PISN. 
Currently in our models it is possible to form BHs with mass greater than 40\,M$_\odot$ directly from stellar evolution and we do see a few of these in our metal-poor models\footnote{The DBH merger we highlight in section~\ref{subsubsec:M-03} produces a merger remnant at the lower edge of the (P)PISN mass gap, around 47.5\,M$_\odot$, but its progenitor BHs formed through stellar evolution were not mass gap objects}. 
In the future we plan to implement (P)PISN mass loss in {\tt NBODY6} using prescriptions  \citep[for example, from][]{Marchant:2016wow, Belczynski_PPSIN:2016jno, Woosley:2016hmi} as was done by \citet{Stevenson:2019rcw} in the population synthesis code COMPAS \citep{Stevenson:2017dlk, Vigna-Gomez:2018dza, Chattopadhyay:2020lff}.
Repeating our study of how cluster initial conditions impacts the properties of the DBH population, after accounting for the (P)PISN mass loss in binary and stellar evolution calculations in {\tt NBODY6} \citep[as was done in different versions of the code, see][]{Banerjee:2020bwk,DiCarlo:2020MNRAS, Kamlah:2021} can aid us understanding the LIGO/Virgo observations in the upper mass gap region (section.~\ref{sec:introduction}). 
Specifically, these studies have shown that massive, low metallicity star clusters may play a crucial role in the formation of BH merger remnants in the (P)PISN mass gap that may eventually lead to IMBH formation \citep{Banerjee:2020bwk,DiCarlo:2020MNRAS}.
We note that the majority of BHs formed in our models have masses less than 40\,M$_\odot$ and so our results will be largely unaffected by this.

We have observed in this study how massive BHs (formed primarily in metal-poor environments) can severely influence a cluster's evolution. 
Though there still remains considerable uncertainty within the field of stellar evolution modelling \citep{Dominik:2012, Vigna-Gomez:2018dza},
implementation of more updated approaches that incorporate recent developments in massive stellar \citep{Agrawal:2020} and binary evolution \citep{Postnov:2014tza} may hence affect the DBH population and consequently the global properties of the cluster. 

We use an artificial uniform supernova kick distribution for NSs and BHs as discussed in section.~\ref{subsec:winds_and_sne}. 
Differentiating various supernova channels \citep{Podsiadlowski:2004, Tauris:2013, Gessner:2018ekd}, applying more updated core-mass to remnant-mass prescriptions and more realistic natal kick prescriptions \citep{MandelMuller:2020qwb} can alter the retention fraction of the compact objects and thereby change the host cluster evolution. 
This will be particularly important for accurately modelling the formation of neutron star binaries in star clusters in future work.

As we have discussed (section.~\ref{sec:introduction}) and shown (section.~\ref{subsec:merger_remnant_properties}), the conservation of (spin and orbital) angular momentum at the time of a BH-BH merger plays an important role in determining the fate of BH merger remnants, whether they are retained in a cluster or ejected. 
Keeping this in mind, we highlight the necessity of modelling BH natal spins by accounting for their formation history (e.g. orbital evolution) and the type of supernova or direct collapse that forms the BH, as well as merger recoils in our future models.

In our models, we find that collisions and mergers between BHs and stars are common.
This can lead to the formation of massive BHs, and in some cases provides a pathway for IMBH formation \citep{Rizzuto:2021MNRAS}. 
However, it is uncertain what fraction of the mass would really be accreted by a BH in such a situation. 
The engulfing of a BH by a giant star is reminiscent of the common envelope phase in isolated binary evolution \citep{Ivanova2013,StevensonNature:2017tfq}. 
Several studies \citep[e.g.,][]{Fryer:1998bh,Schroder:2019xqq} have argued that in the event that the common envelope is not expelled, such a configuration is likely to lead to a supernova or gamma-ray burst like event, with the explosion ejecting most of the stellar mass. 
\citet{Rizzuto:2021MNRAS} explored this uncertainty in the context of IMBH formation in star clusters and found that when restricting the fraction of mass accreted by BHs, the formation of IMBHs was completely suppressed. 
We leave a more thorough exploration of the impact on the masses of the BH population to future work, noting that the fraction of mass accreted by BHs in mergers with other stars follows a prescription based approach in {\tt NBODY6/7} and can be easily varied to accommodate the changing theoretical landscape.

For most situations in dense star clusters, modelling the effect of Newtonian gravity between the $N$-bodies is sufficient, as the typical separations are  large enough, and the velocities low enough, to neglect the impact of relativistic corrections. 
One important exception is during binary-single interactions, in which the dynamics can lead to very small separations between pairs of BHs \citep{Samsing:2013kua}. 
In these situations, the instantaneous energy loss due to gravitational-wave radiation can become large enough to drive a pair of BHs to merge rapidly \citep{Hansen:1972PRD,Samsing:2013kua}, an outcome not predicted in Newtonian dynamics.\footnote{This is the gravitational equivalent of Bremsstrahlung radiation \citep[e.g.,][]{Peters:1970PRD,KovacsThorne:1978ApJ}, and interactions of this form have also been called gravitational-wave captures \citep[e.g.,][]{2018PhRvD..98l3005R}.}
These important relativistic corrections can be included using the post-Newtonian (PN) formalism (see \citealp{Will:2011nz} or \citealp{Blanchet:2014LRR} for a review).
The most important general relativistic correction to the orbital evolution in this context is the dissipative 2.5 PN radiation reaction term \citep{Peters:1964,Damour:1983PRL}.
This term is included in the binary evolution algorithm utilised within {\tt COMPAS} and 
the $N$-body code \citep{Hurley:2000sz,Hurley:2002BSE}, and was first included in modelling dynamics in the {\tt NBODY7} code by \citet{Kupi:2006}. 
Across our suite of models we have experimented with having these PN terms both on and off, finding that there is little impact on the number of mergers produced (for example model M-01 has them off while M-04a and M-04b have them on and all record only 1-2 mergers), although our models do represent a small sample size. 
Recent work has shown that the inclusion of the 2.5 PN term in detailed models of star clusters leads to the prediction that around 5\% of binary BH mergers will retain significant eccentricity ($e > 0.1$) in the LIGO band, at a gravitational-wave frequency of 10\,Hz \citep{Samsing:2017rat,Samsing:2017oij,Samsing:2018PRD,Rodriguez:2017pec,2018PhRvD..98l3005R,Banerjee:2018MNRAS,Banerjee:2020bwk,Zevin:2018kzq}.
It will therefore be important to include these terms in all of our future star cluster models when studying the formation of BH-BH binaries. 

Finally, a more thorough investigation of the parameter space of star clusters (young, open and old globulars) encompassing a broader range of metallicities, initial half-mass radii, initial masses, mass functions, and types of host galaxies is required to fully explore the formation of compact objects in all star clusters. 
Especially for binary BH mergers, massive host clusters with very low metallicity \citep[RBC EXT8][for example]{Larsen:2020Science} and high density are of special importance (as illustrated by our model M-07).
Modelling such clusters with realistic parameters can help us probe more into the interesting correlation between the evolution of star clusters and their BHs.

\section*{Acknowledgements}

We thank Sourav Chatterjee, Krzysztof Belczynski, Fabio Antonini, Sambaran Banerjee and Mark Gieles for their insightful comments and suggestions.
The authors are supported by the Australian Research Council Centre of Excellence for Gravitational Wave Discovery (OzGrav), through project number CE170100004. 
SS is supported by the Australian Research Council Discovery Early Career Research Award DE220100241.
This work made  use  of  the  OzSTAR  high  performance  computer at  Swinburne  University  of  Technology. 
OzSTAR  is funded by Swinburne University of Technology and the National Collaborative Research Infrastructure Strategy(NCRIS). 



\section*{Data Availability}

This work made use of the code \tt{NBODY6}/7 \citep{Aarseth:2003, 2012MNRAS.424..545N}, which is publicly available\footnote{\url{https://people.ast.cam.ac.uk/~sverre/web/pages/nbody.htm}}. 
The results produced for this paper are available from the author upon reasonable request.



\bibliographystyle{mnras}
\bibliography{bib} 


\bsp	
\label{lastpage}
\end{document}